\documentclass[imslayout,preprint]{imsart}

\RequirePackage[OT1]{fontenc}
\RequirePackage{amsthm,amsmath}
\RequirePackage{natbib}
\RequirePackage[colorlinks,citecolor=blue,urlcolor=blue]{hyperref}

\arxiv{arXiv:1912.00434}

\startlocaldefs
\numberwithin{equation}{section}
\theoremstyle{plain}

\newtheorem{remark}{Remark}[section]
\endlocaldefs

\usepackage{graphicx}
\graphicspath{{figures/}}
\usepackage{subcaption}
\usepackage{float}

\usepackage{hyperref}
\usepackage{lscape} 
\usepackage{xcolor} 
\usepackage{adjustbox} 
\graphicspath{{figures/}}
\usepackage{amsfonts}

\newcommand{\rulesep}{\unskip\ {\vrule width .5pt}\ }

\usepackage{geometry}
\usepackage{xr-hyper}
\begin{document}

\begin{frontmatter}
\title{Joint and individual analysis of breast cancer histologic images and genomic covariates} 
\runtitle{Joint image and genetic analysis}

\begin{aug}
\author{\fnms{Iain} \snm{Carmichael}\thanksref{m1}\ead[label=e1]{idc9@uw.edu}},
\author{\fnms{Benjamin C.} \snm{Calhoun}\thanksref{m2}\ead[label=e2]{ben.calhoun@unchealth.unc.edu}},
\author{\fnms{Katherine A.} \snm{Hoadley}\thanksref{m2}\ead[label=e3]{hoadley@med.unc.edu}},
\author{\fnms{Melissa A.} \snm{Troester}\thanksref{m2}\ead[label=e4]{troester@unc.edu}},
\author{\fnms{Joseph} \snm{Geradts}\thanksref{m3}\ead[label=e5]{jgeradts@coh.org}},
\author{\fnms{Heather D.} \snm{Couture}\thanksref{m4}\ead[label=e6]{heather@pixelscientia.com}},
\author{\fnms{Linnea} \snm{Olsson}\thanksref{m2}\ead[label=e7]{lolsson@live.unc.edu}},
\author{\fnms{Charles M.} \snm{Perou}\thanksref{m2}\ead[label=e8]{cperou@med.unc.edu}},
\author{\fnms{Marc} \snm{Niethammer}\thanksref{m2}\ead[label=e9]{mn@cs.unc.edu}},
\author{\fnms{Jan} \snm{Hannig}\thanksref{m2}\ead[label=e10]{jan.hannig@unc.edu}},
\and
\author{\fnms{J.S.} \snm{Marron}\thanksref{m2}\ead[label=e11]{marron@unc.edu}}

\runauthor{I. Carmichael et al.}

\affiliation{University of Washington\thanksmark{m1}, University of North Carolina at Chapel Hill\thanksmark{m2}, City of Hope National Medical Center \thanksmark{m3}, and Pixel Scientia Labs\thanksmark{m4}}

\address{I. Carmichael\\
Department of Statistics\\
University of Washington\\
Seattle, WA, 98195 \\
\phantom{E-mail:\ }\printead*{e1}}

\address{B.C. Calhoun\\
Department of Pathology and Laboratory Medicine\\
University of North Carolina at Chapel Hill\\
Chapel Hill, NC, 27599 \\
\phantom{E-mail:\ }\printead*{e2}}

\address{K.A. Hoadley\\
Department of Genetics\\
Lineberger Comprehensive Cancer Center\\
Computational Medicine Program\\
University of North Carolina at Chapel Hill\\
Chapel Hill, NC, 27599 \\
\phantom{E-mail:\ }\printead*{e3}}

\address{M.A. Troester\\
Department of Epidemiology\\
Department of Pathology and Laboratory Medicine\\
University of North Carolina at Chapel Hill\\
Chapel Hill, NC, 27599 \\
\phantom{E-mail:\ }\printead*{e4}}

\address{J. Geradts\\
Department of Population Sciences \\
City of Hope National Medical Center\\
Duarte, CA 91010\\
\phantom{E-mail:\ }\printead*{e5}}

\address{H.D. Couture\\
Pixel Scientia Labs\\
Raleigh, NC, 27615\\
\phantom{E-mail:\ }\printead*{e6}}

\address{L. Olsson\\
Department of Epidemiology\\
University of North Carolina at Chapel Hill\\
Chapel Hill, NC, 27599 \\
\phantom{E-mail:\ }\printead*{e7}}

\address{C.M. Perou\\
Department of Genetics\\
Department of Pathology and Laboratory Medicine\\
University of North Carolina at Chapel Hill\\
Chapel Hill, NC, 27599 \\
\phantom{E-mail:\ }\printead*{e8}}

\address{M. Niethammer\\
Department of Computer Science\\
University of North Carolina at Chapel Hill\\
Chapel Hill, NC, 27599 \\
\phantom{E-mail:\ }\printead*{e9}}

\address{J. Hannig\\
Department of Statistics\\
University of North Carolina at Chapel Hill\\
Chapel Hill, NC, 27599 \\
\phantom{E-mail:\ }\printead*{e10}}

\address{J.S. Marron\\
Department of Statistics\\
University of North Carolina at Chapel Hill\\
Chapel Hill, NC, 27599\\
\phantom{E-mail:\ }\printead*{e11}}

\end{aug}

\begin{abstract}
The two main approaches in the study of breast cancer are histopathology (analyzing visual characteristics of tumors) and genomics.
While both histopathology and genomics are fundamental to cancer research, the connections between these fields have been relatively superficial.
We bridge this gap investigating the Carolina Breast Cancer Study through the development of an integrative, exploratory analysis framework.
Our analysis gives exciting insights -- some known, some novel -- that are engaging to both pathologists and geneticists.
Our analysis framework is based on Angle-based Joint and Individual Variation Explained (AJIVE) for statistical data integration and exploits Convolutional Neural Networks (CNNs) as a powerful, automatic method for image feature extraction.
CNNs raise interpretability issues that we address by developing novel methods to explore visual modes of variation captured by statistical algorithms (e.g. PCA or AJIVE) applied to CNN features.

\end{abstract}

\begin{keyword}[class=MSC] 
\kwd[Primary ]{62H35}  
\kwd{62P10}
\end{keyword}

\begin{keyword}
Multi-view data, dimensionality reduction, image analysis, deep learning, interpretability, breast cancer histopathology, gene expression
\end{keyword}
\end{frontmatter}

\section{Introduction} \label{s:intro}

\begin{figure}[h]
\centering
\includegraphics[scale=.05]{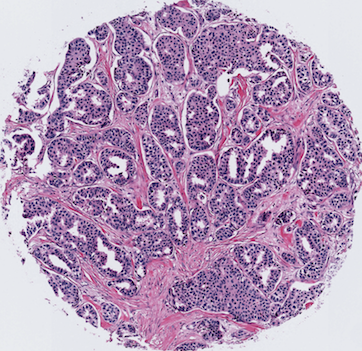}
\caption{
A lower resolution view of a \textit{hematoxylin and eosin} (H\&E) stained 1mm core from a breast cancer tissue microarray.
The darker purple color (hematoxylin) stains nuclear material such as DNA, while the pink (eosin) highlights structures such as connective tissue.
}
\label{fig:he_example}
\end{figure}

Histologic images (Figure \ref{fig:he_example}) of tissue morphology have long been utilized in treatment decisions and prognostics for breast cancer. 
For example, tumor grade is scored by evaluating tumor morphologic features and has high value in predicting outcomes for cancer cases.
Recent years have found that genomic assays can offer a second line of evidence to guide treatment and prognostics.
While both histopathology and genomics are known to be valuable, they are typically assessed separately in both clinical and research settings.
 Pathology data is almost immediately available after a tumor is excised, whereas genomic data may not arrive for many weeks.
 Efforts to integrate these two types of information are typically confined to assessing whether the pathology and the genomic data are concordant in their estimation of progression risk. 
 Unfortunately, this means that information that may be gained by understanding the interaction between histology and genomics has been largely neglected.

The primary goal of our study is to leverage both pathology and genomic data in a more concerted fashion. Our approach, based on Angle-Based Joint and Individual Variation Explained (AJIVE) \cite{feng2018angle}, allows domain experts to explore how information is shared across histopathology and genetic data.
We expect some information to be jointly shared by both data modalities e.g. some tumors may have a high mitotic index in the pathology data and show high expression of proliferation-related genes.
Other information may be contained in one modality, but not the other e.g. over-expression of a particular oncogene may not manifest morphologically, or some microenvironment features that are pronounced in images, such as mucin or adipose content, may have limited effects on gene expression.
By identifying signals that are \textit{joint} and those that are \textit{individual} (present in one data type but not the other), we provide a powerful exploratory analysis framework. 

One historical barrier to histologic image analysis is that morphologic features are often described only by qualitative features (e.g. high, medium, or low nuclear grade), which are determined by pathologists.
Extracting this information is time consuming, expensive, ignores information and is subject to inter-rater variability \citep{elmore2015diagnostic}.
Recent quantitative histopathology approaches have begun to address these issues by developing statistical tools for histological image data \citep{beck2011systematic, ash2018joint}.
For example, \textit{convolutional neural networks} (CNNs)  have proven adept at solving predictive tasks in cancer histology \citep{liu2017detecting, couture2018image, chen2019pathomic}.

Inspired by the success of CNNs for predictive tasks, we leverage CNN features for automatic image feature extraction in our analysis. 
While CNN features offer representational advantages, they are challenging to interpret and are not traditionally used for exploratory/inferential analyses.
This leads to a secondary goal of our study: developing an approach to interpret signals in the histology image data captured by CNN features (\textit{representative patch views}, see Section \ref{s:rep_patch_view}).
These representative patch views are critical for bringing different domains together (Pathology, Genomics and Epidemiology) because they help enable experts from these different domains to engage deeply with each other through our analysis. 

A limiting factor for this type of research is availability of well-annotated datasets with both digital pathology and genomic data.  When both data types are available, studies tend to be of relatively small size. However, the Carolina Breast Cancer Study (CBCS) motivating our investigation includes histologic images and RNA expression data for a large cohort (n=1,191) of breast cancer patients diagnosed with invasive breast cancer between 2008 and 2013 \citep{troester2017racial}. 

Our analysis discovers immediately interpretable joint and individual signals. The first AJIVE joint component uncovers tumor grade in the pathology and the Basal-like subtype in the genetics. The association between high grade and Basal-like tumors is well known in the breast cancer literature \citep{carey2006race}. The fact this association shows up prominently in the joint information is highly encouraging to domain experts. 

The second joint component identifies previously unknown histologic features of Luminal B tumors. Understanding the histology of Luminal B tumors is immediately relevant to clinicians because genetic based testing can be expensive and time consuming.  Histologic grade is one feature that distinguishes Luminal A and B tumors, but it has limited accuracy \cite{allott2018frequency}. Our analysis suggests that retraction artifacts (among other things) can be used by pathologists to identify Luminal B tumors. These histological features are traditionally ignored by pathologists as artifacts of the tumor processing pipeline. 

The histologic individual AJIVE components capture features of the tumor microenvironment including \textit{mucinous carcinoma}, high adipose content, and degraded tissue. These features further validate that we can accurately separate individual information from joint. The genetic individual AJIVE components contain additional genetic subtype information as well as pick up on well-known technical variation.



Section \ref{s:data} presents the data provided by CBCS as well as the patch based, CNN image features extraction approach.
Section \ref{s:rep_patch_view} presents our approach for interpreting visual modes of variation.
Section \ref{s:int_analysis} discusses the integrative analysis including an overview of AJIVE.
Section \ref{s:results} discusses the results of the joint, image individual and genetic individual AJIVE components.
Finally, Section \ref{s:conclusion} concludes with more discussion about transfer learning and exploratory analysis with deep learning.
The supplementary material provides additional results and details including: explanation of important tissue structures discussed in the results section, all figures shown to the pathologists, and more details about statistical procedures.

All correlations and AUC statistics reported in the text of the paper are statistically significant at a level of 0.05 after controlling for multiple testing with the Benjamini-Hochberg procedure \cite{benjamini1995controlling} unless stated otherwise (see Section \ref{ss:clinical_interpretation_methods}).

\subsection{Related literature} \label{ss:intro-lit}

There is a large literature on dimensionality reduction for multi-block data including classical algorithms \citep{hotelling1936relation, wold1985partial, kettenring1971canonical, yang2015non, gaynanova2017structural}.
JIVE \cite{lock2013joint} and AJIVE \cite{feng2018angle} are some of the first methods to look at both joint as well as individual modes of variation.


Interpretability in deep learning is a growing field \citep{vellido2012making, molnar2018interpretable, chen2018looks, kim2018interpretability, olah2018building, holzinger2019causability}.
We explored adapting \textit{saliency map} approaches \citep{zeiler2014visualizing, springenberg2014striving, selvaraju2017grad, sundararajan2017axiomatic, adebayo2018sanity} for interpreting the results of our analysis.
Unfortunately, none of the methods provided consistently interpretable outputs (potentially due to our use of transfer learning) and raised issues which will be explored in a follow up paper.



Deep learning based predictive analysis of histological images is a growing area \citep{komura2018machine, aeffner2019introduction} which includes tasks such as classification/regression \citep{wang2016deep, liu2017detecting, bejnordi2018using, ilse2018attention, liu2018artificial, veta2019predicting}, semantic segmentation \cite{jimenez2019deep, mahmood2019deep}, and microscope augmentation \citep{chen2018microscope}.
CNN architectures that integrate genetic (or other) information are also being explored for these predictive tasks \citep{couture2018image, srivastava2018building, mahmood2018multimodal, chen2019pathomic}.
Other studies used non-deep learning based methods to do exploratory, integrative analysis of histological and genetic data \citep{beck2011systematic, wang2013identifying, cooper2015novel}

A similar joint, exploratory analysis of breast cancer H\&E image and gene expression data was performed by \citep{ash2018joint}.
Our methods differ from theirs in a number of ways: they only examined joint signals while we examine both joint and individual signals; they used a sparse CCA while we use AJIVE;  we develop and use the RPVs for image interpretation; they trained an auto-encoder while we use transfer learning.
An important result of our paper is that even simple transfer learning effectively captures the important signals in the data.



\subsection{Software and data release}\label{ss:intro-software}

The code to reproduce the analysis in this paper can be found at \url{github.com/idc9/breast_cancer_image_analysis}.
The raw data e.g. H\&E images, gene expression data, clinical variables cannot be released publicly due to patient confidentiality concerns.
Researchers may request permission to access the raw data used in this study by visiting \url{https://unclineberger.org/cbcs/for-researchers/}.

We used many of the standard python data science libraries \citep{hunter2007matplotlib, van2011numpy, scikit2011pedregosa, mckinney2011pandas, van2014scikit, jones2014scipy, waskom2018seaborn}.
The PyTorch framework is used for all neural network computations and the pre-trained VGG16 weights are downloaded with the PyTorch vision library  \citep{paszke2017automatic}.
AJIVE computations are done with the py$\_$jive package \cite{carmichael2019pyjive} which was developed for this project.

\section{Data}\label{s:data}

\subsection{Carolina Breast Cancer Study} \label{s:data_cbcs}

The data are from the Carolina Breast Cancer Study, a population-based study of black and white women with invasive breast cancer diagnosed between 2008-2013 in North Carolina. 
Tumor blocks were collected and cores were transferred from the donor paraffin blocks to prepare tissue microarrays as well as to isolate RNA for gene expression analysis. 
The current analysis includes $n=1,191$ patients for whom both image and gene expression data were available.
Additional details about these data (e.g. descriptive statistics, tissue preparation, gene expression processing) are described in \citep{troester2017racial, allott2018frequency}.

For each patient, a pathologist reviewed a paraffin-embedded tumor block and marked the area containing the invasive carcinoma.
Then a lab technician extracted a number of circular ``cores", which were then 
transferred into a recipient TMA paraffin block and eosin (H\&E) and imaged. 
\ref{supp_viz} shows a graphical depiction of this process.
The upshot is that for each patient we have a median\footnote{Minimum of 1 and maximum of 8.} of 4 H\&E stained core images.
The images of these cores are roughly circular with an average width of approximately 2500 pixels\footnote{Min 600, max 3400.}.
An example core image is shown in Figure \ref{fig:he_example}.
It is appealing to work with cores and not the much larger \textit{whole slide images} because the cores provide more concentrated tumor cells and are more computationally tractable\footnote{The whole slide images can be of order 50,000 $\times$ 50,000 pixels or larger.}.

Pathologic evaluation of the tumor (including histologic type and grade) was based on the original whole tissue sections.  
We also compute a number of variables describing image features such as the proportion of white background and the median intensity of the background pixels.

For each patient, we have the PAM50 gene expression measurements, which are 50 genes chosen to distinguish the 5 clinically relevant, genetic subtypes (Basal-like, Luminal A, Luminal B, molecular HER2 and Normal-like) \citep{parker2009supervised}. 
The intrinsic subtype gene list was developed using genes which were consistently expressed within the tumor while minimizing the contribution of the non-tumor microenvironment; therefore the PAM50 genes do not describe the tumor microenvironment \citep{perou2000molecular}.
We also have variables derived from the PAM50 gene expression such as proliferation score and risk of recurrence score (ROR-PT)  \citep{parker2009supervised}.

CBCS provides clinically relevant immunohistochemical variables (ER status, clinical HER2 status and PR status), which are derived from routine methods used in the clinical laboratory.

The $50$ gene expression variables are centering and scaled by their standard deviation resulting in the gene expression data matrix $X^{\text{genes}} \in \mathbb{R}^{1,191 \times 50}$.

\subsection{Image processing and patch representation} \label{ss:image_pro}

In order to achieve uniform visual stain density, the raw H\&E core images are stain normalized using the procedure described in \citep{macenko2009method}.
The set of background pixels of each image (i.e. the whitespace in Figure \ref{fig:he_example}) is then estimated via the following procedure.
Each image is converted to grayscale, then a background pixel intensity threshold is estimated with weighted\footnote{Through exploratory analysis we noticed that off the shelf methods (e.g. Otsu alone) had systemic issues with images which have a high proportion of background (e.g. those with a high fat content or high mucin content). This particular combination was selected by tuning on a visual examination of the 100 images with the highest proportion background.} combination of (0.1)  Otsu's method \cite{otsu1979threshold} and (0.9) the triangle method \citep{zack1977automatic}.
The \textit{background mask} (True/False array saying whether or not a pixel is in the background) is then used for a variety of downstream tasks.
For example, using the background mask we can estimate the channel wise median background pixel and compute the proportion of background in the entire image.

\begin{figure}[H]
\begin{subfigure}[t]{.5\textwidth}
\centering
\includegraphics[width=.4\linewidth]{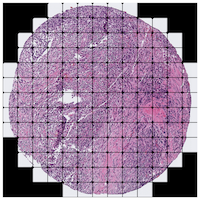}
\caption{
The patch grid for an example core.
}
\label{fig:CBCS3_HE_66088_group_1_patch_grid}
\end{subfigure}%
\begin{subfigure}[t]{.5\textwidth}
\centering
\includegraphics[width=.4\linewidth]{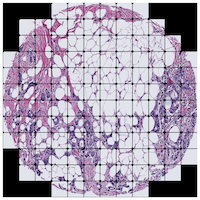}
\caption{
A core with high fat content. 
}
\label{fig:CBCS3_HE_35068_group_1_patch_grid}
\end{subfigure}
\caption{
Patch grid for two example cores, one with low fat content (a) and one with high fat content (b).
Black squares indicate patches with too much background that were excluded.
The background threshold is selected such that the thin pieces of tissue surrounding fat cells (and other structures with lots of white space such as mucin) are included.
}
\label{fig:patch_grid}
\end{figure}

Next we create a \textit{patch-based} representation of each image.
Each core image is broken into a grid of $200 \times 200$ pixel patches.
To make an even grid of patches, the image is first padded with the estimated typical background pixel so its dimensions are divisible by 200.
Using the background mask, patches which are more than 90\% background are thrown out
(Figures \ref{fig:CBCS3_HE_66088_group_1_patch_grid} and \ref{fig:CBCS3_HE_35068_group_1_patch_grid}).
The background threshold (90\%) was selected via manual inspection to be the smallest value such that patches with large amounts of fat and some tissue are still included (Figure \ref{fig:CBCS3_HE_35068_group_1_patch_grid}).

There are a total of $5,970$ core images from the $1,191$ subjects resulting in $761,767$ patches.
We estimate the channel (red, green, blue values for each pixel) mean and standard deviation from the patch dataset.
Before being input into the neural network, each pixel channel is mean centered then scaled by the standard deviation. 

\subsection{CNN feature extraction} \label{ss:cnn_feats}


After the raw images are processed, CNN features are extracted from each patch.
We use the last convolutional layer of the VGG16 architecture \cite{simonyan2014very} with an additional \textit{spatial mean pooling} layer added to the end of this architecture to average out spatial information resulting in 512 CNN features.
In other words, if the output from the original network applied to a $200 \times 200 \times 3$ pixel image is sized $H \times W \times 512$ (where 512 is the depth), the spatial mean pool will output a $512$ dimensional vector. 
The pre-trained weights of the network are downloaded from the torch vision library.
No additional fine-tuning is performed (see Section \ref{sss:transfer_learning} for discussion).


Finally, core-images are represented as an average of their patch features (again, ignoring patches which are over 90\% background).
Patients are then represented by an average of their cores.
\ref{supp_viz} provides a graphical representation of this process.
Each CNN feature is first mean centered then scaled by its standard deviation resulting in the image feature data matrix $X^{\text{image}} \in \mathbb{R}^{1,191 \times 512}$.

\section{Representative patch views} \label{s:rep_patch_view}

A key challenge for doing exploratory/inferential analysis on populations of images using deep learning is interpretability.
In this section we explain the novel, broadly applicable RPV method for interpretation of the visual signals captured by CNN (or other) features.

The RPVs assume images\footnote{In the CBCS study each subject has a number of core-images and subjects are represented as an average of their images. For exposition purposes we pretend each subject has one image in this section, however, the extension to the multi-image case is clear.} are represented via the patch based approach described in Sections \ref{ss:image_pro} and \ref{ss:cnn_feats}.
In other words, images are broken into a collection of patches; image features are extracted for each patch, and then images are represented as an average of their patch features.
Suppose we compute a loadings\footnote{We use the convention that \textit{loadings} correspond to features while \textit{scores} correspond subjects.} vector of image features (e.g. the first PCA component).
The RPVs highlight the visual features in one image associated with one end of a loadings vector (e.g. the positive end of PC 1 for the image with the most positive scores).

\begin{figure}[H]
\centering
\includegraphics[scale=.45, angle=90]{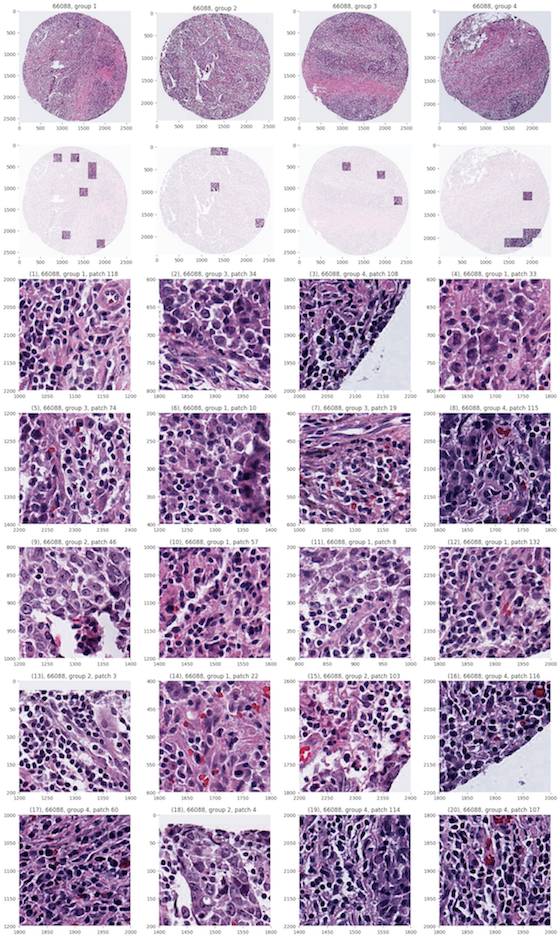}
\caption{
The full representative patch view from Figure \ref{fig:corepatches_common_1_neg_77880}.
The first column shows the 4 cores for this subject (high resolution is needed to see detail).
The 20 patches (Section \ref{ss:patch_based_localization}) in the last 5 columns are representative of the visual features associated with one extreme of a mode of variation.
The second column shows the patches' locations on the cores.
Lymphocytes show up prominently in these patches.
Some images (1, 8, 11, 18) show predominantly tumor cells, and others (10, 14, 15) show a roughly even mixture of lymphocytes and tumor cells.
}
\label{fig:corepatches_common_1_neg_66088}
\end{figure}

Figure \ref{fig:corepatches_common_1_neg_66088} shows an example RPV for one subject (with the most negative joint scores) for the negative end of AJIVE joint component 1.
The leftmost column shows the four cores for this subject. 
The rightmost five columns show the top 20 patches for the negative end of the first joint component from the \textit{patch based localization} approach described below in Section \ref{ss:patch_based_localization}.
The second column shows the location on each core of the top 20 patches.
The RPVs are multi-scale in the sense that they give insights at both the core level and the patch level.

\subsubsection{Patch based localization} \label{ss:patch_based_localization}

Here we describe the general approach used to select the representative patches for the RPVs described in the above section.
Patch based localization is accomplished by projecting patches onto the loadings vector.
Main ideas are illustrated below in the context of PCA, but this approach is applicable to other methods (e.g. AJIVE, linear regression, linear classification, etc).


Let $n$ be the number of subjects (images) in the dataset, $m_i$ be the number of patches for the $i$th image, $p_{ij}$ be the $j$th patch for the $i$th subject, $f: \text{image} \to \mathbb{R}^{k}$ the feature extraction function outputting $k$ features.
Also let $z_{ij} := f(p_{ij})$ be the features for patch $p_{ij}$,  $z_i := \frac{1}{m_i} \sum_{j=1}^{m_i} z_{ij}$ be the average patch features for the $i$th subject, $Z \in \mathbb{R}^{n \times k}$ be the image feature dataset, and $\widetilde{Z} \in \mathbb{R}^{n \times k}$ be $Z$ after processing (e.g. centering and scaling). 
Let $\mathbf{v} \in \mathbb{R}^k$ be a loadings vector and $\mathbf{u} = \widetilde{Z}  \mathbf{v} \in \mathbb{R}^n$ the scores vector\footnote{We assume that the scores are the projection of the data onto the loadings vector.} computed from $\widetilde{Z}$ (e.g. PC component 1).

Consider the positive end of this component and let $i^*$ be the index of a particular image. 
We perform \textit{patch based localization}\footnote{We use the term localization because this method helps identify which regions in the image are playing an important role in the given mode of variation.} by projecting every patch of subject $i^*$ onto $\mathbf{v}$. 
In detail, let $\widetilde{z}_{i^*, j}$ be the features of the $j$th patch for subject $i^*$ after the processing.
Let $s_{i^*, j} := \widetilde{z}_{i^*, j}^T \mathbf{v}$ be the scores of this patch for $j = 1, \dots, m_{i^*}$.
Now let $j_{(1)}, \dots, j_{(20)}$ be the indices of the patches with the 20 most positive patch scores (i.e. $s_{i^*, j_{(1)}}  \ge s_{i^*, j_{(2)}} \dots$).
We call these the \textit{representative patches} for the positive end of this component.


\section{Integrative analysis}\label{s:int_analysis}

This section gives an overview of the integrative image and genetic analysis.
We assume image features for each patient have been extracted as described in Section \ref{s:data_cbcs}.
The gene expression data have been processed as in \cite{troester2017racial, allott2018frequency} with additional mean centering and scaling.
The first subsection describes the AJIVE analysis and the section subsection describes the pathology review process for interpreting the AJIVE image modes of variation.

\subsection{Angle-based joint and individual variation explained} \label{ss:ajive}

AJIVE is a statistical feature extraction/dimensionality reduction algorithm for multi-block data \citep{feng2018angle}.
The goal of AJIVE is to find joint signals, if any exist, which are common to all data blocks as well as individual signals which are specific to each block, if they exist.
Here we give a brief overview of AJIVE for two data blocks.

Consider two data blocks $X \in \mathbb{R}^{n \times d_x}$, $Y \in \mathbb{R}^{n \times d_y}$ on the same set of $n$ observations.
AJIVE estimates what variation is \textit{joint} to both data blocks as well as what variation is \textit{\text{individual}} to each block.
In particular, each matrix is decomposed into a sum of \textit{joint}, \textit{\text{individual}}, and \textit{error} terms,
\begin{align*}
X = J^x + I^x + E^x \text{  and }Y  = J^y + I^y + E^y
\end{align*}
while imposing the following constraints
\begin{itemize}
\item $\text{col-span} (J^x) = \text{col-span} (J^y)  := \mathcal{J} \subseteq \mathbb{R}^n$
\item $\text{col-span} (J^x) \perp \text{col-span} (I^x)$ and $\text{col-span} (J^y) \perp \text{col-span} (I^y)$
\item $\text{col-span} (I^x) \cap \text{col-span} (I^y) = \emptyset$
\end{itemize}
All subspaces live in $\mathbb{R}^n$ where $n$ is the number of observations.
The two joint matrices span the same \textit{joint subspace}, $\mathcal{J}$.
The two individual matrices span subspaces which are orthogonal to the joint subspace.
We refer to the rank of the joint subspace as the \textit{joint rank}, $R_J$, and the rank of the two individual subspaces as the $X$ and $Y$ individual ranks, $R_x, R_y$.

The mechanics of AJIVE are outlined below for the case of $B=2$ data blocks\footnote{The original paper describes the procedure some what differently, but this description is equivalent.}.
The properties of the common normalized scores discussed below follow from the fact they are the subspace flag mean of the PCA scores subspaces \cite{draper2014flag}.
We use a different estimate of the block common loadings, $L^x, L^y$ than in the original paper. 
One of the key statistical procedures in AJIVE is to estimate the joint rank\footnote{This is accomplished by estimating which principal angles between $\text{col-span}(U_{\text{init}}^x) $and $\text{col-span}(U_{\text{init}}^y)$ are smaller than random in an appropriate sense.} which is achieved using the Wedin bound and the random direction bound detailed in \citep{feng2018angle}.

\begin{enumerate}

\item \textbf{Initial signal extraction}: 
Estimate low rank PCAs of $X, Y$ with ranks $r_{\text{init}}^x, r_{\text{init}}^y$  (e.g. selected by inspecting the PCA scree plots).
Denote this initial PCA of $X$ by $U_{\text{init}}^x , D_{\text{init}}^x, V_{\text{init}}^x$ where $U_{\text{init}}^x \in \mathbb{R}^{n \times r_{\text{init}}^x}$,  $V_{\text{init}}^x \in \mathbb{R}^{d_b \times r_{\text{init}}^x}$.
Similarly for $y$.

\item \textbf{Signal space extraction}: 
Perform CCA on the PCA scores, $U^x_{\text{init}}, U^y_{\text{init}}$.
Using the \textit{random direction bound} and the \textit{Wedin bound} estimate the CCA rank, $R_J$.
Let $S^x, S^y \in \mathbb{R}^{n \times R_J}$ be the matrices whose columns are the x/y CCA scores with unit norm.
Let $W^x \in \mathbb{R}^{ r_{\text{init}}^x \times R_J}, W^y \in \mathbb{R}^{ r_{\text{init}}^y \times R_J}$ be the matrices whose columns are the CCA x/y loadings.
Let $C \in \mathbb{R}^{n \times R_J}$ be the \textit{common normalized scores} which have the property of being proportional to the average of the $x/y$ CCA scores.
In other words, the $j$th column of $C$ is unit norm and is proportional to the average of the j$th$ columns of $S^x$ and $S^y$.
Additionally the common normalized scores are orthonormal i.e. $C^TC = I_{R_J}$.
Finally, let\footnote{Note the $j$th column of $V^x$ is equivalent to the rank $r_{\text{init}}^x$ principal components regression coefficient of the $j$th column of the common normalized scores, $C$, regressed on the $X$ matrix. } $L^x := V_{\text{init}}^x D_{\text{init}}^{x^{-1}} W^x \in \mathbb{R}^{d_x \times R_J}$ be the x-\textit{common loadings}.
Similarly for $y$.

\item  \textbf{Signal space extraction}: 
Let $J^x := CC^TX$ be the estimated joint matrix (i.e. projection onto the joint subspace).
Let $\tilde{I}^x := (I - CC^T)X \in \mathbb{R}^{n \times d_x}$ (i.e. projection onto the orthogonal complement of the joint subspace).
Let $R_x$ be the number of singular values of $\tilde{I}^x$ above the threshold discussed in Section 2.4 of \cite{feng2018angle} and let $I^x  \in \mathbb{R}^{n \times d_x}$ be the rank $R_x$ SVD approximation of $\tilde{I}^x$.
We denote the PCA of the individual matrix $I^x$ by $U_{\text{individual}}^x \in \mathbb{R}^{n \times R_x}$, $D_{\text{individual}}^x  \in \mathbb{R}^{R_x \times R_x}$, $V_{\text{individual}}^{x}  \in \mathbb{R}^{d_x \times R_x}$ which is also of interest.
Similarly for $y$.
\end{enumerate}

The outputs of interest in this paper are the following
\begin{itemize}
\item The joint rank, $R_J$.
\item The \textit{common normalized scores}, $C \in \mathbb{R}^{n \times R_J}$.
\item The \textit{common loadings}\footnote{These were not given names in the \cite{feng2018angle} and were computed slightly differently.}, $L^x \in \mathbb{R}^{d_x \times R_J}, L_y \in \mathbb{R}^{d_y \times R_J}$.
\item The $U_{\text{individual}}^x \in \mathbb{R}^{n \times R_x}$ and $V_{\text{individual}}^x \in \mathbb{R}^{d_x \times R_x}$, which are referred to as the \textit{block specific, individual scores and loadings}. Similarly for $y$.
\end{itemize}

The common loadings, $L^x, L^y$ are different than those in \citep{feng2018angle}.
The loadings computed here are the loadings such that $XL^x + YL^y \propto C$ i.e. the average of the resulting scores are proportional to the common normalized scores.
Computing the loadings in this way ensures that they incorporate joint information only.
\begin{remark}
It can be checked that the random direction bound is equivalent to the classical Roy's largest root test CCA rank selection method \citep{johnstone2008multivariate}.
\end{remark}

\subsubsection{AJIVE analysis of CBCS data} \label{sss:ajive_cbcs}

The only variables used in the AJIVE analysis are the $512$ CNN image features and expressions for 50 genes from PAM50; the other variables are used to interpret the AJIVE results.
The initial signal ranks are 81 (image features) and 30 (genes) and were selected by inspection of the difference of the log-singular values and airing on the side of picking too high a rank.
AJIVE estimates a joint rank of 7, image individual rank of 76 and genetic individual rank of 25.
The AJIVE diagnostic plot, detailed in \cite{feng2018angle}, is provided in Section \ref{app:ajive_diagnostics}.

\subsection{Pathology review of images} \label{ss:pathology_review}

In close collaboration with pathologists (B.C. and J.G.), we reviewed the first three joint and image individual components at two levels of granularity.

In the first approach, which we refer to as \textit{global sort}, all core images are reviewed in sequence after sorting by the patient scores. 
Joint components are sorted by common normalized scores, $C$, and individual components are sorted by block specific scores, $U^{\text{image}}_{\text{\text{individual}}}$ (see Section \ref{ss:ajive}).
After sorting, the images are reviewed in sequence (e.g. from the negative to the positive end) to explore the visual signals captured by a given component.
The benefits of the global sort method are i) a large number of images are inspected  ii) we get a sense of the high level changes\footnote{In a preliminary analysis where image patches with a large amount of background were not excluded (see Section \ref{ss:image_pro}), the global sort method on the first few principal components revealed that the primary modes of variation in the data are driven by the raw amount of background.
This motivated the exclusion of patches with too much background.} as we move along a component from the extreme negative to the middle then to the extreme positive end and iii) we can see if the trends found in the RPVs (see next paragraph) hold broadly for the entire component.
The downsides of this method are that it is time intensive and does not provide explicit information about what visual signals are important in a given image.
The H\&E images are quite large and complex and finding patterns across a set of images is challenging.

The RPV approach developed in Section \ref{s:rep_patch_view} extracts more fine-grained information at the patient level.
The RPVs of the 15 most negative and 15 most positive subjects are inspected for each component.
The RPVs are created with the common loadings $L^{\text{image}}$ for the joint components and the block specific individual loadings $V^{\text{image}}_{\text{\text{individual}}}$ (see Section \ref{ss:ajive}).
The number 
 15 was selected to balance showing ``enough" information without taking too much time.
The RPVs have the benefit of highlighting a more focused set of visual patterns.

Tables \ref{tab:path_review_joint} and \ref{tab:path_review_image_indiv} display the pathologist's observations based on the RPVs at each end of each component.
Each column summarizes the pathologist's impression of a clinically relevant histological feature.
The \textit{homogeneous} column indicates whether or not there appeared to be a consistent pattern across the reviewed RPVs.
The global sort review shows these trends hold for more than just the 15 most extreme images.
These observations are key to understanding the connections between the pathology and the genetics.

\section{Results}\label{s:results}

This section discusses the results for the joint AJIVE components (Section \ref{ss:results_joint}), the image individual (Section \ref{ss:results_image_indiv}) and genetic individual  (Section \ref{ss:results_genetic_indiv}).
For the sake of time -- both the readers' and the pathologists' -- we focus on the top 3 components from each of the joint, image individual and genetic individual.

The pathology review of the images from the joint and image individual components is described in Section \ref{ss:pathology_review}.
While the pathologist reviewed the full RPVs (Figure \ref{fig:corepatches_common_1_neg_66088}), only mini-RPVs (e.g. Figure \ref{fig:corepatches_common_2_neg_67309}) displaying 8 patches are shown below in the text of the paper.
The full RPVs shown to the pathologists, all AJIVE genetic loadings, and all clinical data comparisons are provided in \ref{supp}.
The methodology for clinical data comparisons (e.g. multiple testing control) is discussed in Section \ref{ss:clinical_interpretation_methods}.

\subsection{Joint image and genetic information} \label{ss:results_joint}
Table \ref{tab:path_review_joint} summarizes the pathologist's observations based on the RPVs of the first three joint components. 
Table \ref{tab:er_her2_joint} shows the association between immunohistochemical status (ER and clinical Her2) and the first three joint components.

\begin{table}[H]  
\centering
\begin{adjustbox}{width=1\textwidth}
\begin{tabular}{|l|l|l|l|l|l|l|l|l|l|}
\hline
component & end & homogeneous & \begin{tabular}[c]{@{}l@{}}tumor\\ cellularity\end{tabular} & \begin{tabular}[c]{@{}l@{}}tubule\\ formation\end{tabular} & \begin{tabular}[c]{@{}l@{}}nuclear\\ grade\end{tabular} & \begin{tabular}[c]{@{}l@{}}adipocytic\\ stroma\end{tabular} & \begin{tabular}[c]{@{}l@{}}collagenous\\ stroma\end{tabular} & lymphocytes & necrosis \\ \hline

1         & positive & no          & low               & yes              & 1, 2          & yes               & yes                & no          & no       \\ \hline
          & negative & yes         & high              & no               & 3             & no                & limited            & yes         & yes      \\ \hline
2         & positive & no          & variable          & yes              & 3             & focal             & yes                & few         & no       \\ \hline
          & negative & yes         & moderate/high     & yes              & 2             & focal             & yes                & no          & no       \\ \hline
3         & positive & no          & variable          & yes              & 3             & yes               & limited            & yes         & no       \\ \hline
          & negative & yes         & moderate/high     & no               & 3             & no                & yes                & no          & no       \\ \hline
\end{tabular}
\end{adjustbox}
\caption{
A pathologist's summary of important clinical features based on the RPVs of the 15 most extreme subjects for each end of the first three joint components.
The ``homogeneous" column indicates whether or not the patterns shown in the RPVs were consistent across the 15 subjects.
}
\label{tab:path_review_joint}
\end{table}

\begin{table}[H]
\centering
\begin{tabular}{|l|l|l|}
\hline
Component & ER status & Clinical HER2  status\\ \hline
1 & 0.883 ($+$) & 0.558 \\ \hline
2 & 0.752 ($-$) & 0.617 ($+$)\\ \hline
3 & 0.551 & 0.777 ($-$)\\ \hline
\end{tabular}
\caption{
AUC scores for two immunohistochemical (IHC) variables, ER status and clinical HER2 status (positive vs. negative classes), for first three joint components based on AJIVE common normalized scores.
All six of of these comparisons are statistically significant.
The sign in parentheses indicates which end of the component the IHC positive status cluster on if there was a clear visual separation in the histogram (see \ref{supp}).
For example, ER status positive tumors cluster on the positive end of component 1.
}
\label{tab:er_her2_joint}
\end{table}

\subsubsection{First AJIVE joint component} \label{ss:results_joint_1}
We initially consider the negative and positive extremes of the first joint component separately.

\begin{figure}[H]
\centering
\includegraphics[scale=.5]{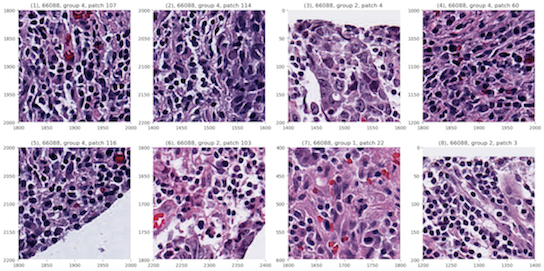}
\caption{
Several representative patches from the most negative subject of joint component 1.
The dark, round, purple structures are lymphocyte nuclei; these patches show dense tumor infiltrating lymphocytes which are characteristic of high central grade tumors.
The third images in both rows show a significant number of tumor cells, mixed with some lymphocytes.
This tumor is a genetically Basal-like tumor like most of the other tumors on the negative end of this component (Figure \ref{fig:common_1_scores_v_pam50}).
}
\label{fig:mini_corepatches_common_1_neg_66088}
\end{figure}

\begin{figure}[H]
\centering
\includegraphics[scale=.5]{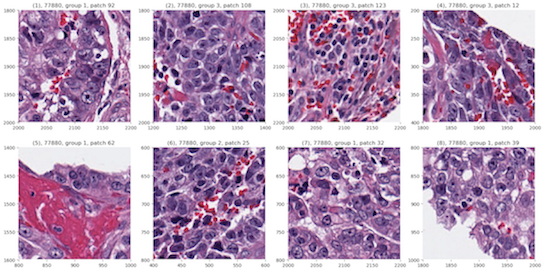}
\caption{
The third most negative subject from joint component 1.
The patches selected for the RPV of this tumor show high nuclear grade cells which are large and irregularly shaped.
These are also characteristic of high grade tumors.
This tumor is also genetically Basal-like.
}
\label{fig:corepatches_common_1_neg_77880}
\end{figure}

From the pathology perspective, two distinct visual patterns show up in the negative end of the first joint component (Figures \ref{fig:mini_corepatches_common_1_neg_66088} and \ref{fig:corepatches_common_1_neg_77880}).
Section \ref{app:tumor_structure_examples} has a brief explanation of the various tumor structures which are relevant to this paper.
The first pattern is dense \textit{tumor infiltrating lymphocytes} (TILs) and is illustrated in Figure \ref{fig:mini_corepatches_common_1_neg_66088} which shows the RPV of the most negative subject of the first joint component.
The smaller cells which have hyperchromatic round nuclei and relatively scant cytoplasm (i.e dark, round, purple structures), are lymphocytes.
In particular types of breast cancer, TILs in the intratumoral stroma are associated with prognosis and may be associated with response to immunomodulatory therapy \citep{wein2017clinical}.

The second visual pattern in the negative end of the first joint component is dense, \textit{high nuclear grade tumor cells} and is illustrated in  Figure \ref{fig:corepatches_common_1_neg_77880}.
Nuclear grade describes how abnormal the tumor cells look: ``low grade" means the tumor cells look similar to regular cells (``well-differentiated") and ``high grade" means the tumor cells look markedly abnormal (``poorly-differentiated") e.g. are enlarged and irregularly shaped \citep{rosen2001rosen}.

\begin{figure}[H]
\centering
\includegraphics[scale=.5]{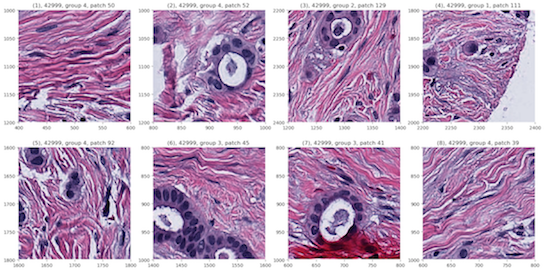}
\caption{
The subject with the most positive scores.
These RPV patches are comprised primarily of normal breast tissue and contain few tumor cells.
The eosinophilic, fibrillar (light pink, stringy) tissue is collagenous stroma which is found in normal breast tissue while the white circles surrounded by purple dots are neoplastic ducts.
This tumor is genetically Luminal A.
}
\label{fig:corepatches_common_1_pos_42999}
\end{figure}

On the positive end of the first joint component, the pathology review shows subjects whose cores contain mostly normal breast tissue i.e. little tumor tissue.
This pattern is illustrated by Figure \ref{fig:corepatches_common_1_pos_42999}, which shows the subject with the most positive scores for the first joint component.
These patches contain few tumor cells and are mostly normal breast structures such as \textit{collagenous stroma} (the light pink, stringy tissue) and ducts.

\begin{figure}[H]
\begin{subfigure}[t]{.32\textwidth}
\centering
\includegraphics[width=1\linewidth]{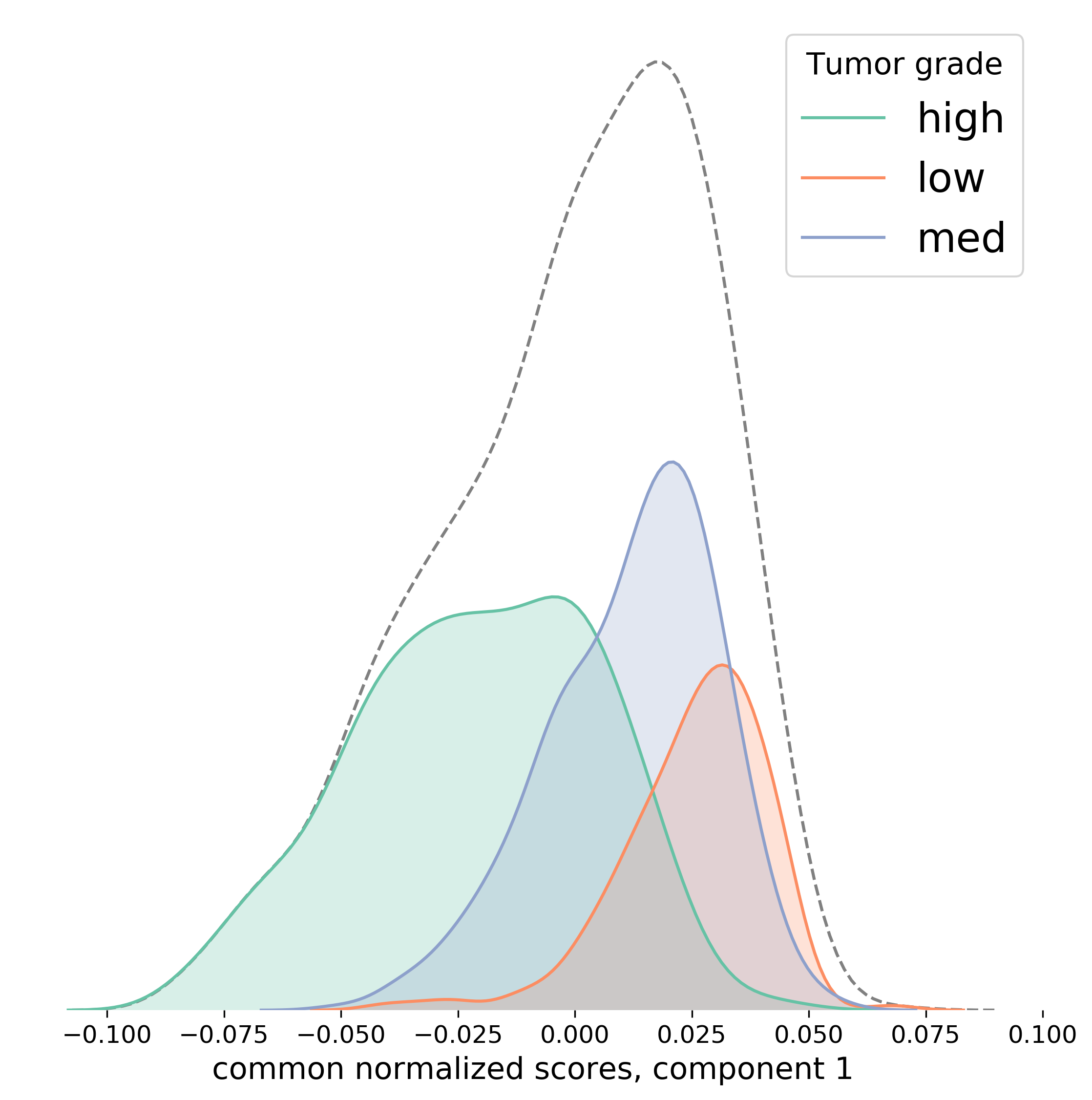}
\caption{
Tumors on the negative end tend to be high grade while those on positive end tend to be low grade.
}
\label{fig:common_1_scores_v_cgrade}
\end{subfigure}%
\hfill
\begin{subfigure}[t]{.32\textwidth}
\centering
\includegraphics[width=1\linewidth]{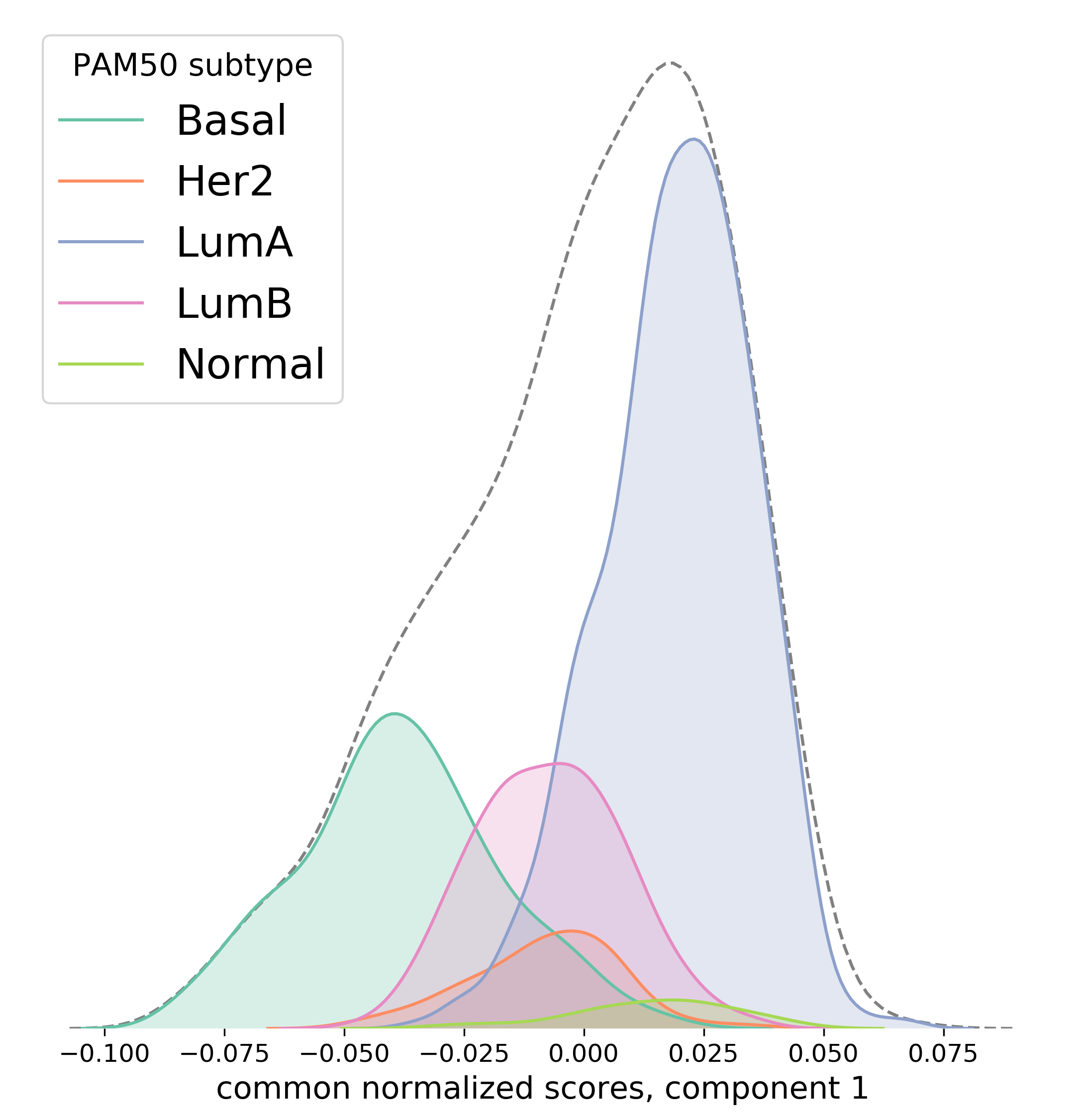}
\caption{
The Basal-like class is on the negative end of the scores, the HER2/Luminal B classes are in the middle and the Normal/Luminal A classes are on the positive end.
}
\label{fig:common_1_scores_v_pam50}
\end{subfigure}%
\hfill
\begin{subfigure}[t]{.32\textwidth}
\centering
\includegraphics[width=1\linewidth]{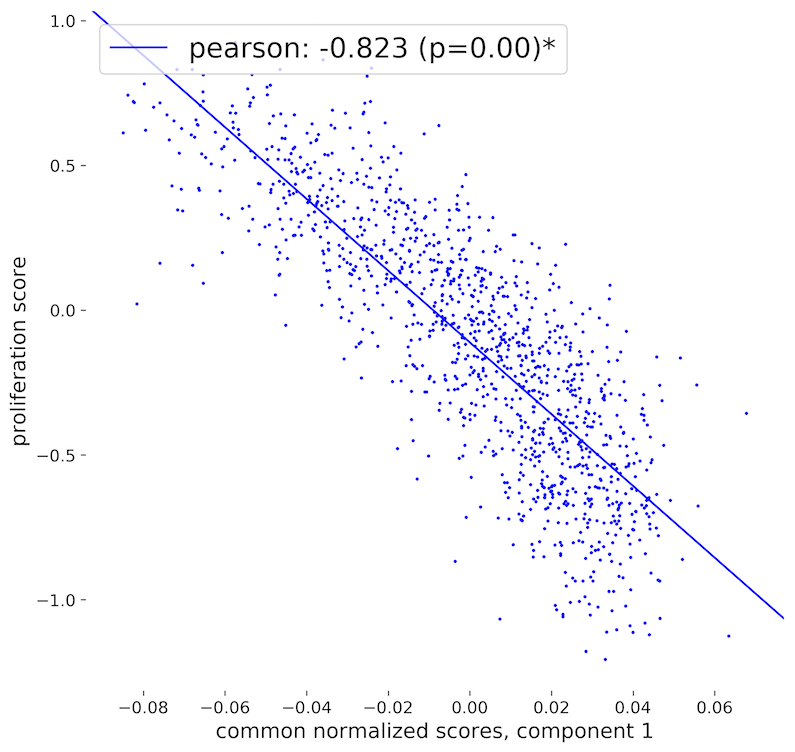}
\caption{
Proliferation score is strongly associated with the common normalized scores with a Pearson correlation of -0.82.
}
\label{fig:common_1_scores_v_proliferation}
\end{subfigure}
\caption{
Joint component 1 scores vs. PAM50 molecular subtype (\ref{fig:common_1_scores_v_pam50}), central grade (\ref{fig:common_1_scores_v_cgrade}) and proliferation score (\ref{fig:common_1_scores_v_proliferation}).
The x-axis in each plot shows the subjects' common normalized scores.
These figures tell a complementary story.
For example, Basal-like tumors tend to be more aggressive, high grade and have a high proliferation score.
}
\label{fig:common_1_scores_v}
\end{figure}

The first joint component is related to histopathological features including \textit{tumor grade} and \textit{histological type} (ductal vs. lobular). 
For example, Figure \ref{fig:common_1_scores_v_cgrade} shows that high \textit{grade} tumors cluster on the negative end of the first joint component while low grade tumors cluster on the positive end (AUC = 0.945).
Tumor grade incorporates cellular differentiation and other architectural features as an indicator of aggressiveness \citep{elston2002pathological}.
This first component is also statistically significantly related to histological type with \textit{ductal}  on the negative end and \textit{lobular} on the positive end (AUC = 0.785). 


\begin{figure}[H]
\centering
\includegraphics[scale=.38]{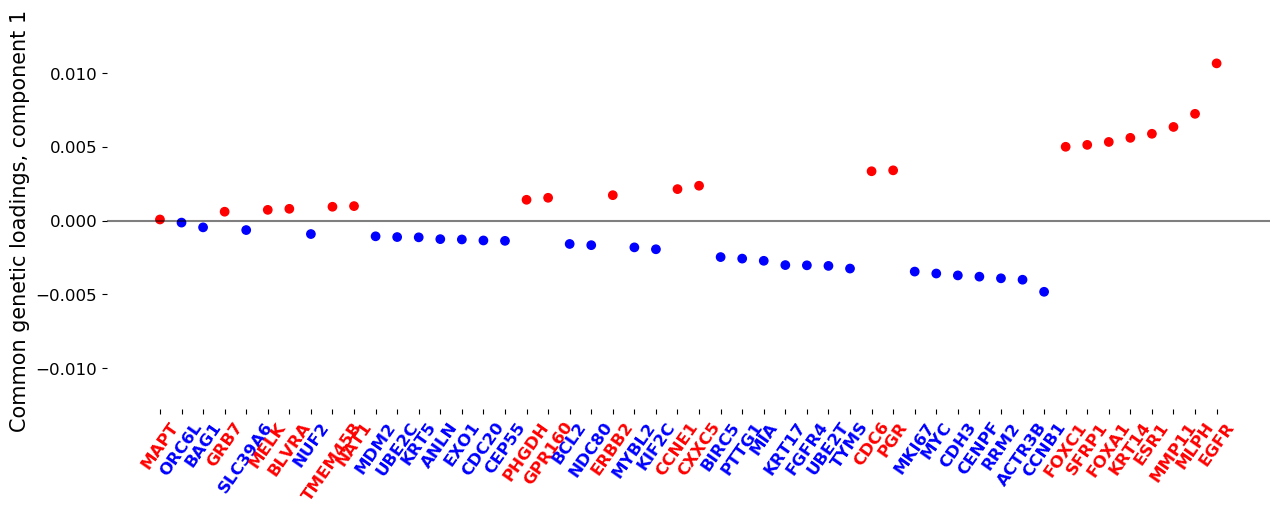}
\caption{
PAM50 loadings vector from joint component 1.
Several of the top negative genes are associated with high tumor cellularity and typically have low expression levels in normal breast tissue (consistent with Figures \ref{fig:corepatches_common_1_neg_77880} and \ref{fig:mini_corepatches_common_1_neg_66088}).
Several of the top positive genes tend to have high expression levels in normal breast tissue (consistent with Figure \ref{fig:corepatches_common_1_pos_42999}).
}
\label{fig:common_loadings_comp_1_intro}
\end{figure}

From the genetics perspective, the first joint component strongly tracks the \textit{proliferation score} as well as the contrast between Basal-like vs. Luminal A tumors.
Figure \ref{fig:common_loadings_comp_1_intro} shows the PAM50 joint loadings vector for the first component.
Several of the top negative genes (e.g. CCNB1, CENPF, MYC, MKI67) are associated with high tumor cell proliferation and tend to have low expression levels in normal breast tissue.
Several of the top positive genes (e.g. MLPH, MMP11) tend to have high expression levels in normal breast tissue.
Note that FOXC1 is highly expressed in both basal-like and normal-like breast myoepithelium.

Figure \ref{fig:common_1_scores_v_pam50} shows that Basal-like tumors cluster on the negative end of the first joint component, molecular HER2 and Luminal B cluster in the middle while Luminal A and normal tumors cluster on the positive end.
Note the AUC score for Basal-like vs. Lum A is 0.984 which is quite high. 
Luminal B and molecular HER2 are separated from Basal-like (AUCs of 0.886 and 0.876).
The separation indicates that this joint component is distinguishing more subtle histopathological and molecular features beyond proliferation and cellularity.
Figure \ref{fig:common_1_scores_v_proliferation} shows a strong, negative correlation between the first joint component scores and the \textit{proliferation score}, which is a genetic measure indicative of how fast tumor cells grow \citep{whitfield2002identification}.

Strikingly, the first joint component almost perfectly separates ROR-PT, which is a combined genetic and pathology based \textit{risk of recurrence score} \citep{parker2009supervised}.
Patients with high ROR-PT are clustered on the negative end while patients with a low ROR-PT are clustered on the positive end with an AUC of 0.999.

In addition to genetic phenotypes measured by RNA expression data as just discussed, we also have immunohistochemistry (IHC) data, a surrogate measure of RNA subtypes and the most common way of classifying tumors in a clinical setting.
From the IHC perspective, the first joint component is strongly related to ER status and weakly related to clinical HER2 status (see Table \ref{tab:er_her2_joint}).
Clinical ER negative tumors cluster on the negative end of this component with an AUC of 0.883.

In this first joint component, the pathology and genetics tell complementary stories that are familiar to breast cancer experts.
The data raise the possibility that this joint component separates tumors based on one or more histologic features associated with tumor grade.
These features could include aspects of nuclear atypia (i.e increased nuclear size, irregular shape, altered chromatin pattern, multiple nucleoli) which are reflected in the nuclear grade.
Tumors with a high combined histologic grade also tend to be more cellular and show less tubule or gland formation as compared to low-grade tumors.

From the genetics perspective, Basal-like tumors are on the negative end, molecular HER2/Luminal B tumors are in the middle, and Luminal A/Normal like tumors are on the positive end.
The joint scores are strongly negatively correlated with the proliferation score.
The negative genes in Figure \ref{fig:common_loadings_comp_1_intro} are predominantly proliferation regulated genes; however, we note several of the positive genes are often considered basal-specific genes.
These genes are also expressed in normal myoepthelieum and are representative of the normal ducts still observed within slides of the low grade tumors \citep{livasy2006phenotypic, heng2017molecular}. 


Aggressive tumors tend to have high tumor cellularity and little benign tissue. 
In less aggressive tumors, there is typically more normal breast tissue.
Basal-like tumors tend to be more aggressive and are generally associated with  high tumor grade, ER negativity, ductal histology, and high proliferation score \citep{livasy2006phenotypic, troester2017racial, williams2019differences}.
Luminal B and molecular HER2 tumors tend to be moderately aggressive.
Luminal A and Normal like tumors are less aggressive and it is known these tend to be low grade.

It is promising that this mode of variation turned up in the first joint component.
These connections between the underlying genetic drivers and the pathological impressions have both geneticists and pathologists excited about the potential of AJIVE to quantitatively integrate these different aspects of cancer.



\subsubsection{Joint component 2} \label{sss:results_joint_2}

\begin{figure}[H]
\begin{subfigure}{.48\textwidth}
\centering
\includegraphics[width=1\linewidth]{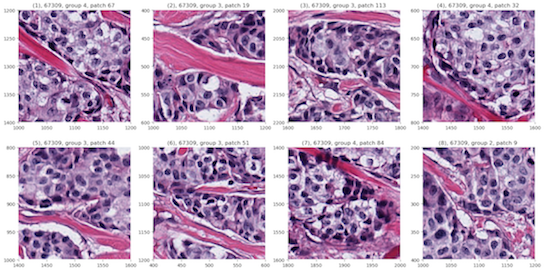}
\caption{
}
\label{fig:corepatches_common_2_neg_67309}
\end{subfigure}%
\rulesep
\begin{subfigure}{.48\textwidth}
\centering
\includegraphics[width=1\linewidth]{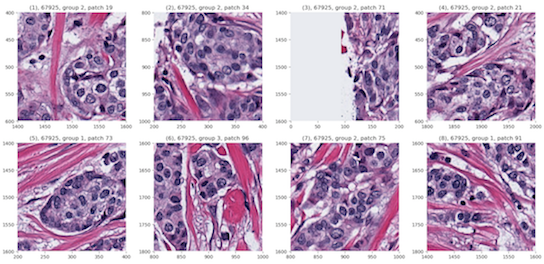}
\caption{
}
\label{fig:corepatches_common_2_neg_67925}
\end{subfigure}
\caption{
Two tumors from the negative end of joint component 2.
Both RPVs show collagenous stroma (light pink, stringy tissue) surrounded by moderate nuclear grade tumor cells and moderate tumor cellularity.
The tumor in (a) is genetically Luminal B and the tumor in (b) is Luminal A.
}
\label{fig:joint_2_neg}
\end{figure}

From the pathology perspective, the tumors on the negative end of joint component 2 show mostly collagenous stroma surrounded by moderate nuclear grade tumor cells. 
Figure \ref{fig:joint_2_neg} shows the mini-RPVs of two subjects from the negative end of the second joint component.
The positive end of this component was not homogeneous (Table \ref{tab:path_review_joint}).

\begin{figure}[H]
\begin{subfigure}[t]{.39\textwidth}
\centering
\includegraphics[scale=.4]{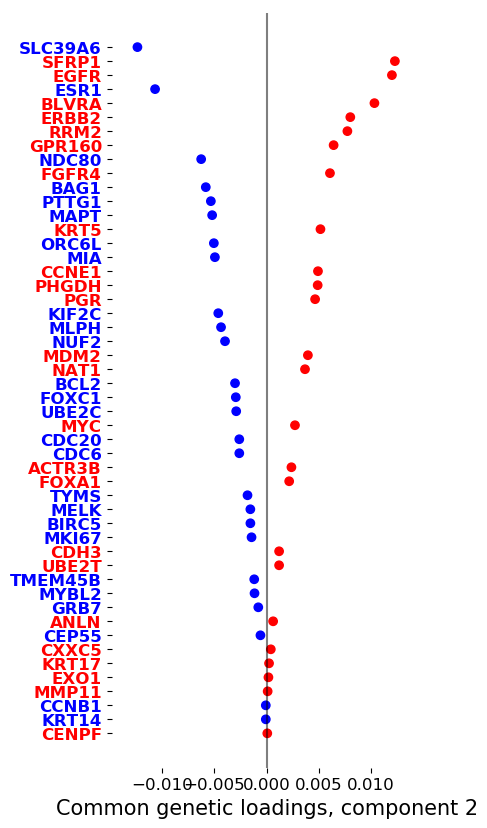}
\caption{
Joint component 2, PAM50 loadings.
}
\label{fig:common_loadings_comp_2}
\end{subfigure}%
\hfill
\begin{subfigure}[t]{.6\textwidth}
\centering
\includegraphics[scale=.4]{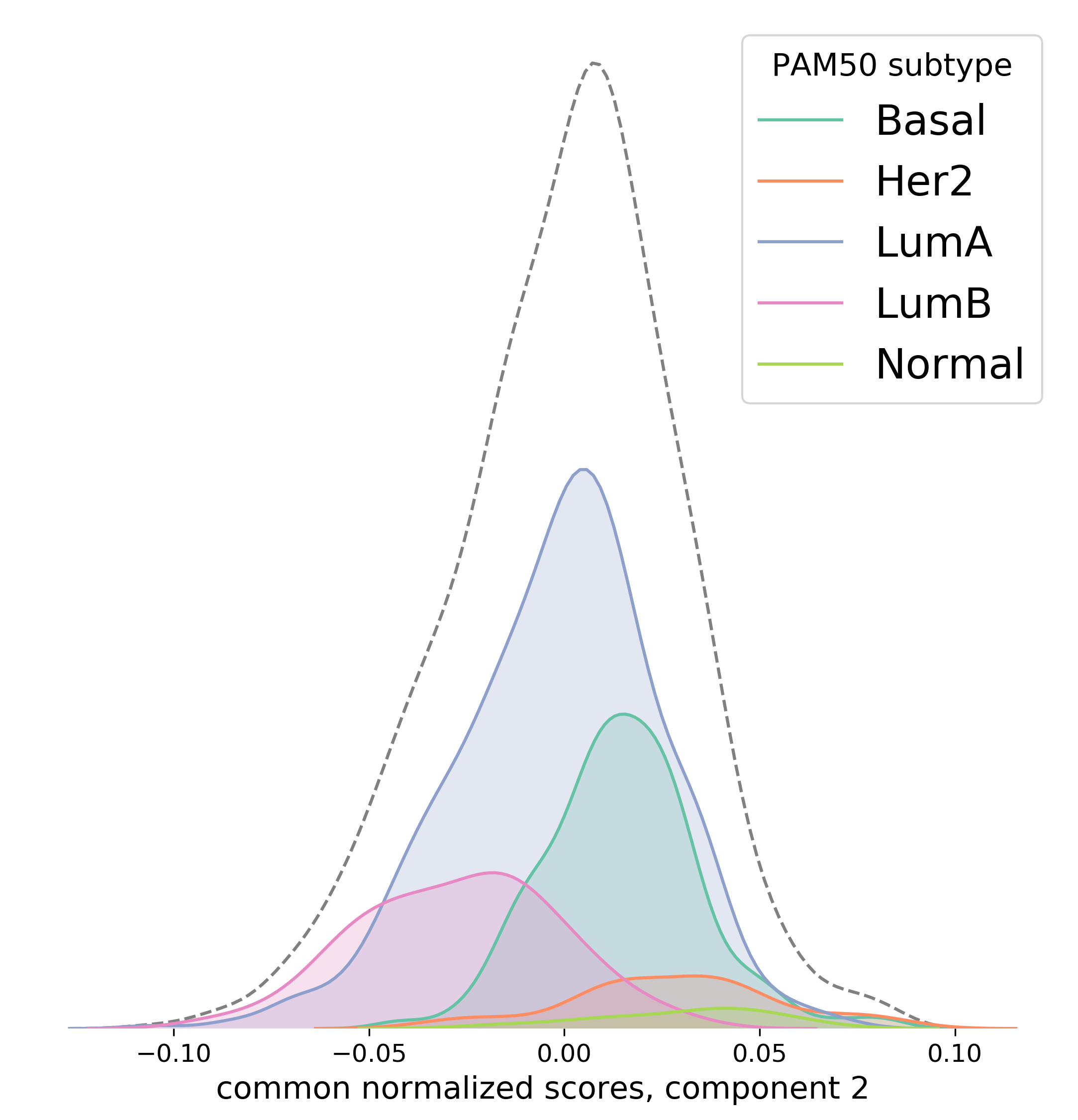}
\caption{
Joint component 2 scores conditioned on PAM50 subtype.
}
\label{fig:common_2_scores_v_pam50}
\end{subfigure}
\caption{
The Luminal B and Luminal A classes are associated with the negative end of joint component 2.
Not much is known about this histology of Luminal cancers.
}
\label{fig:joint_2_load_and_cd}
\end{figure}

From the genetics perspective, the negative end of joint component 2 picks out the Luminal B subtype (Figure \ref{fig:common_2_scores_v_pam50}). 
Looking at the PAM50 loadings vector, ESR1, SLC39A6 are the two most negative genes in the PAM50 loadings (Figure \ref{fig:common_loadings_comp_2}) and are known to be high in clinically ER+ cancers \citep{parker2009supervised}.
The Luminal B observations cluster on the negative end of this direction and are statistical significantly separated from the other PAM50 subtypes with AUC scores of: Basal = 0.905, HER2 = 0.933, Luminal A = 0.760, Normal = 0.950 (Figure \ref{fig:common_2_scores_v_pam50}).

From the immunohistochemical perspective, the second joint component is moderately related to ER status while weakly related to clinical HER2 status (Table \ref{tab:er_her2_joint}).
Clinical ER positive tumors cluster on the negative end of this component with an AUC of 0.752.

The pathology perspective of this second joint component appears to pick up on morphological features of Luminal B tumors i.e. intratumoral channels of stromal cells which are surrounded by moderate nuclear grade cancer cells.
To our knowledge, little is known about the histological features of Luminal B tumors.

Pathologists do not currently use stromal features in the diagnosis and classification of tumors.
However, tumor stroma and microenvironment \citep{eiro2019breast} and the stromal features of benign and tumor-adjacent normal tissue \citep{roman2012gene, chollet2018stroma} are areas of active investigation.
Interestingly, \cite{beck2011systematic} used image analysis approaches to demonstrate connections between certain stroma morphological features and patient survival.
Recent studies using CNNs have shown that breast biopsies may be accurately classified as malignant solely based on stromal features \citep{bejnordi2018using}.

\subsubsection{Joint component 3} \label{sss:results_joint_3}

\begin{figure}[H]
\begin{subfigure}[t]{.39\textwidth}
\centering
\includegraphics[scale=.4]{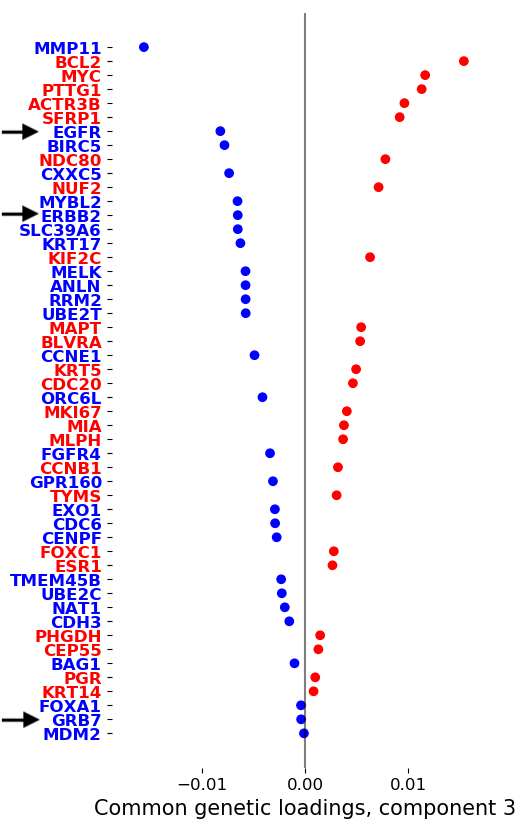}
\caption{
Joint component 3, PAM50 loadings.
}
\label{fig:common_loadings_comp_3}
\end{subfigure}%
\hfill
\begin{subfigure}[t]{.6\textwidth}
\centering
\includegraphics[scale=.4]{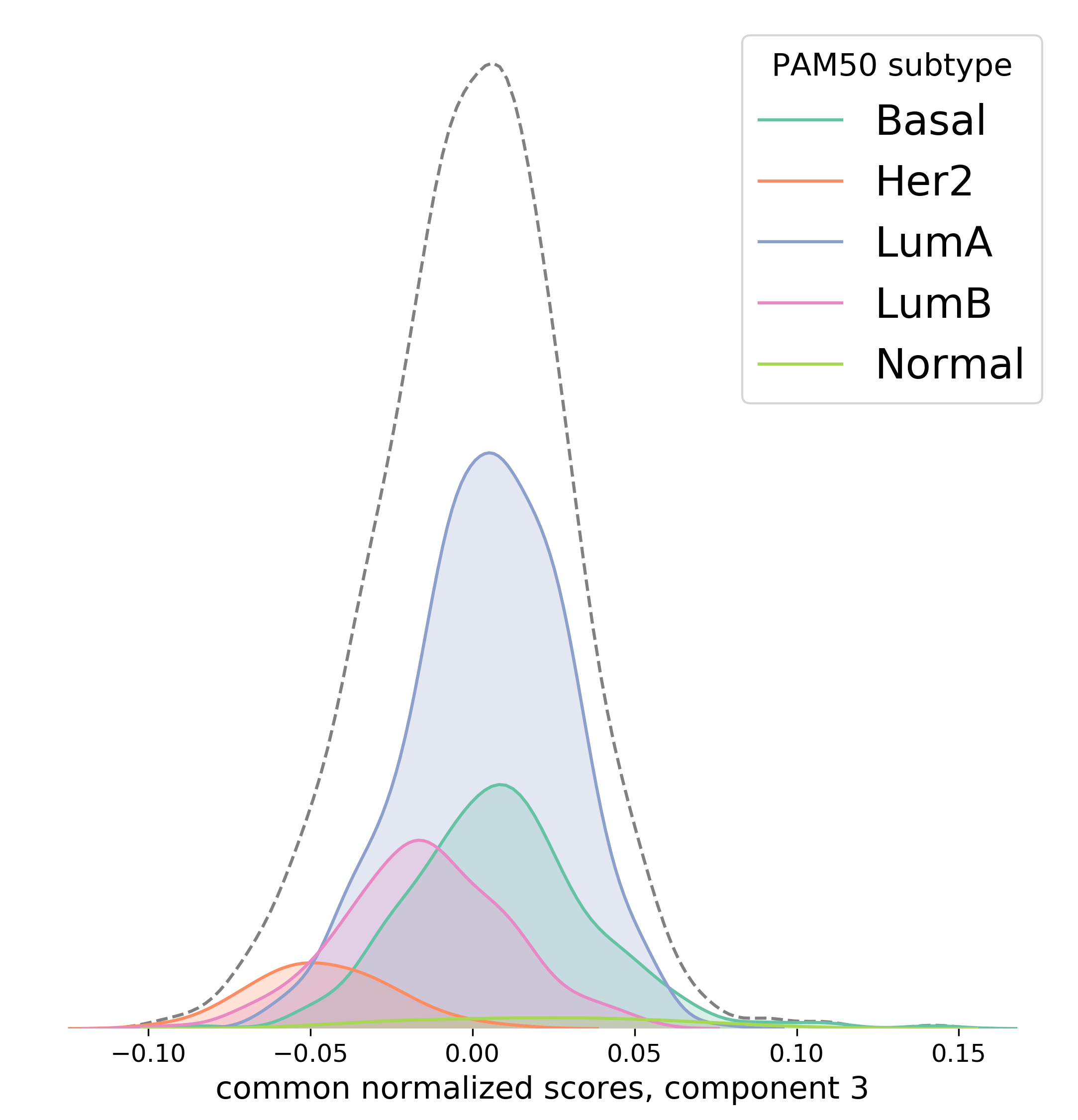}
\caption{
Joint component 3 scores conditioned on PAM50 subtype.
}
\label{fig:common_3_scores_v_pam50}
\end{subfigure}
\caption{
The molecular HER2 subtype are associated with the negative end of joint component 3.
}
\label{fig:joint_3_load_and_cd}
\end{figure}

From the genetics perspective, the negative end of joint component 3 picks out molecular HER2. 
The HER2 observations are separated from the other PAM50 subtypes with AUC scores of: Basal = 0.947, Luminal A = 0.940, Luminal B= 0.833, Normal = 0.950.
Interestingly, ERBB2 and EGFR have large negative values in the joint loadings vector while GRB7, which is on the same \textit{amplicon} as ERBB2, is almost 0 (Figure \ref{fig:common_loadings_comp_3}).
The negative end of this component is also moderately related to clinical HER2 status with an AUC of 0.777 (Table \ref{tab:er_her2_joint}).
This component is identifying not only clinical HER2 samples (as determined by IHC staining) but more strongly the molecular HER2 subtype of samples (as determined by gene expression).
Previous work \citep{cancer2012comprehensive} has shown both gene expression and protein and phosphoprotein levels of ERBB2 and EGFR are significantly enriched in clinically HER2 samples that are also the molecular HER2 subtype compared to clinical HER2 samples that are Luminal subtypes.
This is consistent with the separations we see in joint component 3.

\begin{figure}[H]
\begin{subfigure}{.48\textwidth}
\centering
\includegraphics[width=1\linewidth]{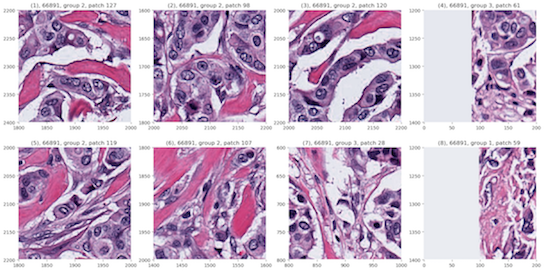}
\caption{
}
\label{fig:corepatches_common_3_neg_66891}
\end{subfigure}%
\rulesep
\begin{subfigure}{.48\textwidth}
\centering
\includegraphics[width=1\linewidth]{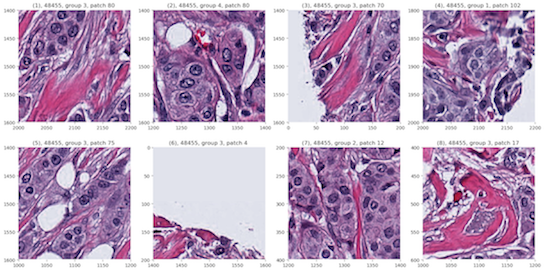}
\caption{
}
\label{fig:corepatches_common_3_neg_48455}
\end{subfigure}
\caption{
Tumors from the negative end of joint component 3 showing tumor cells surrounded collagenous stroma.
}
\label{fig:joint_3_neg}
\end{figure}

From the pathology perspective, the negative end joint component 3 again shows collagenous stroma, but this time surrounded by high nuclear grade tumor cells (Figure \ref{fig:joint_3_neg}).
Recall joint component 2 was similar but with moderate grade tumor cells.
This third joint component appears to be picking up on morphological features of molecular HER2 tumors. 
Similar to joint component 2, it is interesting that the stroma appear to play an important role in this component.

\subsection{Image individual information}\label{ss:results_image_indiv}

As mentioned in Section \ref{s:data_cbcs}, the PAM50 genes were selected to emphasize genes expressed in tumor epithelium, not genes highly expressed in tumor microenvironment features such as fat cells, collagenous stroma, and in some cases mucin.
Several of these microenvironment features have clear visual signals (e.g. high fat content images have round, clear adipose cells) and show up prominently in the AJIVE individual components.

\begin{table}[H] 
\centering
\begin{adjustbox}{width=1\textwidth}
\begin{tabular}{|l|l|l|l|l|l|l|l|l|l|}
\hline
component & end & homogeneous & \begin{tabular}[c]{@{}l@{}}tumor\\ cellularity\end{tabular} & \begin{tabular}[c]{@{}l@{}}tubule\\ formation\end{tabular} & \begin{tabular}[c]{@{}l@{}}nuclear\\ grade\end{tabular} & \begin{tabular}[c]{@{}l@{}}adipocytic\\ stroma\end{tabular} & \begin{tabular}[c]{@{}l@{}}collagenous\\ stroma\end{tabular} & lymphocytes & necrosis \\ \hline
1         & positive & yes         & low               & focal            & 1, 2          & yes               & limited            & few         & no       \\ \hline
          & negative & yes         & variable          & no               & 2             & no                & yes                & few         & some     \\ \hline
2         & positive & yes         & moderate          & no               & 2, 3          & focal             & yes                & yes         & no       \\ \hline
          & negative & yes         & low               & no               & 1             & focal             & limited            & no          & no       \\ \hline
3         & positive & yes         & low               & yes              & 2             & focal             & yes                & few         & no       \\ \hline
          & negative & yes         & low               & no               & 2             & yes               & no                 & no          & no       \\ \hline
\end{tabular}
\end{adjustbox}
\caption{
A pathologist's observations of first three image individual components from RPVs of 15 most extreme subjects on either end of the component.
}
 \label{tab:path_review_image_indiv}
\end{table}

\subsubsection{Image individual component 1} \label{sss:results_image_indiv_1}

\begin{figure}[H]
\begin{subfigure}{.48\textwidth}
\centering
\includegraphics[width=1\linewidth]{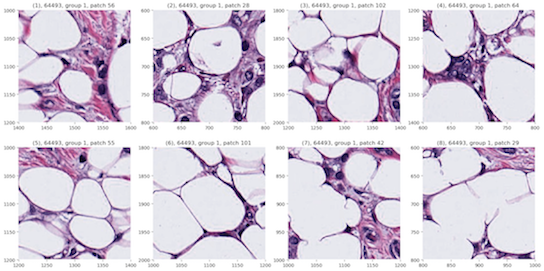}
\caption{
}
\label{fig:corepatches_image_indiv_1_pos_64493}
\end{subfigure}%
\rulesep
\begin{subfigure}{.48\textwidth}
\centering
\includegraphics[width=1\linewidth]{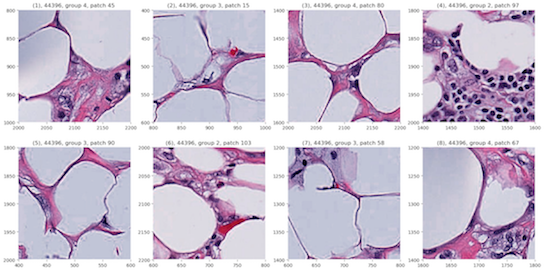}
\caption{
}
\label{fig:corepatches_image_indiv_1_pos_44396}
\end{subfigure}
\caption{
Two tumors from the positive end of image individual component 1 showing high fat content.
}
\label{fig:image_indiv_1_pos}
\end{figure}

\begin{figure}[H]
\begin{subfigure}{.48\textwidth}
\centering
\includegraphics[width=1\linewidth]{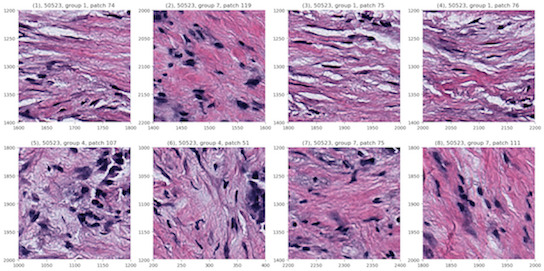}
\caption{
}
\label{fig:corepatches_image_indiv_1_neg_50523}
\end{subfigure}%
\rulesep
\begin{subfigure}{.48\textwidth}
\centering
\includegraphics[width=1\linewidth]{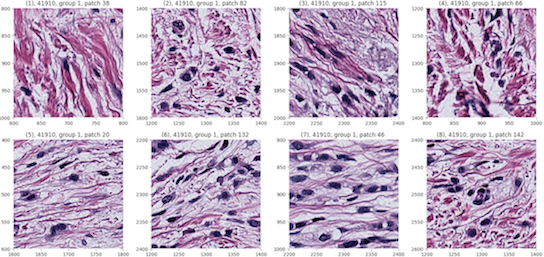}
\caption{
}
\label{fig:mini_corepatches_image_indiv_1_neg_70774}
\end{subfigure}
\caption{
Two tumors from the negative end of image individual component 1 showing moderate nuclear grade, variable tumor cellularity, collagenous stroma.
}
\label{fig:image_indiv_1_neg}
\end{figure}

All of the images on the positive end of the first image individual component shows a very clear theme of tumors with high fat content (Figure \ref{fig:image_indiv_1_pos}).
High fat content is a strong visual signal so it makes sense that it shows up as an early individual mode of variation for image data.
The negative end of the first image individual component shows tumors with low tumor cellularity and low/moderate grade nuclei (Table \ref{tab:path_review_image_indiv} and Figure \ref{fig:image_indiv_1_neg}).

\subsubsection{Image individual component 2} \label{sss:results_image_indiv_2}

\begin{figure}[H]
\begin{subfigure}{.48\textwidth}
\centering
\includegraphics[width=1\linewidth]{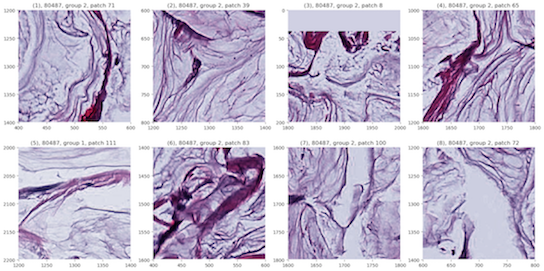}
\caption{
}
\label{fig:corepatches_image_indiv_2_neg_80487}
\end{subfigure}%
\rulesep
\begin{subfigure}{.48\textwidth}
\centering
\includegraphics[width=1\linewidth]{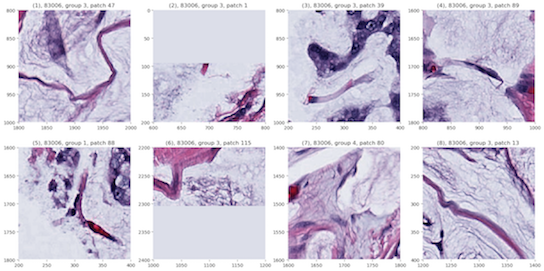}
\caption{
}
\label{fig:corepatches_image_indiv_2_neg_83006}
\end{subfigure}
\caption{
Two tumors from the negative end of image individual component 2 both displaying mucinous carcinomas.
}
\label{fig:image_indiv_2_neg}
\end{figure}

The negative end of image individual component 2 clearly picks out \textit{mucinous carcinoma} tumors (Figure \ref{fig:image_indiv_2_neg}).
Mucinous carcinomas are characterized by tumor cells floating in pools of mucin. 
These cancers presents a very clear visual pattern of dark purple tumor cells surrounded by wispy looking mucin.
Mucinous carcinoma is a rare histological subtype which the PAM50 genes \cite{perou2000molecular} are not designed to identify.

Mucinous carcinomas are typically low-grade, hormone receptor-positive, have a good prognosis and appear to be genetically different from invasive ductal carcinomas of no special type \citep{diab1999tumor, di2008retrospective, weigelt2009mucinous, lacroix2010mucinous}.
Mucinous carcinomas are usually genetically Luminal-type (typically Luminal A) \citep{colleoni2011outcome, caldarella2013invasive, weigelt2009mucinous}.
All of the top 15 tumors on the negative end of this component are genetically Luminal (12 are Luminal A and 3 are Luminal B).
Interestingly, neither the Luminal A nor B classes are strongly associated with the individual scores for this component overall; none of the difference in distribution tests (Section \ref{ss:clinical_interpretation_methods}) for Luminal A vs. another class were statistically significant (similarly for Luminal B).
This is consistent with variation appearing in an image individual component.

The positive end of individual component 2 picks out images with moderate cellularity and collagenous stroma surrounded by moderate nuclear grade tumor cells.



\begin{figure}[H]
\begin{subfigure}{.48\textwidth}
\centering
\includegraphics[width=1\linewidth]{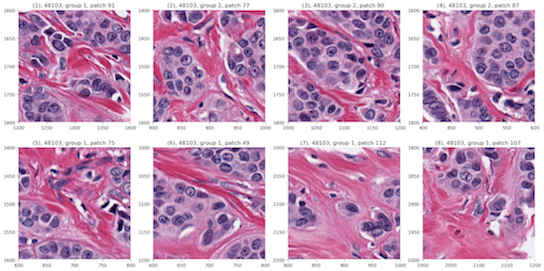}
\caption{
}
\label{fig:corepatches_image_indiv_2_pos_48103}
\end{subfigure}%
\rulesep
\begin{subfigure}{.48\textwidth}
\centering
\includegraphics[width=1\linewidth]{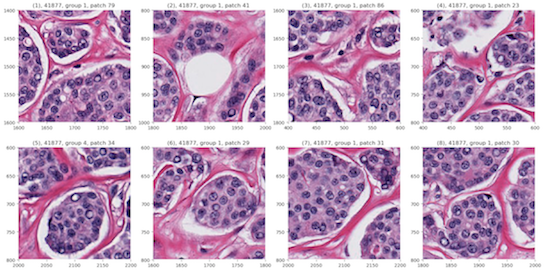}
\caption{
}
\label{fig:corepatches_image_indiv_2_pos_41877}
\end{subfigure}
\caption{
Two tumors from the positive end of image individual component 2 both showing: collagenous stroma, moderate to high nuclear grade, moderate tumor cellularity.
}
\label{fig:image_indiv_2_pos}
\end{figure}

The positive and negative ends of image individual component 2 show contrasting histologic features.
The patches from the tumors on the positive end are entirely filled with a combination of tumor cell aggregates separated by areas of dense collagenous stroma.
Adipocytic stroma is absent and the only optically clear space is in areas of retraction artifact where tumor cell groups appear to be pulled away from adjacent stroma (a known artifact of histologic preparation in some invasive tumors).
The patches from the negative end show extracellular mucin from mucinous carcinomas with low or no tumor cellularity and just a few wispy bands of stromal collagen.
The contrasting histology raises the possibility that this component may separate tumors based on one or more of the following features: tumor cellularity, tumor grade, extracellular stromal composition.

\subsubsection{Image individual component 3} \label{sss:results_image_indiv_3}

\begin{figure}[H]
\begin{subfigure}{.32\textwidth}
\centering
\includegraphics[width=1\linewidth]{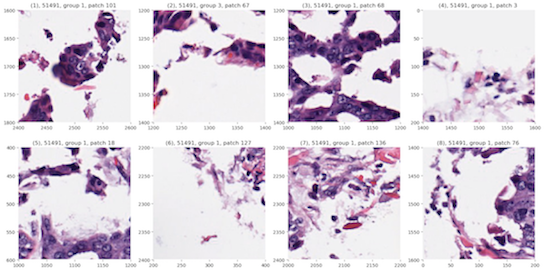}
\caption{
Disrupted.
}
\label{fig:corepatches_image_indiv_3_neg_51491}
\end{subfigure}%
\rulesep
\begin{subfigure}{.32\textwidth}
\centering
\includegraphics[width=1\linewidth]{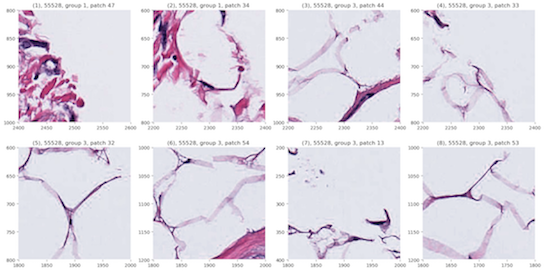}
\caption{
Fat cells.
}
\label{fig:corepatches_image_indiv_3_neg_55528}
\end{subfigure}%
\rulesep
\begin{subfigure}{.32\textwidth}
\centering
\includegraphics[width=1\linewidth]{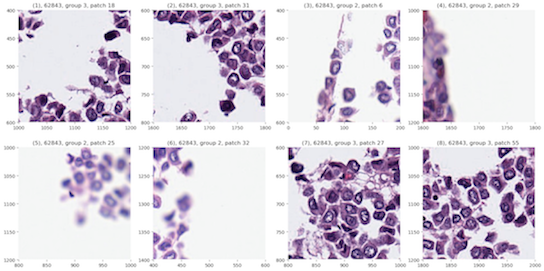}
\caption{
Discohesion.
}
\label{fig:corepatches_image_indiv_3_neg_62843}
\end{subfigure}
\caption{
Three tumors from the negative end of image individual component 3 all showing the clear visual pattern of a large amount of optically clear space.
}
\label{fig:image_indiv_3_neg}
\end{figure}

The negative end of image individual component 3 picks up on tumors whose patches contain a large amount of optically clear space.
This includes tumors with: with high fat content (Figure \ref{fig:corepatches_image_indiv_3_neg_55528}), where the cells discohesive (Figure \ref{fig:corepatches_image_indiv_3_neg_62843}) and disrupted tissue sections (Figure \ref{fig:corepatches_image_indiv_3_neg_51491}). 
Recall (Section \ref{ss:image_pro}) that patches with too much background (over 90\%) are removed.
Therefore white space surrounding the tumors and large white spaces in the core are unlikely to influence the amount of white space in the patches representing the image.
Some of the features seen in the images in Figure 17 \ref{fig:corepatches_image_indiv_3_neg_51491} and \ref{fig:corepatches_image_indiv_3_neg_62843} are likely related to technical variation in the tumor fixation/preservation and the quality of the histologic preparation. 
While the high fat content pattern seen in the positive end of the first image individual component is similar to this component (i.e. it picks up on large amounts of white space) the first component uniformly contains high fat content images in the top 15 images which is unlike this third component.

\begin{figure}[H]
\begin{subfigure}{.48\textwidth}
\centering
\includegraphics[width=1\linewidth]{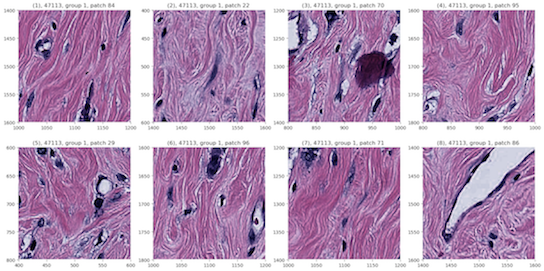}
\caption{
}
\label{fig:corepatches_image_indiv_3_pos_47113}
\end{subfigure}%
\rulesep
\begin{subfigure}{.48\textwidth}
\centering
\includegraphics[width=1\linewidth]{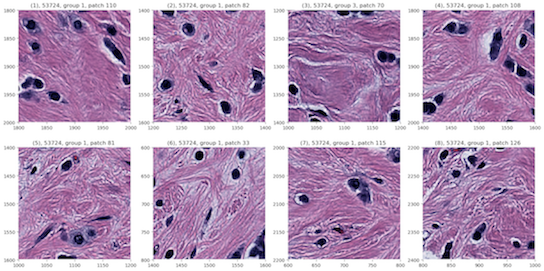}
\caption{
}
\label{fig:corepatches_image_indiv_3_pos_53724}
\end{subfigure}
\caption{
Two tumors from the positive end of image individual component 3.
Both show a visually clear pattern of dense collagenous stroma, low tumor cellularity, moderate nuclear grade, some lymphocytes.
}
\label{fig:image_indiv_3_pos}
\end{figure}

The positive end of this third component picks on images with a large amount of dense collagenous stroma (Figure \ref{fig:image_indiv_3_pos}).
This pattern is very clear in all 15 of the most positive subjects' representative images views (see \ref{supp}).
These tumors have lower tumor cellularity, moderate nuclear grade and have a moderate number of lymphocytes.
Similar to the amount of white space, the dense collagenous stroma is a clear visual pattern.

\subsection{Genetic individual information}\label{ss:results_genetic_indiv}

\begin{figure}[H]
\begin{subfigure}[t]{.35\textwidth}
\centering
\includegraphics[scale=.4]{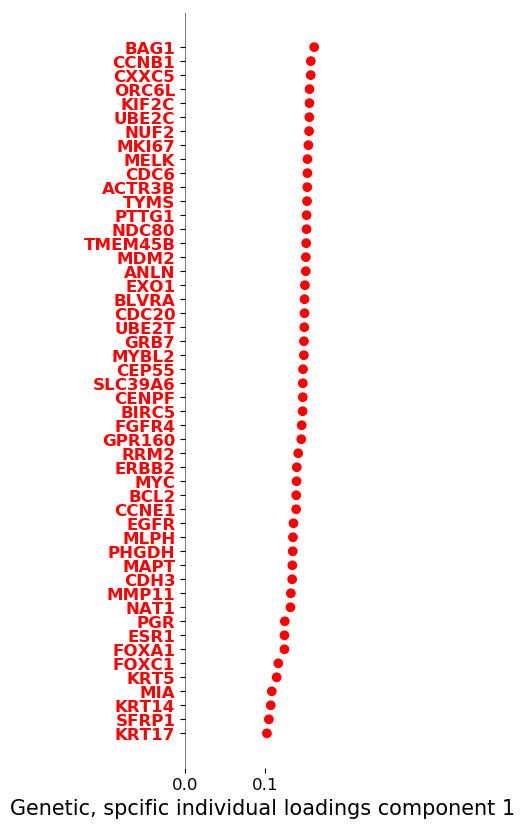}
\caption{
Genetic individual component 1 loadings shows common genetic technical mode of variation not expected to be associated with the images.
}
\label{fig:gene_indiv_loadings_comp_1}
\end{subfigure}%
\hfill
\begin{subfigure}[t]{.6\textwidth}
\includegraphics[scale=.5]{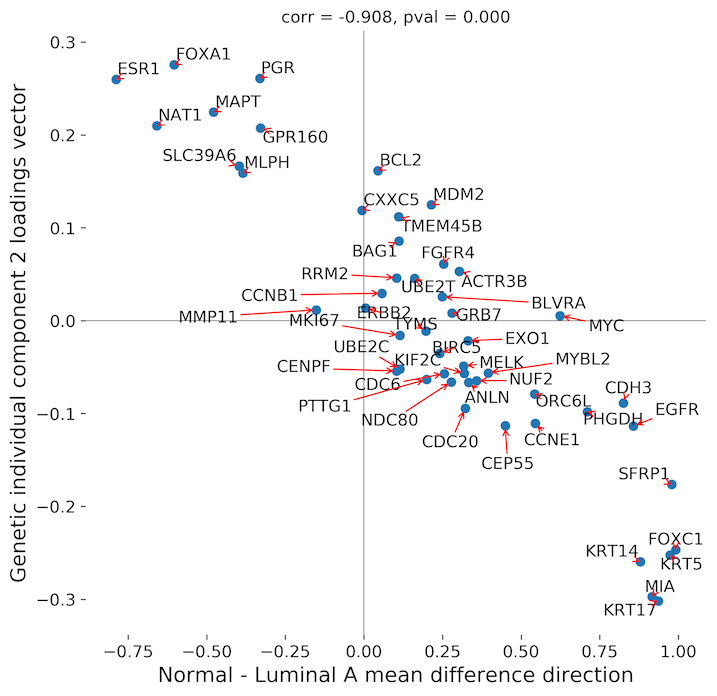}
\caption{
Genetic individual component 2 loadings vector compared to Normal-Luminal A mean difference direction.
Luminal A is on the left (negative) while Normal-like is on the right (positive).
}
\label{fig:gene_indiv_comp_2_loading_v_normal-lumA}
\end{subfigure}
\caption{
Genetic individual components show technical variation as well as additional PAM50 subtype information.
}
\label{fig:gene_indiv_loadings}
\end{figure}

Figure \ref{fig:gene_indiv_loadings_comp_1} shows the PAM50 loadings vector of the first genetic individual component.
This component picks up on overall gene expression levels which is a common source of technical variation. 
Both the second and third genetic individual components show connections to the PAM50 subtypes based on the clinical data comparisons given in \ref{supp} albeit with weaker separations than the joint components.

The second genetic individual component identifies additional information which varies between Luminal A and Normal that is not dependent on cell proliferation and seems to be more related to features such as estrogen receptor signaling and keratin expression status.
The scores for this second component separate Normal-like from Luminal A with an AUC of 0.801.

Figure \ref{fig:gene_indiv_comp_2_loading_v_normal-lumA} shows a scatter plot of the loadings vector from genetic individual component 2 compared to the Normal-Luminal A mean difference direction\footnote{The genes were first scaled by their standard deviation so this is the naive Bayes classification direction.}.
Several of the genes on the top left of \ref{fig:gene_indiv_comp_2_loading_v_normal-lumA} (ESR1, FOXA1, PGR) are all part of the estrogen signaling pathway \citep{oh2006estrogen}.
Several of the genes in the middle (CCNB1, MYC, MKI67, TYMS, MYBL2, CCNE1) are related to proliferation suggesting this component is unrelated to proliferation.
Several of the genes in the bottom right (KRT5, KRT14, KRT17) are characteristic of normal myoepithelium as well as Basal-like like breast cancer \citep{lazard1993expression}.

\section{Conclusion}\label{s:conclusion}

This paper develops methods that, using deep learning and AJIVE give interpretable, simultaneous image and genetic results.
Inferential and exploratory analysis leveraging deep learning is a promising area which presents many interesting, open questions -- some of which are discussed below.
These analytical tools enable simultaneous engagement from both the pathology and genetic communities which is critical for the fundamental biomedical interpretations.

Future research should evaluate whether the features learned in this paper can be reproducibly identified by pathologists and/or automated computer vision systems as well as whether these features can be validated in external test sets.

Scaling histological image analysis pipelines to gigapixel \textit{whole slide images} (WSI) is an important future direction.
In clinical practice, pathologists use WSIs; the core images used in this paper require additional preparation, are typically only available in some research settings and may ignore important tumor information (e.g. spatial heterogeneity across the tumor, particularly histological patterns not observed in the sampled region).
Analyses of WSIs presents computational challenges as these images are orders of magnitude larger than the core images.

\subsubsection{Patch representation} \label{sss:patch_rep}

Patch based approaches have shown promise for predictive tasks using deep learning \cite{ilse2018attention}. 
The patch based approach taken in this paper was selected because it i) will scale to whole slide images ii) can identify localized image information e.g. with the RPVs and iii) creates a smaller feature set\footnote{Passing the full core images through the CNN resulted in $10^6$ features.}.
The approach of averaging of patch features ignores some within image heterogeneity.
For image-only analysis, methods such as \cite{bishop1998hierarchical, backenroth2018modeling} may be able to capture additional within-image heterogeneity.
In the context of multi-view data, additional methodology needs to be developed to account for grouped observations (e.g. \cite{pourzanjani2017understanding}).

\subsubsection{Transfer learning} \label{sss:transfer_learning}
Training a neural network can be time and resource intensive.
Furthermore, CNNs often require a large amount of training data to be fit effectively.
Transfer learning allows the data analyst to use more powerful neural networks with less data and less time spent tuning CNN parameters \citep{yosinski2014transferable, sharif2014cnn}.
First, a CNN is trained to solve a different predictive task on a large, external dataset  -- typically the famous ImageNet classification task \citep{deng2009imagenet}.
Then the pre-trained network parameters may be \textit{fine-tuned} on the dataset of interest to solve the predictive problem of interest.

The setting of this paper is a bit different.
First, we are doing exploratory analysis, not predictive analysis.
Second, while we do have labels which could be used to fine-tune the CNN (e.g. the PAM50 subtypes) we do not want to use these labels because then the 
network would be aware of information which we might want to (re)discover and/or validate in the following analysis.
This leaves us with a couple of options to still use transfer learning including: training an unsupervised algorithm (e.g. auto-encoders \cite{kingma2013auto}, \textit{generative adversarial networks} \cite{goodfellow2014generative} or self-supervised learning algorithms \cite{oord2018representation, lu2019semi}) or not doing any fine tuning.
 \cite{ash2018joint} trains auto-encoder which has some disadvantages: a CNN is required to be trained which is time/resource consuming, a number of new hyper-parameters are introduced into the problem and either the data are used twice or external data are needed.
We explore the latter option and show, perhaps surprisingly, that pre-trained CNN features with no fine tuning are able to capture complex visual signals in a domain vastly different than the one they were originally trained on.

Even in the context of transfer learning, there are many choices to be made about how to extract neural network features from an image including: network architecture, layer (or layers) of the network, and feature aggregation (e.g. spatial mean pooling discussed in Section \ref{ss:cnn_feats}).
For predictive modeling these hyper-parameters can be set using an error metric and methods such as cross-validation, however, as discussed in the above paragraph, we do not have such error metrics readily available to guide hyper-parameter choices.
Preliminary sensitivity analysis showed that the results of our analysis are not particularly sensitive to mild differences in architecture choices.
Better methods to select these CNN hyper-parameters is an open area of research.



\section*{Acknowledgements}
We thank the Carolina Breast Cancer Study participants and staff.
We also want to acknowledge Robert C. Millikan, founder of the Carolina Breast Cancer Study Phase 3.
Research reported in this publication was supported by a Specialized Program of Research Excellence (SPORE) in breast cancer (P50 CA058223), an award from the Susan G. Komen Foundation (OGUNC1202), the North Carolina University Cancer Research Fund, and a Cancer Center Support Grant (P30 CA016086).
Iain Carmichael and J. S. Marron were partially supported by NSF Grant IIS-1633074, BIG DATA 2016-2019.
Iain Carmichael is currently supported by NSF MSPRF DMS-1902440.
Katherine Hoadley was supported by Komen Career Catalyst Grant (CCR16376756).

\begin{supplement}
\sname{Supplement A}\label{supp}
\stitle{Supplementary results and important biological background}
\slink[url]{https://marronwebfiles.sites.oasis.unc.edu/AJIVE-Hist-Gene/}
\sdescription{
See supplement\_A.pdf.
Discussion of important tissue structures discussed in the results, AJIVE diagnostic plot and clinical data comparison methods (e.g. multiple testing control).
}
\end{supplement}

\begin{supplement}
\sname{Supplement B}\label{supp_viz}
\stitle{Supplementary vizualizations}
\slink[url]{https://marronwebfiles.sites.oasis.unc.edu/AJIVE-Hist-Gene/}
\sdescription{
Additional figures can be found in the zipped folder located at the above link (this file is large -- approximately 1.5 Gb).
These include all representative patch views shown to pathologists, all AJIVE genetic loadings vectors and all clinical data comparisons.
See the file readme.txt for details.
}
\end{supplement}

%
%
%
%
%
%
%

\bibliographystyle{imsart-nameyear}
\bibliography{refs}

\begin{thebibliography}{88}

\bibitem[\protect\citeauthoryear{Adebayo et~al.}{2018}]{adebayo2018sanity}
\begin{binproceedings}[author]
\bauthor{\bsnm{Adebayo},~\bfnm{Julius}\binits{J.}},
  \bauthor{\bsnm{Gilmer},~\bfnm{Justin}\binits{J.}},
  \bauthor{\bsnm{Muelly},~\bfnm{Michael}\binits{M.}},
  \bauthor{\bsnm{Goodfellow},~\bfnm{Ian}\binits{I.}},
  \bauthor{\bsnm{Hardt},~\bfnm{Moritz}\binits{M.}} \AND
  \bauthor{\bsnm{Kim},~\bfnm{Been}\binits{B.}}
(\byear{2018}).
\btitle{Sanity checks for saliency maps}.
In \bbooktitle{Advances in Neural Information Processing Systems}
\bpages{9505--9515}.
\end{binproceedings}
\endbibitem

\bibitem[\protect\citeauthoryear{Aeffner
  et~al.}{2019}]{aeffner2019introduction}
\begin{barticle}[author]
\bauthor{\bsnm{Aeffner},~\bfnm{Famke}\binits{F.}},
  \bauthor{\bsnm{Zarella},~\bfnm{Mark~D}\binits{M.~D.}},
  \bauthor{\bsnm{Buchbinder},~\bfnm{Nathan}\binits{N.}},
  \bauthor{\bsnm{Bui},~\bfnm{Marilyn~M}\binits{M.~M.}},
  \bauthor{\bsnm{Goodman},~\bfnm{Matthew~R}\binits{M.~R.}},
  \bauthor{\bsnm{Hartman},~\bfnm{Douglas~J}\binits{D.~J.}},
  \bauthor{\bsnm{Lujan},~\bfnm{Giovanni~M}\binits{G.~M.}},
  \bauthor{\bsnm{Molani},~\bfnm{Mariam~A}\binits{M.~A.}},
  \bauthor{\bsnm{Parwani},~\bfnm{Anil~V}\binits{A.~V.}},
  \bauthor{\bsnm{Lillard},~\bfnm{Kate}\binits{K.}} \betal{et~al.}
(\byear{2019}).
\btitle{Introduction to digital image analysis in whole-slide imaging: A white
  paper from the digital pathology association}.
\bjournal{Journal of pathology informatics}
\bvolume{10}.
\end{barticle}
\endbibitem

\bibitem[\protect\citeauthoryear{Allott et~al.}{2018}]{allott2018frequency}
\begin{barticle}[author]
\bauthor{\bsnm{Allott},~\bfnm{Emma~H}\binits{E.~H.}},
  \bauthor{\bsnm{Geradts},~\bfnm{Joseph}\binits{J.}},
  \bauthor{\bsnm{Cohen},~\bfnm{Stephanie~M}\binits{S.~M.}},
  \bauthor{\bsnm{Khoury},~\bfnm{Thaer}\binits{T.}},
  \bauthor{\bsnm{Zirpoli},~\bfnm{Gary~R}\binits{G.~R.}},
  \bauthor{\bsnm{Bshara},~\bfnm{Wiam}\binits{W.}},
  \bauthor{\bsnm{Davis},~\bfnm{Warren}\binits{W.}},
  \bauthor{\bsnm{Omilian},~\bfnm{Angela}\binits{A.}},
  \bauthor{\bsnm{Nair},~\bfnm{Priya}\binits{P.}},
  \bauthor{\bsnm{Ondracek},~\bfnm{Rochelle~P}\binits{R.~P.}} \betal{et~al.}
(\byear{2018}).
\btitle{Frequency of breast cancer subtypes among African American women in the
  AMBER consortium}.
\bjournal{Breast Cancer Research}
\bvolume{20}
\bpages{12}.
\end{barticle}
\endbibitem

\bibitem[\protect\citeauthoryear{Ash et~al.}{2018}]{ash2018joint}
\begin{barticle}[author]
\bauthor{\bsnm{Ash},~\bfnm{Jordan}\binits{J.}},
  \bauthor{\bsnm{Darnell},~\bfnm{Gregory}\binits{G.}},
  \bauthor{\bsnm{Munro},~\bfnm{Daniel}\binits{D.}} \AND
  \bauthor{\bsnm{Engelhardt},~\bfnm{Barbara}\binits{B.}}
(\byear{2018}).
\btitle{Joint analysis of gene expression levels and histological images
  identifies genes associated with tissue morphology}.
\bjournal{bioRxiv}
\bpages{458711}.
\end{barticle}
\endbibitem

\bibitem[\protect\citeauthoryear{Backenroth
  et~al.}{2018}]{backenroth2018modeling}
\begin{barticle}[author]
\bauthor{\bsnm{Backenroth},~\bfnm{Daniel}\binits{D.}},
  \bauthor{\bsnm{Goldsmith},~\bfnm{Jeff}\binits{J.}},
  \bauthor{\bsnm{Harran},~\bfnm{Michelle~D}\binits{M.~D.}},
  \bauthor{\bsnm{Cortes},~\bfnm{Juan~C}\binits{J.~C.}},
  \bauthor{\bsnm{Krakauer},~\bfnm{John~W}\binits{J.~W.}} \AND
  \bauthor{\bsnm{Kitago},~\bfnm{Tomoko}\binits{T.}}
(\byear{2018}).
\btitle{Modeling Motor Learning Using Heteroscedastic Functional Principal
  Components Analysis}.
\bjournal{Journal of the American Statistical Association}
\bvolume{113}
\bpages{1003--1015}.
\end{barticle}
\endbibitem

\bibitem[\protect\citeauthoryear{Beck et~al.}{2011}]{beck2011systematic}
\begin{barticle}[author]
\bauthor{\bsnm{Beck},~\bfnm{Andrew~H}\binits{A.~H.}},
  \bauthor{\bsnm{Sangoi},~\bfnm{Ankur~R}\binits{A.~R.}},
  \bauthor{\bsnm{Leung},~\bfnm{Samuel}\binits{S.}},
  \bauthor{\bsnm{Marinelli},~\bfnm{Robert~J}\binits{R.~J.}},
  \bauthor{\bsnm{Nielsen},~\bfnm{Torsten~O}\binits{T.~O.}}, \bauthor{\bsnm{Van
  De~Vijver},~\bfnm{Marc~J}\binits{M.~J.}},
  \bauthor{\bsnm{West},~\bfnm{Robert~B}\binits{R.~B.}}, \bauthor{\bsnm{Van
  De~Rijn},~\bfnm{Matt}\binits{M.}} \AND
  \bauthor{\bsnm{Koller},~\bfnm{Daphne}\binits{D.}}
(\byear{2011}).
\btitle{Systematic analysis of breast cancer morphology uncovers stromal
  features associated with survival}.
\bjournal{Science translational medicine}
\bvolume{3}
\bpages{108ra113--108ra113}.
\end{barticle}
\endbibitem

\bibitem[\protect\citeauthoryear{Bejnordi et~al.}{2018}]{bejnordi2018using}
\begin{barticle}[author]
\bauthor{\bsnm{Bejnordi},~\bfnm{Babak~Ehteshami}\binits{B.~E.}},
  \bauthor{\bsnm{Mullooly},~\bfnm{Maeve}\binits{M.}},
  \bauthor{\bsnm{Pfeiffer},~\bfnm{Ruth~M}\binits{R.~M.}},
  \bauthor{\bsnm{Fan},~\bfnm{Shaoqi}\binits{S.}},
  \bauthor{\bsnm{Vacek},~\bfnm{Pamela~M}\binits{P.~M.}},
  \bauthor{\bsnm{Weaver},~\bfnm{Donald~L}\binits{D.~L.}},
  \bauthor{\bsnm{Herschorn},~\bfnm{Sally}\binits{S.}},
  \bauthor{\bsnm{Brinton},~\bfnm{Louise~A}\binits{L.~A.}},
  \bauthor{\bparticle{van} \bsnm{Ginneken},~\bfnm{Bram}\binits{B.}},
  \bauthor{\bsnm{Karssemeijer},~\bfnm{Nico}\binits{N.}} \betal{et~al.}
(\byear{2018}).
\btitle{Using deep convolutional neural networks to identify and classify
  tumor-associated stroma in diagnostic breast biopsies}.
\bjournal{Modern Pathology}
\bvolume{31}
\bpages{1502}.
\end{barticle}
\endbibitem

\bibitem[\protect\citeauthoryear{Benjamini and
  Hochberg}{1995}]{benjamini1995controlling}
\begin{barticle}[author]
\bauthor{\bsnm{Benjamini},~\bfnm{Yoav}\binits{Y.}} \AND
  \bauthor{\bsnm{Hochberg},~\bfnm{Yosef}\binits{Y.}}
(\byear{1995}).
\btitle{Controlling the false discovery rate: a practical and powerful approach
  to multiple testing}.
\bjournal{Journal of the Royal statistical society: series B (Methodological)}
\bvolume{57}
\bpages{289--300}.
\end{barticle}
\endbibitem

\bibitem[\protect\citeauthoryear{Bishop and
  Tipping}{1998}]{bishop1998hierarchical}
\begin{barticle}[author]
\bauthor{\bsnm{Bishop},~\bfnm{Christopher~M}\binits{C.~M.}} \AND
  \bauthor{\bsnm{Tipping},~\bfnm{Michael~E}\binits{M.~E.}}
(\byear{1998}).
\btitle{A hierarchical latent variable model for data visualization}.
\bjournal{IEEE Transactions on Pattern Analysis and Machine Intelligence}
\bvolume{20}
\bpages{281--293}.
\end{barticle}
\endbibitem

\bibitem[\protect\citeauthoryear{Caldarella
  et~al.}{2013}]{caldarella2013invasive}
\begin{barticle}[author]
\bauthor{\bsnm{Caldarella},~\bfnm{A}\binits{A.}},
  \bauthor{\bsnm{Buzzoni},~\bfnm{C}\binits{C.}},
  \bauthor{\bsnm{Crocetti},~\bfnm{E}\binits{E.}},
  \bauthor{\bsnm{Bianchi},~\bfnm{S}\binits{S.}},
  \bauthor{\bsnm{Vezzosi},~\bfnm{V}\binits{V.}},
  \bauthor{\bsnm{Apicella},~\bfnm{P}\binits{P.}},
  \bauthor{\bsnm{Biancalani},~\bfnm{M}\binits{M.}},
  \bauthor{\bsnm{Giannini},~\bfnm{A}\binits{A.}},
  \bauthor{\bsnm{Urso},~\bfnm{C}\binits{C.}},
  \bauthor{\bsnm{Zolfanelli},~\bfnm{F}\binits{F.}} \betal{et~al.}
(\byear{2013}).
\btitle{Invasive breast cancer: a significant correlation between histological
  types and molecular subgroups}.
\bjournal{Journal of cancer research and clinical oncology}
\bvolume{139}
\bpages{617--623}.
\end{barticle}
\endbibitem

\bibitem[\protect\citeauthoryear{Carey et~al.}{2006}]{carey2006race}
\begin{barticle}[author]
\bauthor{\bsnm{Carey},~\bfnm{Lisa~A}\binits{L.~A.}},
  \bauthor{\bsnm{Perou},~\bfnm{Charles~M}\binits{C.~M.}},
  \bauthor{\bsnm{Livasy},~\bfnm{Chad~A}\binits{C.~A.}},
  \bauthor{\bsnm{Dressler},~\bfnm{Lynn~G}\binits{L.~G.}},
  \bauthor{\bsnm{Cowan},~\bfnm{David}\binits{D.}},
  \bauthor{\bsnm{Conway},~\bfnm{Kathleen}\binits{K.}},
  \bauthor{\bsnm{Karaca},~\bfnm{Gamze}\binits{G.}},
  \bauthor{\bsnm{Troester},~\bfnm{Melissa~A}\binits{M.~A.}},
  \bauthor{\bsnm{Tse},~\bfnm{Chiu~Kit}\binits{C.~K.}},
  \bauthor{\bsnm{Edmiston},~\bfnm{Sharon}\binits{S.}} \betal{et~al.}
(\byear{2006}).
\btitle{Race, breast cancer subtypes, and survival in the Carolina Breast
  Cancer Study}.
\bjournal{Jama}
\bvolume{295}
\bpages{2492--2502}.
\end{barticle}
\endbibitem

\bibitem[\protect\citeauthoryear{Carmichael}{2019}]{carmichael2019pyjive}
\begin{bmisc}[author]
\bauthor{\bsnm{Carmichael},~\bfnm{Iain}\binits{I.}}
(\byear{2019}).
\btitle{PyJive: an python library implementing AJIVE}.
\bhowpublished{\url{https://github.com/idc9/py_jive}}.
\end{bmisc}
\endbibitem

\bibitem[\protect\citeauthoryear{Chen et~al.}{2018a}]{chen2018looks}
\begin{barticle}[author]
\bauthor{\bsnm{Chen},~\bfnm{Chaofan}\binits{C.}},
  \bauthor{\bsnm{Li},~\bfnm{Oscar}\binits{O.}},
  \bauthor{\bsnm{Tao},~\bfnm{Chaofan}\binits{C.}},
  \bauthor{\bsnm{Barnett},~\bfnm{Alina~Jade}\binits{A.~J.}},
  \bauthor{\bsnm{Su},~\bfnm{Jonathan}\binits{J.}} \AND
  \bauthor{\bsnm{Rudin},~\bfnm{Cynthia}\binits{C.}}
(\byear{2018}a).
\btitle{This looks like that: deep learning for interpretable image
  recognition}.
\bjournal{arXiv preprint arXiv:1806.10574}.
\end{barticle}
\endbibitem

\bibitem[\protect\citeauthoryear{Chen et~al.}{2018b}]{chen2018microscope}
\begin{barticle}[author]
\bauthor{\bsnm{Chen},~\bfnm{Po-Hsuan~Cameron}\binits{P.-H.~C.}},
  \bauthor{\bsnm{Gadepalli},~\bfnm{Krishna}\binits{K.}},
  \bauthor{\bsnm{MacDonald},~\bfnm{Robert}\binits{R.}},
  \bauthor{\bsnm{Liu},~\bfnm{Yun}\binits{Y.}},
  \bauthor{\bsnm{Nagpal},~\bfnm{Kunal}\binits{K.}},
  \bauthor{\bsnm{Kohlberger},~\bfnm{Timo}\binits{T.}},
  \bauthor{\bsnm{Dean},~\bfnm{Jeffrey}\binits{J.}},
  \bauthor{\bsnm{Corrado},~\bfnm{Greg~S}\binits{G.~S.}},
  \bauthor{\bsnm{Hipp},~\bfnm{Jason~D}\binits{J.~D.}} \AND
  \bauthor{\bsnm{Stumpe},~\bfnm{Martin~C}\binits{M.~C.}}
(\byear{2018}b).
\btitle{Microscope 2.0: an augmented reality microscope with real-time
  artificial intelligence integration}.
\bjournal{arXiv preprint arXiv:1812.00825}.
\end{barticle}
\endbibitem

\bibitem[\protect\citeauthoryear{Chen et~al.}{2019}]{chen2019pathomic}
\begin{barticle}[author]
\bauthor{\bsnm{Chen},~\bfnm{Richard~J}\binits{R.~J.}},
  \bauthor{\bsnm{Lu},~\bfnm{Ming~Y}\binits{M.~Y.}},
  \bauthor{\bsnm{Wang},~\bfnm{Jingwen}\binits{J.}},
  \bauthor{\bsnm{Williamson},~\bfnm{Drew~FK}\binits{D.~F.}},
  \bauthor{\bsnm{Rodig},~\bfnm{Scott~J}\binits{S.~J.}},
  \bauthor{\bsnm{Lindeman},~\bfnm{Neal~I}\binits{N.~I.}} \AND
  \bauthor{\bsnm{Mahmood},~\bfnm{Faisal}\binits{F.}}
(\byear{2019}).
\btitle{Pathomic Fusion: An Integrated Framework for Fusing Histopathology and
  Genomic Features for Cancer Diagnosis and Prognosis}.
\bjournal{arXiv preprint arXiv:1912.08937}.
\end{barticle}
\endbibitem

\bibitem[\protect\citeauthoryear{Chollet-Hinton
  et~al.}{2018}]{chollet2018stroma}
\begin{barticle}[author]
\bauthor{\bsnm{Chollet-Hinton},~\bfnm{Lynn}\binits{L.}},
  \bauthor{\bsnm{Puvanesarajah},~\bfnm{Samantha}\binits{S.}},
  \bauthor{\bsnm{Sandhu},~\bfnm{Rupninder}\binits{R.}},
  \bauthor{\bsnm{Kirk},~\bfnm{Erin~L}\binits{E.~L.}},
  \bauthor{\bsnm{Midkiff},~\bfnm{Bentley~R}\binits{B.~R.}},
  \bauthor{\bsnm{Ghosh},~\bfnm{Karthik}\binits{K.}},
  \bauthor{\bsnm{Brandt},~\bfnm{Kathleen~R}\binits{K.~R.}},
  \bauthor{\bsnm{Scott},~\bfnm{Christopher~G}\binits{C.~G.}},
  \bauthor{\bsnm{Gierach},~\bfnm{Gretchen~L}\binits{G.~L.}},
  \bauthor{\bsnm{Sherman},~\bfnm{Mark~E}\binits{M.~E.}} \betal{et~al.}
(\byear{2018}).
\btitle{Stroma modifies relationships between risk factor exposure and
  age-related epithelial involution in benign breast}.
\bjournal{Modern Pathology}
\bvolume{31}
\bpages{1085}.
\end{barticle}
\endbibitem

\bibitem[\protect\citeauthoryear{Colleoni et~al.}{2011}]{colleoni2011outcome}
\begin{barticle}[author]
\bauthor{\bsnm{Colleoni},~\bfnm{M}\binits{M.}},
  \bauthor{\bsnm{Rotmensz},~\bfnm{N}\binits{N.}},
  \bauthor{\bsnm{Maisonneuve},~\bfnm{P}\binits{P.}},
  \bauthor{\bsnm{Mastropasqua},~\bfnm{MG}\binits{M.}},
  \bauthor{\bsnm{Luini},~\bfnm{A}\binits{A.}},
  \bauthor{\bsnm{Veronesi},~\bfnm{P}\binits{P.}},
  \bauthor{\bsnm{Intra},~\bfnm{M}\binits{M.}},
  \bauthor{\bsnm{Montagna},~\bfnm{E}\binits{E.}},
  \bauthor{\bsnm{Cancello},~\bfnm{G}\binits{G.}},
  \bauthor{\bsnm{Cardillo},~\bfnm{A}\binits{A.}} \betal{et~al.}
(\byear{2011}).
\btitle{Outcome of special types of luminal breast cancer}.
\bjournal{Annals of oncology}
\bvolume{23}
\bpages{1428--1436}.
\end{barticle}
\endbibitem

\bibitem[\protect\citeauthoryear{Cooper et~al.}{2015}]{cooper2015novel}
\begin{barticle}[author]
\bauthor{\bsnm{Cooper},~\bfnm{Lee~AD}\binits{L.~A.}},
  \bauthor{\bsnm{Kong},~\bfnm{Jun}\binits{J.}},
  \bauthor{\bsnm{Gutman},~\bfnm{David~A}\binits{D.~A.}},
  \bauthor{\bsnm{Dunn},~\bfnm{William~D}\binits{W.~D.}},
  \bauthor{\bsnm{Nalisnik},~\bfnm{Michael}\binits{M.}} \AND
  \bauthor{\bsnm{Brat},~\bfnm{Daniel~J}\binits{D.~J.}}
(\byear{2015}).
\btitle{Novel genotype-phenotype associations in human cancers enabled by
  advanced molecular platforms and computational analysis of whole slide
  images}.
\bjournal{Laboratory investigation}
\bvolume{95}
\bpages{366}.
\end{barticle}
\endbibitem

\bibitem[\protect\citeauthoryear{Couture et~al.}{2018}]{couture2018image}
\begin{barticle}[author]
\bauthor{\bsnm{Couture},~\bfnm{Heather~D}\binits{H.~D.}},
  \bauthor{\bsnm{Williams},~\bfnm{Lindsay~A}\binits{L.~A.}},
  \bauthor{\bsnm{Geradts},~\bfnm{Joseph}\binits{J.}},
  \bauthor{\bsnm{Nyante},~\bfnm{Sarah~J}\binits{S.~J.}},
  \bauthor{\bsnm{Butler},~\bfnm{Ebonee~N}\binits{E.~N.}},
  \bauthor{\bsnm{Marron},~\bfnm{James~Stephen}\binits{J.~S.}},
  \bauthor{\bsnm{Perou},~\bfnm{Charles~M}\binits{C.~M.}},
  \bauthor{\bsnm{Troester},~\bfnm{Melissa~A}\binits{M.~A.}} \AND
  \bauthor{\bsnm{Niethammer},~\bfnm{Marc}\binits{M.}}
(\byear{2018}).
\btitle{Image analysis with deep learning to predict breast cancer grade, ER
  status, histologic subtype, and intrinsic subtype}.
\bjournal{NPJ breast cancer}
\bvolume{4}
\bpages{30}.
\end{barticle}
\endbibitem

\bibitem[\protect\citeauthoryear{Deng et~al.}{2009}]{deng2009imagenet}
\begin{binproceedings}[author]
\bauthor{\bsnm{Deng},~\bfnm{Jia}\binits{J.}},
  \bauthor{\bsnm{Dong},~\bfnm{Wei}\binits{W.}},
  \bauthor{\bsnm{Socher},~\bfnm{Richard}\binits{R.}},
  \bauthor{\bsnm{Li},~\bfnm{Li-Jia}\binits{L.-J.}},
  \bauthor{\bsnm{Li},~\bfnm{Kai}\binits{K.}} \AND
  \bauthor{\bsnm{Fei-Fei},~\bfnm{Li}\binits{L.}}
(\byear{2009}).
\btitle{Imagenet: A large-scale hierarchical image database}.
In \bbooktitle{2009 IEEE conference on computer vision and pattern recognition}
\bpages{248--255}.
\bpublisher{Ieee}.
\end{binproceedings}
\endbibitem

\bibitem[\protect\citeauthoryear{Di~Saverio, Gutierrez and
  Avisar}{2008}]{di2008retrospective}
\begin{barticle}[author]
\bauthor{\bsnm{Di~Saverio},~\bfnm{Salomone}\binits{S.}},
  \bauthor{\bsnm{Gutierrez},~\bfnm{Juan}\binits{J.}} \AND
  \bauthor{\bsnm{Avisar},~\bfnm{Eli}\binits{E.}}
(\byear{2008}).
\btitle{A retrospective review with long term follow up of 11,400 cases of pure
  mucinous breast carcinoma}.
\bjournal{Breast cancer research and treatment}
\bvolume{111}
\bpages{541--547}.
\end{barticle}
\endbibitem

\bibitem[\protect\citeauthoryear{Diab et~al.}{1999}]{diab1999tumor}
\begin{barticle}[author]
\bauthor{\bsnm{Diab},~\bfnm{Sami~G}\binits{S.~G.}},
  \bauthor{\bsnm{Clark},~\bfnm{Gary~M}\binits{G.~M.}},
  \bauthor{\bsnm{Osborne},~\bfnm{C~Kent}\binits{C.~K.}},
  \bauthor{\bsnm{Libby},~\bfnm{Arlene}\binits{A.}},
  \bauthor{\bsnm{Allred},~\bfnm{D~Craig}\binits{D.~C.}} \AND
  \bauthor{\bsnm{Elledge},~\bfnm{Richard~M}\binits{R.~M.}}
(\byear{1999}).
\btitle{Tumor characteristics and clinical outcome of tubular and mucinous
  breast carcinomas}.
\bjournal{Journal of clinical oncology}
\bvolume{17}
\bpages{1442--1442}.
\end{barticle}
\endbibitem

\bibitem[\protect\citeauthoryear{Draper et~al.}{2014}]{draper2014flag}
\begin{barticle}[author]
\bauthor{\bsnm{Draper},~\bfnm{Bruce}\binits{B.}},
  \bauthor{\bsnm{Kirby},~\bfnm{Michael}\binits{M.}},
  \bauthor{\bsnm{Marks},~\bfnm{Justin}\binits{J.}},
  \bauthor{\bsnm{Marrinan},~\bfnm{Tim}\binits{T.}} \AND
  \bauthor{\bsnm{Peterson},~\bfnm{Chris}\binits{C.}}
(\byear{2014}).
\btitle{A flag representation for finite collections of subspaces of mixed
  dimensions}.
\bjournal{Linear Algebra and its Applications}
\bvolume{451}
\bpages{15--32}.
\end{barticle}
\endbibitem

\bibitem[\protect\citeauthoryear{Eiro et~al.}{2019}]{eiro2019breast}
\begin{barticle}[author]
\bauthor{\bsnm{Eiro},~\bfnm{Noemi}\binits{N.}},
  \bauthor{\bsnm{Gonzalez},~\bfnm{Luis~O}\binits{L.~O.}},
  \bauthor{\bsnm{Fraile},~\bfnm{Mar{\'\i}a}\binits{M.}},
  \bauthor{\bsnm{Cid},~\bfnm{Sandra}\binits{S.}},
  \bauthor{\bsnm{Schneider},~\bfnm{Jose}\binits{J.}} \AND
  \bauthor{\bsnm{Vizoso},~\bfnm{Francisco~J}\binits{F.~J.}}
(\byear{2019}).
\btitle{Breast cancer tumor stroma: cellular components, phenotypic
  heterogeneity, intercellular communication, prognostic implications and
  therapeutic opportunities}.
\bjournal{Cancers}
\bvolume{11}
\bpages{664}.
\end{barticle}
\endbibitem

\bibitem[\protect\citeauthoryear{Elmore et~al.}{2015}]{elmore2015diagnostic}
\begin{barticle}[author]
\bauthor{\bsnm{Elmore},~\bfnm{Joann~G}\binits{J.~G.}},
  \bauthor{\bsnm{Longton},~\bfnm{Gary~M}\binits{G.~M.}},
  \bauthor{\bsnm{Carney},~\bfnm{Patricia~A}\binits{P.~A.}},
  \bauthor{\bsnm{Geller},~\bfnm{Berta~M}\binits{B.~M.}},
  \bauthor{\bsnm{Onega},~\bfnm{Tracy}\binits{T.}},
  \bauthor{\bsnm{Tosteson},~\bfnm{Anna~NA}\binits{A.~N.}},
  \bauthor{\bsnm{Nelson},~\bfnm{Heidi~D}\binits{H.~D.}},
  \bauthor{\bsnm{Pepe},~\bfnm{Margaret~S}\binits{M.~S.}},
  \bauthor{\bsnm{Allison},~\bfnm{Kimberly~H}\binits{K.~H.}},
  \bauthor{\bsnm{Schnitt},~\bfnm{Stuart~J}\binits{S.~J.}} \betal{et~al.}
(\byear{2015}).
\btitle{Diagnostic concordance among pathologists interpreting breast biopsy
  specimens}.
\bjournal{Jama}
\bvolume{313}
\bpages{1122--1132}.
\end{barticle}
\endbibitem

\bibitem[\protect\citeauthoryear{Elston and
  Ellis}{2002}]{elston2002pathological}
\begin{barticle}[author]
\bauthor{\bsnm{Elston},~\bfnm{Christopher~W}\binits{C.~W.}} \AND
  \bauthor{\bsnm{Ellis},~\bfnm{Ian~O}\binits{I.~O.}}
(\byear{2002}).
\btitle{Pathological prognostic factors in breast cancer. I. The value of
  histological grade in breast cancer: experience from a large study with
  long-term follow-up. CW Elston \& IO Ellis. Histopathology 1991; 19;
  403--410: AUTHOR COMMENTARY}.
\bjournal{Histopathology}
\bvolume{41}
\bpages{151--151}.
\end{barticle}
\endbibitem

\bibitem[\protect\citeauthoryear{Feng et~al.}{2018}]{feng2018angle}
\begin{barticle}[author]
\bauthor{\bsnm{Feng},~\bfnm{Qing}\binits{Q.}},
  \bauthor{\bsnm{Jiang},~\bfnm{Meilei}\binits{M.}},
  \bauthor{\bsnm{Hannig},~\bfnm{Jan}\binits{J.}} \AND
  \bauthor{\bsnm{Marron},~\bfnm{JS}\binits{J.}}
(\byear{2018}).
\btitle{Angle-based joint and individual variation explained}.
\bjournal{Journal of multivariate analysis}
\bvolume{166}
\bpages{241--265}.
\end{barticle}
\endbibitem

\bibitem[\protect\citeauthoryear{Gaynanova and
  Li}{2017}]{gaynanova2017structural}
\begin{barticle}[author]
\bauthor{\bsnm{Gaynanova},~\bfnm{Irina}\binits{I.}} \AND
  \bauthor{\bsnm{Li},~\bfnm{Gen}\binits{G.}}
(\byear{2017}).
\btitle{Structural learning and integrative decomposition of multi-view data}.
\bjournal{arXiv preprint arXiv:1707.06573}.
\end{barticle}
\endbibitem

\bibitem[\protect\citeauthoryear{Goodfellow
  et~al.}{2014}]{goodfellow2014generative}
\begin{binproceedings}[author]
\bauthor{\bsnm{Goodfellow},~\bfnm{Ian}\binits{I.}},
  \bauthor{\bsnm{Pouget-Abadie},~\bfnm{Jean}\binits{J.}},
  \bauthor{\bsnm{Mirza},~\bfnm{Mehdi}\binits{M.}},
  \bauthor{\bsnm{Xu},~\bfnm{Bing}\binits{B.}},
  \bauthor{\bsnm{Warde-Farley},~\bfnm{David}\binits{D.}},
  \bauthor{\bsnm{Ozair},~\bfnm{Sherjil}\binits{S.}},
  \bauthor{\bsnm{Courville},~\bfnm{Aaron}\binits{A.}} \AND
  \bauthor{\bsnm{Bengio},~\bfnm{Yoshua}\binits{Y.}}
(\byear{2014}).
\btitle{Generative adversarial nets}.
In \bbooktitle{Advances in neural information processing systems}
\bpages{2672--2680}.
\end{binproceedings}
\endbibitem

\bibitem[\protect\citeauthoryear{Heng et~al.}{2017}]{heng2017molecular}
\begin{barticle}[author]
\bauthor{\bsnm{Heng},~\bfnm{Yujing~J}\binits{Y.~J.}},
  \bauthor{\bsnm{Lester},~\bfnm{Susan~C}\binits{S.~C.}},
  \bauthor{\bsnm{Tse},~\bfnm{Gary~MK}\binits{G.~M.}},
  \bauthor{\bsnm{Factor},~\bfnm{Rachel~E}\binits{R.~E.}},
  \bauthor{\bsnm{Allison},~\bfnm{Kimberly~H}\binits{K.~H.}},
  \bauthor{\bsnm{Collins},~\bfnm{Laura~C}\binits{L.~C.}},
  \bauthor{\bsnm{Chen},~\bfnm{Yunn-Yi}\binits{Y.-Y.}},
  \bauthor{\bsnm{Jensen},~\bfnm{Kristin~C}\binits{K.~C.}},
  \bauthor{\bsnm{Johnson},~\bfnm{Nicole~B}\binits{N.~B.}},
  \bauthor{\bsnm{Jeong},~\bfnm{Jong~Cheol}\binits{J.~C.}} \betal{et~al.}
(\byear{2017}).
\btitle{The molecular basis of breast cancer pathological phenotypes}.
\bjournal{The Journal of pathology}
\bvolume{241}
\bpages{375--391}.
\end{barticle}
\endbibitem

\bibitem[\protect\citeauthoryear{Holzinger
  et~al.}{2019}]{holzinger2019causability}
\begin{barticle}[author]
\bauthor{\bsnm{Holzinger},~\bfnm{Andreas}\binits{A.}},
  \bauthor{\bsnm{Langs},~\bfnm{Georg}\binits{G.}},
  \bauthor{\bsnm{Denk},~\bfnm{Helmut}\binits{H.}},
  \bauthor{\bsnm{Zatloukal},~\bfnm{Kurt}\binits{K.}} \AND
  \bauthor{\bsnm{M{\"u}ller},~\bfnm{Heimo}\binits{H.}}
(\byear{2019}).
\btitle{Causability and explainabilty of artificial intelligence in medicine}.
\bjournal{Wiley Interdisciplinary Reviews: Data Mining and Knowledge Discovery}
\bpages{e1312}.
\end{barticle}
\endbibitem

\bibitem[\protect\citeauthoryear{Hotelling}{1936}]{hotelling1936relation}
\begin{barticle}[author]
\bauthor{\bsnm{Hotelling},~\bfnm{Harold}\binits{H.}}
(\byear{1936}).
\btitle{Relation between two sets of variates}.
\bjournal{Biometrica}.
\end{barticle}
\endbibitem

\bibitem[\protect\citeauthoryear{Hunter}{2007}]{hunter2007matplotlib}
\begin{barticle}[author]
\bauthor{\bsnm{Hunter},~\bfnm{John~D}\binits{J.~D.}}
(\byear{2007}).
\btitle{Matplotlib: A 2D graphics environment}.
\bjournal{Computing in science \& engineering}
\bvolume{9}
\bpages{90}.
\end{barticle}
\endbibitem

\bibitem[\protect\citeauthoryear{Ilse, Tomczak and
  Welling}{2018}]{ilse2018attention}
\begin{barticle}[author]
\bauthor{\bsnm{Ilse},~\bfnm{Maximilian}\binits{M.}},
  \bauthor{\bsnm{Tomczak},~\bfnm{Jakub~M}\binits{J.~M.}} \AND
  \bauthor{\bsnm{Welling},~\bfnm{Max}\binits{M.}}
(\byear{2018}).
\btitle{Attention-based deep multiple instance learning}.
\bjournal{arXiv preprint arXiv:1802.04712}.
\end{barticle}
\endbibitem

\bibitem[\protect\citeauthoryear{Jim{\'e}nez and
  Racoceanu}{2019}]{jimenez2019deep}
\begin{barticle}[author]
\bauthor{\bsnm{Jim{\'e}nez},~\bfnm{Gabriel}\binits{G.}} \AND
  \bauthor{\bsnm{Racoceanu},~\bfnm{Daniel}\binits{D.}}
(\byear{2019}).
\btitle{Deep Learning for Semantic Segmentation versus Classification in
  Computational Pathology: Application to mitosis analysis in Breast Cancer
  grading}.
\bjournal{Frontiers in Bioengineering and Biotechnology}
\bvolume{7}
\bpages{145}.
\end{barticle}
\endbibitem

\bibitem[\protect\citeauthoryear{Johnstone}{2008}]{johnstone2008multivariate}
\begin{barticle}[author]
\bauthor{\bsnm{Johnstone},~\bfnm{Iain~M}\binits{I.~M.}}
(\byear{2008}).
\btitle{Multivariate analysis and Jacobi ensembles: Largest eigenvalue,
  Tracy--Widom limits and rates of convergence}.
\bjournal{Annals of statistics}
\bvolume{36}
\bpages{2638}.
\end{barticle}
\endbibitem

\bibitem[\protect\citeauthoryear{Jones, Oliphant and
  Peterson}{2014}]{jones2014scipy}
\begin{barticle}[author]
\bauthor{\bsnm{Jones},~\bfnm{Eric}\binits{E.}},
  \bauthor{\bsnm{Oliphant},~\bfnm{Travis}\binits{T.}} \AND
  \bauthor{\bsnm{Peterson},~\bfnm{Pearu}\binits{P.}}
(\byear{2014}).
\btitle{SciPy: Open source scientific tools for Python}.
\end{barticle}
\endbibitem

\bibitem[\protect\citeauthoryear{Kettenring}{1971}]{kettenring1971canonical}
\begin{barticle}[author]
\bauthor{\bsnm{Kettenring},~\bfnm{Jon~R}\binits{J.~R.}}
(\byear{1971}).
\btitle{Canonical analysis of several sets of variables}.
\bjournal{Biometrika}
\bvolume{58}
\bpages{433--451}.
\end{barticle}
\endbibitem

\bibitem[\protect\citeauthoryear{Kim et~al.}{2018}]{kim2018interpretability}
\begin{binproceedings}[author]
\bauthor{\bsnm{Kim},~\bfnm{Been}\binits{B.}},
  \bauthor{\bsnm{Wattenberg},~\bfnm{Martin}\binits{M.}},
  \bauthor{\bsnm{Gilmer},~\bfnm{Justin}\binits{J.}},
  \bauthor{\bsnm{Cai},~\bfnm{Carrie}\binits{C.}},
  \bauthor{\bsnm{Wexler},~\bfnm{James}\binits{J.}},
  \bauthor{\bsnm{Viegas},~\bfnm{Fernanda}\binits{F.}} \betal{et~al.}
(\byear{2018}).
\btitle{Interpretability Beyond Feature Attribution: Quantitative Testing with
  Concept Activation Vectors (TCAV)}.
In \bbooktitle{International Conference on Machine Learning}
\bpages{2673--2682}.
\end{binproceedings}
\endbibitem

\bibitem[\protect\citeauthoryear{Kingma and Welling}{2013}]{kingma2013auto}
\begin{barticle}[author]
\bauthor{\bsnm{Kingma},~\bfnm{Diederik~P}\binits{D.~P.}} \AND
  \bauthor{\bsnm{Welling},~\bfnm{Max}\binits{M.}}
(\byear{2013}).
\btitle{Auto-encoding variational bayes}.
\bjournal{arXiv preprint arXiv:1312.6114}.
\end{barticle}
\endbibitem

\bibitem[\protect\citeauthoryear{Komura and Ishikawa}{2018}]{komura2018machine}
\begin{barticle}[author]
\bauthor{\bsnm{Komura},~\bfnm{Daisuke}\binits{D.}} \AND
  \bauthor{\bsnm{Ishikawa},~\bfnm{Shumpei}\binits{S.}}
(\byear{2018}).
\btitle{Machine learning methods for histopathological image analysis}.
\bjournal{Computational and structural biotechnology journal}
\bvolume{16}
\bpages{34--42}.
\end{barticle}
\endbibitem

\bibitem[\protect\citeauthoryear{Lacroix-Triki
  et~al.}{2010}]{lacroix2010mucinous}
\begin{barticle}[author]
\bauthor{\bsnm{Lacroix-Triki},~\bfnm{Magali}\binits{M.}},
  \bauthor{\bsnm{Suarez},~\bfnm{Paula~H}\binits{P.~H.}},
  \bauthor{\bsnm{MacKay},~\bfnm{Alan}\binits{A.}},
  \bauthor{\bsnm{Lambros},~\bfnm{Maryou~B}\binits{M.~B.}},
  \bauthor{\bsnm{Natrajan},~\bfnm{Rachael}\binits{R.}},
  \bauthor{\bsnm{Savage},~\bfnm{Kay}\binits{K.}},
  \bauthor{\bsnm{Geyer},~\bfnm{Felipe~C}\binits{F.~C.}},
  \bauthor{\bsnm{Weigelt},~\bfnm{Britta}\binits{B.}},
  \bauthor{\bsnm{Ashworth},~\bfnm{Alan}\binits{A.}} \AND
  \bauthor{\bsnm{Reis-Filho},~\bfnm{Jorge~S}\binits{J.~S.}}
(\byear{2010}).
\btitle{Mucinous carcinoma of the breast is genomically distinct from invasive
  ductal carcinomas of no special type}.
\bjournal{The Journal of pathology}
\bvolume{222}
\bpages{282--298}.
\end{barticle}
\endbibitem

\bibitem[\protect\citeauthoryear{Lazard et~al.}{1993}]{lazard1993expression}
\begin{barticle}[author]
\bauthor{\bsnm{Lazard},~\bfnm{Daniel}\binits{D.}},
  \bauthor{\bsnm{Sastre},~\bfnm{Xavier}\binits{X.}},
  \bauthor{\bsnm{Frid},~\bfnm{Maria~G}\binits{M.~G.}},
  \bauthor{\bsnm{Glukhova},~\bfnm{Marina~A}\binits{M.~A.}},
  \bauthor{\bsnm{Thiery},~\bfnm{Jean-Paul}\binits{J.-P.}} \AND
  \bauthor{\bsnm{Koteliansky},~\bfnm{Victor~E}\binits{V.~E.}}
(\byear{1993}).
\btitle{Expression of smooth muscle-specific proteins in myoepithelium and
  stromal myofibroblasts of normal and malignant human breast tissue.}
\bjournal{Proceedings of the National Academy of Sciences}
\bvolume{90}
\bpages{999--1003}.
\end{barticle}
\endbibitem

\bibitem[\protect\citeauthoryear{Liu et~al.}{2017}]{liu2017detecting}
\begin{barticle}[author]
\bauthor{\bsnm{Liu},~\bfnm{Yun}\binits{Y.}},
  \bauthor{\bsnm{Gadepalli},~\bfnm{Krishna}\binits{K.}},
  \bauthor{\bsnm{Norouzi},~\bfnm{Mohammad}\binits{M.}},
  \bauthor{\bsnm{Dahl},~\bfnm{George~E}\binits{G.~E.}},
  \bauthor{\bsnm{Kohlberger},~\bfnm{Timo}\binits{T.}},
  \bauthor{\bsnm{Boyko},~\bfnm{Aleksey}\binits{A.}},
  \bauthor{\bsnm{Venugopalan},~\bfnm{Subhashini}\binits{S.}},
  \bauthor{\bsnm{Timofeev},~\bfnm{Aleksei}\binits{A.}},
  \bauthor{\bsnm{Nelson},~\bfnm{Philip~Q}\binits{P.~Q.}},
  \bauthor{\bsnm{Corrado},~\bfnm{Greg~S}\binits{G.~S.}} \betal{et~al.}
(\byear{2017}).
\btitle{Detecting cancer metastases on gigapixel pathology images}.
\bjournal{arXiv preprint arXiv:1703.02442}.
\end{barticle}
\endbibitem

\bibitem[\protect\citeauthoryear{Liu et~al.}{2018}]{liu2018artificial}
\begin{barticle}[author]
\bauthor{\bsnm{Liu},~\bfnm{Yun}\binits{Y.}},
  \bauthor{\bsnm{Kohlberger},~\bfnm{Timo}\binits{T.}},
  \bauthor{\bsnm{Norouzi},~\bfnm{Mohammad}\binits{M.}},
  \bauthor{\bsnm{Dahl},~\bfnm{George~E}\binits{G.~E.}},
  \bauthor{\bsnm{Smith},~\bfnm{Jenny~L}\binits{J.~L.}},
  \bauthor{\bsnm{Mohtashamian},~\bfnm{Arash}\binits{A.}},
  \bauthor{\bsnm{Olson},~\bfnm{Niels}\binits{N.}},
  \bauthor{\bsnm{Peng},~\bfnm{Lily~H}\binits{L.~H.}},
  \bauthor{\bsnm{Hipp},~\bfnm{Jason~D}\binits{J.~D.}} \AND
  \bauthor{\bsnm{Stumpe},~\bfnm{Martin~C}\binits{M.~C.}}
(\byear{2018}).
\btitle{Artificial Intelligence--Based Breast Cancer Nodal Metastasis
  Detection: Insights Into the Black Box for Pathologists}.
\bjournal{Archives of pathology \& laboratory medicine}.
\end{barticle}
\endbibitem

\bibitem[\protect\citeauthoryear{Livasy et~al.}{2006}]{livasy2006phenotypic}
\begin{barticle}[author]
\bauthor{\bsnm{Livasy},~\bfnm{Chad~A}\binits{C.~A.}},
  \bauthor{\bsnm{Karaca},~\bfnm{Gamze}\binits{G.}},
  \bauthor{\bsnm{Nanda},~\bfnm{Rita}\binits{R.}},
  \bauthor{\bsnm{Tretiakova},~\bfnm{Maria~S}\binits{M.~S.}},
  \bauthor{\bsnm{Olopade},~\bfnm{Olufunmilayo~I}\binits{O.~I.}},
  \bauthor{\bsnm{Moore},~\bfnm{Dominic~T}\binits{D.~T.}} \AND
  \bauthor{\bsnm{Perou},~\bfnm{Charles~M}\binits{C.~M.}}
(\byear{2006}).
\btitle{Phenotypic evaluation of the basal-like subtype of invasive breast
  carcinoma}.
\bjournal{Modern pathology}
\bvolume{19}
\bpages{264}.
\end{barticle}
\endbibitem

\bibitem[\protect\citeauthoryear{Lock et~al.}{2013}]{lock2013joint}
\begin{barticle}[author]
\bauthor{\bsnm{Lock},~\bfnm{Eric~F}\binits{E.~F.}},
  \bauthor{\bsnm{Hoadley},~\bfnm{Katherine~A}\binits{K.~A.}},
  \bauthor{\bsnm{Marron},~\bfnm{James~Stephen}\binits{J.~S.}} \AND
  \bauthor{\bsnm{Nobel},~\bfnm{Andrew~B}\binits{A.~B.}}
(\byear{2013}).
\btitle{Joint and individual variation explained (JIVE) for integrated analysis
  of multiple data types}.
\bjournal{The annals of applied statistics}
\bvolume{7}
\bpages{523}.
\end{barticle}
\endbibitem

\bibitem[\protect\citeauthoryear{Lu et~al.}{2019}]{lu2019semi}
\begin{barticle}[author]
\bauthor{\bsnm{Lu},~\bfnm{Ming~Y}\binits{M.~Y.}},
  \bauthor{\bsnm{Chen},~\bfnm{Richard~J}\binits{R.~J.}},
  \bauthor{\bsnm{Wang},~\bfnm{Jingwen}\binits{J.}},
  \bauthor{\bsnm{Dillon},~\bfnm{Debora}\binits{D.}} \AND
  \bauthor{\bsnm{Mahmood},~\bfnm{Faisal}\binits{F.}}
(\byear{2019}).
\btitle{Semi-Supervised Histology Classification using Deep Multiple Instance
  Learning and Contrastive Predictive Coding}.
\bjournal{arXiv preprint arXiv:1910.10825}.
\end{barticle}
\endbibitem

\bibitem[\protect\citeauthoryear{Macenko et~al.}{2009}]{macenko2009method}
\begin{binproceedings}[author]
\bauthor{\bsnm{Macenko},~\bfnm{Marc}\binits{M.}},
  \bauthor{\bsnm{Niethammer},~\bfnm{Marc}\binits{M.}},
  \bauthor{\bsnm{Marron},~\bfnm{James~S}\binits{J.~S.}},
  \bauthor{\bsnm{Borland},~\bfnm{David}\binits{D.}},
  \bauthor{\bsnm{Woosley},~\bfnm{John~T}\binits{J.~T.}},
  \bauthor{\bsnm{Guan},~\bfnm{Xiaojun}\binits{X.}},
  \bauthor{\bsnm{Schmitt},~\bfnm{Charles}\binits{C.}} \AND
  \bauthor{\bsnm{Thomas},~\bfnm{Nancy~E}\binits{N.~E.}}
(\byear{2009}).
\btitle{A method for normalizing histology slides for quantitative analysis}.
In \bbooktitle{2009 IEEE International Symposium on Biomedical Imaging: From
  Nano to Macro}
\bpages{1107--1110}.
\bpublisher{IEEE}.
\end{binproceedings}
\endbibitem

\bibitem[\protect\citeauthoryear{Mahmood et~al.}{2018}]{mahmood2018multimodal}
\begin{barticle}[author]
\bauthor{\bsnm{Mahmood},~\bfnm{Faisal}\binits{F.}},
  \bauthor{\bsnm{Yang},~\bfnm{Ziyun}\binits{Z.}},
  \bauthor{\bsnm{Ashley},~\bfnm{Thomas}\binits{T.}} \AND
  \bauthor{\bsnm{Durr},~\bfnm{Nicholas~J}\binits{N.~J.}}
(\byear{2018}).
\btitle{Multimodal densenet}.
\bjournal{arXiv preprint arXiv:1811.07407}.
\end{barticle}
\endbibitem

\bibitem[\protect\citeauthoryear{Mahmood et~al.}{2019}]{mahmood2019deep}
\begin{barticle}[author]
\bauthor{\bsnm{Mahmood},~\bfnm{Faisal}\binits{F.}},
  \bauthor{\bsnm{Borders},~\bfnm{Daniel}\binits{D.}},
  \bauthor{\bsnm{Chen},~\bfnm{Richard}\binits{R.}},
  \bauthor{\bsnm{McKay},~\bfnm{Gregory~N}\binits{G.~N.}},
  \bauthor{\bsnm{Salimian},~\bfnm{Kevan~J}\binits{K.~J.}},
  \bauthor{\bsnm{Baras},~\bfnm{Alexander}\binits{A.}} \AND
  \bauthor{\bsnm{Durr},~\bfnm{Nicholas~J}\binits{N.~J.}}
(\byear{2019}).
\btitle{Deep adversarial training for multi-organ nuclei segmentation in
  histopathology images}.
\bjournal{IEEE transactions on medical imaging}.
\end{barticle}
\endbibitem

\bibitem[\protect\citeauthoryear{McKinney}{2011}]{mckinney2011pandas}
\begin{barticle}[author]
\bauthor{\bsnm{McKinney},~\bfnm{Wes}\binits{W.}}
(\byear{2011}).
\btitle{Pandas: a foundational Python library for data analysis and
  statistics}.
\bjournal{Python for High Performance and Scientific Computing}
\bvolume{14}.
\end{barticle}
\endbibitem

\bibitem[\protect\citeauthoryear{Molnar et~al.}{2018}]{molnar2018interpretable}
\begin{barticle}[author]
\bauthor{\bsnm{Molnar},~\bfnm{Christoph}\binits{C.}} \betal{et~al.}
(\byear{2018}).
\btitle{Interpretable machine learning: A guide for making black box models
  explainable}.
\bjournal{E-book at< https://christophm. github. io/interpretable-ml-book/>,
  version dated}
\bvolume{10}.
\end{barticle}
\endbibitem

\bibitem[\protect\citeauthoryear{Network
  et~al.}{2012}]{cancer2012comprehensive}
\begin{barticle}[author]
\bauthor{\bsnm{Network},~\bfnm{Cancer Genome~Atlas}\binits{C.~G.~A.}}
  \betal{et~al.}
(\byear{2012}).
\btitle{Comprehensive molecular portraits of human breast tumours}.
\bjournal{Nature}
\bvolume{490}
\bpages{61}.
\end{barticle}
\endbibitem

\bibitem[\protect\citeauthoryear{Oh et~al.}{2006}]{oh2006estrogen}
\begin{barticle}[author]
\bauthor{\bsnm{Oh},~\bfnm{Daniel~S}\binits{D.~S.}},
  \bauthor{\bsnm{Troester},~\bfnm{Melissa~A}\binits{M.~A.}},
  \bauthor{\bsnm{Usary},~\bfnm{Jerry}\binits{J.}},
  \bauthor{\bsnm{Hu},~\bfnm{Zhiyuan}\binits{Z.}},
  \bauthor{\bsnm{He},~\bfnm{Xiaping}\binits{X.}},
  \bauthor{\bsnm{Fan},~\bfnm{Cheng}\binits{C.}},
  \bauthor{\bsnm{Wu},~\bfnm{Junyuan}\binits{J.}},
  \bauthor{\bsnm{Carey},~\bfnm{Lisa~A}\binits{L.~A.}} \AND
  \bauthor{\bsnm{Perou},~\bfnm{Charles~M}\binits{C.~M.}}
(\byear{2006}).
\btitle{Estrogen-regulated genes predict survival in hormone receptor-positive
  breast cancers}.
\bjournal{J Clin Oncol}
\bvolume{24}
\bpages{1656--1664}.
\end{barticle}
\endbibitem

\bibitem[\protect\citeauthoryear{Olah et~al.}{2018}]{olah2018building}
\begin{barticle}[author]
\bauthor{\bsnm{Olah},~\bfnm{Chris}\binits{C.}},
  \bauthor{\bsnm{Satyanarayan},~\bfnm{Arvind}\binits{A.}},
  \bauthor{\bsnm{Johnson},~\bfnm{Ian}\binits{I.}},
  \bauthor{\bsnm{Carter},~\bfnm{Shan}\binits{S.}},
  \bauthor{\bsnm{Schubert},~\bfnm{Ludwig}\binits{L.}},
  \bauthor{\bsnm{Ye},~\bfnm{Katherine}\binits{K.}} \AND
  \bauthor{\bsnm{Mordvintsev},~\bfnm{Alexander}\binits{A.}}
(\byear{2018}).
\btitle{The building blocks of interpretability}.
\bjournal{Distill}
\bvolume{3}
\bpages{e10}.
\end{barticle}
\endbibitem

\bibitem[\protect\citeauthoryear{Oord, Li and
  Vinyals}{2018}]{oord2018representation}
\begin{barticle}[author]
\bauthor{\bsnm{Oord},~\bfnm{Aaron van~den}\binits{A.~v.~d.}},
  \bauthor{\bsnm{Li},~\bfnm{Yazhe}\binits{Y.}} \AND
  \bauthor{\bsnm{Vinyals},~\bfnm{Oriol}\binits{O.}}
(\byear{2018}).
\btitle{Representation learning with contrastive predictive coding}.
\bjournal{arXiv preprint arXiv:1807.03748}.
\end{barticle}
\endbibitem

\bibitem[\protect\citeauthoryear{Otsu}{1979}]{otsu1979threshold}
\begin{barticle}[author]
\bauthor{\bsnm{Otsu},~\bfnm{Nobuyuki}\binits{N.}}
(\byear{1979}).
\btitle{A threshold selection method from gray-level histograms}.
\bjournal{IEEE transactions on systems, man, and cybernetics}
\bvolume{9}
\bpages{62--66}.
\end{barticle}
\endbibitem

\bibitem[\protect\citeauthoryear{Parker et~al.}{2009}]{parker2009supervised}
\begin{barticle}[author]
\bauthor{\bsnm{Parker},~\bfnm{Joel~S}\binits{J.~S.}},
  \bauthor{\bsnm{Mullins},~\bfnm{Michael}\binits{M.}},
  \bauthor{\bsnm{Cheang},~\bfnm{Maggie~CU}\binits{M.~C.}},
  \bauthor{\bsnm{Leung},~\bfnm{Samuel}\binits{S.}},
  \bauthor{\bsnm{Voduc},~\bfnm{David}\binits{D.}},
  \bauthor{\bsnm{Vickery},~\bfnm{Tammi}\binits{T.}},
  \bauthor{\bsnm{Davies},~\bfnm{Sherri}\binits{S.}},
  \bauthor{\bsnm{Fauron},~\bfnm{Christiane}\binits{C.}},
  \bauthor{\bsnm{He},~\bfnm{Xiaping}\binits{X.}},
  \bauthor{\bsnm{Hu},~\bfnm{Zhiyuan}\binits{Z.}} \betal{et~al.}
(\byear{2009}).
\btitle{Supervised risk predictor of breast cancer based on intrinsic
  subtypes}.
\bjournal{Journal of clinical oncology}
\bvolume{27}
\bpages{1160}.
\end{barticle}
\endbibitem

\bibitem[\protect\citeauthoryear{Paszke et~al.}{2017}]{paszke2017automatic}
\begin{barticle}[author]
\bauthor{\bsnm{Paszke},~\bfnm{Adam}\binits{A.}},
  \bauthor{\bsnm{Gross},~\bfnm{Sam}\binits{S.}},
  \bauthor{\bsnm{Chintala},~\bfnm{Soumith}\binits{S.}},
  \bauthor{\bsnm{Chanan},~\bfnm{Gregory}\binits{G.}},
  \bauthor{\bsnm{Yang},~\bfnm{Edward}\binits{E.}},
  \bauthor{\bsnm{DeVito},~\bfnm{Zachary}\binits{Z.}},
  \bauthor{\bsnm{Lin},~\bfnm{Zeming}\binits{Z.}},
  \bauthor{\bsnm{Desmaison},~\bfnm{Alban}\binits{A.}},
  \bauthor{\bsnm{Antiga},~\bfnm{Luca}\binits{L.}} \AND
  \bauthor{\bsnm{Lerer},~\bfnm{Adam}\binits{A.}}
(\byear{2017}).
\btitle{Automatic differentiation in pytorch}.
\end{barticle}
\endbibitem

\bibitem[\protect\citeauthoryear{Pedregosa et~al.}{2011}]{scikit2011pedregosa}
\begin{barticle}[author]
\bauthor{\bsnm{Pedregosa},~\bfnm{F.}\binits{F.}},
  \bauthor{\bsnm{Varoquaux},~\bfnm{G.}\binits{G.}},
  \bauthor{\bsnm{Gramfort},~\bfnm{A.}\binits{A.}},
  \bauthor{\bsnm{Michel},~\bfnm{V.}\binits{V.}},
  \bauthor{\bsnm{Thirion},~\bfnm{B.}\binits{B.}},
  \bauthor{\bsnm{Grisel},~\bfnm{O.}\binits{O.}},
  \bauthor{\bsnm{Blondel},~\bfnm{M.}\binits{M.}},
  \bauthor{\bsnm{Prettenhofer},~\bfnm{P.}\binits{P.}},
  \bauthor{\bsnm{Weiss},~\bfnm{R.}\binits{R.}},
  \bauthor{\bsnm{Dubourg},~\bfnm{V.}\binits{V.}},
  \bauthor{\bsnm{Vanderplas},~\bfnm{J.}\binits{J.}},
  \bauthor{\bsnm{Passos},~\bfnm{A.}\binits{A.}},
  \bauthor{\bsnm{Cournapeau},~\bfnm{D.}\binits{D.}},
  \bauthor{\bsnm{Brucher},~\bfnm{M.}\binits{M.}},
  \bauthor{\bsnm{Perrot},~\bfnm{M.}\binits{M.}} \AND
  \bauthor{\bsnm{Duchesnay},~\bfnm{E.}\binits{E.}}
(\byear{2011}).
\btitle{{Scikit-learn: Machine Learning in Python }}.
\bjournal{Journal of Machine Learning Research}
\bvolume{12}
\bpages{2825--2830}.
\end{barticle}
\endbibitem

\bibitem[\protect\citeauthoryear{Perou et~al.}{2000}]{perou2000molecular}
\begin{barticle}[author]
\bauthor{\bsnm{Perou},~\bfnm{Charles~M}\binits{C.~M.}},
  \bauthor{\bsnm{S{\o}rlie},~\bfnm{Therese}\binits{T.}},
  \bauthor{\bsnm{Eisen},~\bfnm{Michael~B}\binits{M.~B.}}, \bauthor{\bsnm{Van
  De~Rijn},~\bfnm{Matt}\binits{M.}},
  \bauthor{\bsnm{Jeffrey},~\bfnm{Stefanie~S}\binits{S.~S.}},
  \bauthor{\bsnm{Rees},~\bfnm{Christian~A}\binits{C.~A.}},
  \bauthor{\bsnm{Pollack},~\bfnm{Jonathan~R}\binits{J.~R.}},
  \bauthor{\bsnm{Ross},~\bfnm{Douglas~T}\binits{D.~T.}},
  \bauthor{\bsnm{Johnsen},~\bfnm{Hilde}\binits{H.}},
  \bauthor{\bsnm{Akslen},~\bfnm{Lars~A}\binits{L.~A.}} \betal{et~al.}
(\byear{2000}).
\btitle{Molecular portraits of human breast tumours}.
\bjournal{nature}
\bvolume{406}
\bpages{747}.
\end{barticle}
\endbibitem

\bibitem[\protect\citeauthoryear{Pourzanjani
  et~al.}{2017}]{pourzanjani2017understanding}
\begin{binproceedings}[author]
\bauthor{\bsnm{Pourzanjani},~\bfnm{Arya~A}\binits{A.~A.}},
  \bauthor{\bsnm{Wu},~\bfnm{Tie~Bo}\binits{T.~B.}},
  \bauthor{\bsnm{Jiang},~\bfnm{Richard~M}\binits{R.~M.}},
  \bauthor{\bsnm{Cohen},~\bfnm{Mitchell~J}\binits{M.~J.}} \AND
  \bauthor{\bsnm{Petzold},~\bfnm{Linda~R}\binits{L.~R.}}
(\byear{2017}).
\btitle{Understanding Coagulopathy using Multi-view Data in the Presence of
  Sub-Cohorts: A Hierarchical Subspace Approach}.
In \bbooktitle{Machine Learning for Healthcare Conference}
\bpages{338--351}.
\end{binproceedings}
\endbibitem

\bibitem[\protect\citeauthoryear{Rom{\'a}n-P{\'e}rez
  et~al.}{2012}]{roman2012gene}
\begin{barticle}[author]
\bauthor{\bsnm{Rom{\'a}n-P{\'e}rez},~\bfnm{Erick}\binits{E.}},
  \bauthor{\bsnm{Casbas-Hern{\'a}ndez},~\bfnm{Patricia}\binits{P.}},
  \bauthor{\bsnm{Pirone},~\bfnm{Jason~R}\binits{J.~R.}},
  \bauthor{\bsnm{Rein},~\bfnm{Jessica}\binits{J.}},
  \bauthor{\bsnm{Carey},~\bfnm{Lisa~A}\binits{L.~A.}},
  \bauthor{\bsnm{Lubet},~\bfnm{Ronald~A}\binits{R.~A.}},
  \bauthor{\bsnm{Mani},~\bfnm{Sendurai~A}\binits{S.~A.}},
  \bauthor{\bsnm{Amos},~\bfnm{Keith~D}\binits{K.~D.}} \AND
  \bauthor{\bsnm{Troester},~\bfnm{Melissa~A}\binits{M.~A.}}
(\byear{2012}).
\btitle{Gene expression in extratumoral microenvironment predicts clinical
  outcome in breast cancer patients}.
\bjournal{Breast Cancer Research}
\bvolume{14}
\bpages{R51}.
\end{barticle}
\endbibitem

\bibitem[\protect\citeauthoryear{Rosen}{2001}]{rosen2001rosen}
\begin{bbook}[author]
\bauthor{\bsnm{Rosen},~\bfnm{Paul~Peter}\binits{P.~P.}}
(\byear{2001}).
\btitle{Rosen's breast pathology}.
\bpublisher{Lippincott Williams \& Wilkins}.
\end{bbook}
\endbibitem

\bibitem[\protect\citeauthoryear{Selvaraju et~al.}{2017}]{selvaraju2017grad}
\begin{binproceedings}[author]
\bauthor{\bsnm{Selvaraju},~\bfnm{Ramprasaath~R}\binits{R.~R.}},
  \bauthor{\bsnm{Cogswell},~\bfnm{Michael}\binits{M.}},
  \bauthor{\bsnm{Das},~\bfnm{Abhishek}\binits{A.}},
  \bauthor{\bsnm{Vedantam},~\bfnm{Ramakrishna}\binits{R.}},
  \bauthor{\bsnm{Parikh},~\bfnm{Devi}\binits{D.}} \AND
  \bauthor{\bsnm{Batra},~\bfnm{Dhruv}\binits{D.}}
(\byear{2017}).
\btitle{Grad-cam: Visual explanations from deep networks via gradient-based
  localization}.
In \bbooktitle{2017 IEEE International Conference on Computer Vision (ICCV)}
\bpages{618--626}.
\bpublisher{IEEE}.
\end{binproceedings}
\endbibitem

\bibitem[\protect\citeauthoryear{Sharif~Razavian et~al.}{2014}]{sharif2014cnn}
\begin{binproceedings}[author]
\bauthor{\bsnm{Sharif~Razavian},~\bfnm{Ali}\binits{A.}},
  \bauthor{\bsnm{Azizpour},~\bfnm{Hossein}\binits{H.}},
  \bauthor{\bsnm{Sullivan},~\bfnm{Josephine}\binits{J.}} \AND
  \bauthor{\bsnm{Carlsson},~\bfnm{Stefan}\binits{S.}}
(\byear{2014}).
\btitle{CNN features off-the-shelf: an astounding baseline for recognition}.
In \bbooktitle{Proceedings of the IEEE conference on computer vision and
  pattern recognition workshops}
\bpages{806--813}.
\end{binproceedings}
\endbibitem

\bibitem[\protect\citeauthoryear{Simonyan and
  Zisserman}{2014}]{simonyan2014very}
\begin{barticle}[author]
\bauthor{\bsnm{Simonyan},~\bfnm{Karen}\binits{K.}} \AND
  \bauthor{\bsnm{Zisserman},~\bfnm{Andrew}\binits{A.}}
(\byear{2014}).
\btitle{Very deep convolutional networks for large-scale image recognition}.
\bjournal{arXiv preprint arXiv:1409.1556}.
\end{barticle}
\endbibitem

\bibitem[\protect\citeauthoryear{Springenberg
  et~al.}{2014}]{springenberg2014striving}
\begin{barticle}[author]
\bauthor{\bsnm{Springenberg},~\bfnm{Jost~Tobias}\binits{J.~T.}},
  \bauthor{\bsnm{Dosovitskiy},~\bfnm{Alexey}\binits{A.}},
  \bauthor{\bsnm{Brox},~\bfnm{Thomas}\binits{T.}} \AND
  \bauthor{\bsnm{Riedmiller},~\bfnm{Martin}\binits{M.}}
(\byear{2014}).
\btitle{Striving for simplicity: The all convolutional net}.
\bjournal{arXiv preprint arXiv:1412.6806}.
\end{barticle}
\endbibitem

\bibitem[\protect\citeauthoryear{Srivastava
  et~al.}{2018}]{srivastava2018building}
\begin{binproceedings}[author]
\bauthor{\bsnm{Srivastava},~\bfnm{Arunima}\binits{A.}},
  \bauthor{\bsnm{Kulkarni},~\bfnm{Chaitanya}\binits{C.}},
  \bauthor{\bsnm{Mallick},~\bfnm{Parag}\binits{P.}},
  \bauthor{\bsnm{Huang},~\bfnm{Kun}\binits{K.}} \AND
  \bauthor{\bsnm{Machiraju},~\bfnm{Raghu}\binits{R.}}
(\byear{2018}).
\btitle{Building trans-omics evidence: using imaging and'omics' to characterize
  cancer profiles.}
In \bbooktitle{PSB}
\bpages{377--388}.
\bpublisher{World Scientific}.
\end{binproceedings}
\endbibitem

\bibitem[\protect\citeauthoryear{Sundararajan, Taly and
  Yan}{2017}]{sundararajan2017axiomatic}
\begin{binproceedings}[author]
\bauthor{\bsnm{Sundararajan},~\bfnm{Mukund}\binits{M.}},
  \bauthor{\bsnm{Taly},~\bfnm{Ankur}\binits{A.}} \AND
  \bauthor{\bsnm{Yan},~\bfnm{Qiqi}\binits{Q.}}
(\byear{2017}).
\btitle{Axiomatic attribution for deep networks}.
In \bbooktitle{Proceedings of the 34th International Conference on Machine
  Learning-Volume 70}
\bpages{3319--3328}.
\bpublisher{JMLR.org}.
\end{binproceedings}
\endbibitem

\bibitem[\protect\citeauthoryear{Troester et~al.}{2017}]{troester2017racial}
\begin{barticle}[author]
\bauthor{\bsnm{Troester},~\bfnm{Melissa~A}\binits{M.~A.}},
  \bauthor{\bsnm{Sun},~\bfnm{Xuezheng}\binits{X.}},
  \bauthor{\bsnm{Allott},~\bfnm{Emma~H}\binits{E.~H.}},
  \bauthor{\bsnm{Geradts},~\bfnm{Joseph}\binits{J.}},
  \bauthor{\bsnm{Cohen},~\bfnm{Stephanie~M}\binits{S.~M.}},
  \bauthor{\bsnm{Tse},~\bfnm{Chiu-Kit}\binits{C.-K.}},
  \bauthor{\bsnm{Kirk},~\bfnm{Erin~L}\binits{E.~L.}},
  \bauthor{\bsnm{Thorne},~\bfnm{Leigh~B}\binits{L.~B.}},
  \bauthor{\bsnm{Mathews},~\bfnm{Michelle}\binits{M.}},
  \bauthor{\bsnm{Li},~\bfnm{Yan}\binits{Y.}} \betal{et~al.}
(\byear{2017}).
\btitle{Racial differences in PAM50 subtypes in the Carolina Breast Cancer
  Study}.
\bjournal{JNCI: Journal of the National Cancer Institute}
\bvolume{110}
\bpages{176--182}.
\end{barticle}
\endbibitem

\bibitem[\protect\citeauthoryear{Van Der~Walt, Colbert and
  Varoquaux}{2011}]{van2011numpy}
\begin{barticle}[author]
\bauthor{\bsnm{Van Der~Walt},~\bfnm{Stefan}\binits{S.}},
  \bauthor{\bsnm{Colbert},~\bfnm{S~Chris}\binits{S.~C.}} \AND
  \bauthor{\bsnm{Varoquaux},~\bfnm{Gael}\binits{G.}}
(\byear{2011}).
\btitle{The NumPy array: a structure for efficient numerical computation}.
\bjournal{Computing in Science \& Engineering}
\bvolume{13}
\bpages{22}.
\end{barticle}
\endbibitem

\bibitem[\protect\citeauthoryear{Van~der Walt et~al.}{2014}]{van2014scikit}
\begin{barticle}[author]
\bauthor{\bparticle{Van~der} \bsnm{Walt},~\bfnm{Stefan}\binits{S.}},
  \bauthor{\bsnm{Sch{\"o}nberger},~\bfnm{Johannes~L}\binits{J.~L.}},
  \bauthor{\bsnm{Nunez-Iglesias},~\bfnm{Juan}\binits{J.}},
  \bauthor{\bsnm{Boulogne},~\bfnm{Fran{\c{c}}ois}\binits{F.}},
  \bauthor{\bsnm{Warner},~\bfnm{Joshua~D}\binits{J.~D.}},
  \bauthor{\bsnm{Yager},~\bfnm{Neil}\binits{N.}},
  \bauthor{\bsnm{Gouillart},~\bfnm{Emmanuelle}\binits{E.}} \AND
  \bauthor{\bsnm{Yu},~\bfnm{Tony}\binits{T.}}
(\byear{2014}).
\btitle{scikit-image: image processing in Python}.
\bjournal{PeerJ}
\bvolume{2}
\bpages{e453}.
\end{barticle}
\endbibitem

\bibitem[\protect\citeauthoryear{Vellido, Mart{\'\i}n-Guerrero and
  Lisboa}{2012}]{vellido2012making}
\begin{binproceedings}[author]
\bauthor{\bsnm{Vellido},~\bfnm{Alfredo}\binits{A.}},
  \bauthor{\bsnm{Mart{\'\i}n-Guerrero},~\bfnm{Jos{\'e}~David}\binits{J.~D.}}
  \AND \bauthor{\bsnm{Lisboa},~\bfnm{Paulo~JG}\binits{P.~J.}}
(\byear{2012}).
\btitle{Making machine learning models interpretable.}
In \bbooktitle{ESANN}
\bvolume{12}
\bpages{163--172}.
\bpublisher{Citeseer}.
\end{binproceedings}
\endbibitem

\bibitem[\protect\citeauthoryear{Veta et~al.}{2019}]{veta2019predicting}
\begin{barticle}[author]
\bauthor{\bsnm{Veta},~\bfnm{Mitko}\binits{M.}},
  \bauthor{\bsnm{Heng},~\bfnm{Yujing~J}\binits{Y.~J.}},
  \bauthor{\bsnm{Stathonikos},~\bfnm{Nikolas}\binits{N.}},
  \bauthor{\bsnm{Bejnordi},~\bfnm{Babak~Ehteshami}\binits{B.~E.}},
  \bauthor{\bsnm{Beca},~\bfnm{Francisco}\binits{F.}},
  \bauthor{\bsnm{Wollmann},~\bfnm{Thomas}\binits{T.}},
  \bauthor{\bsnm{Rohr},~\bfnm{Karl}\binits{K.}},
  \bauthor{\bsnm{Shah},~\bfnm{Manan~A}\binits{M.~A.}},
  \bauthor{\bsnm{Wang},~\bfnm{Dayong}\binits{D.}},
  \bauthor{\bsnm{Rousson},~\bfnm{Mikael}\binits{M.}} \betal{et~al.}
(\byear{2019}).
\btitle{Predicting breast tumor proliferation from whole-slide images: the
  TUPAC16 challenge}.
\bjournal{Medical Image Analysis}.
\end{barticle}
\endbibitem

\bibitem[\protect\citeauthoryear{Wang et~al.}{2013}]{wang2013identifying}
\begin{barticle}[author]
\bauthor{\bsnm{Wang},~\bfnm{Chao}\binits{C.}},
  \bauthor{\bsnm{P{\'e}cot},~\bfnm{Thierry}\binits{T.}},
  \bauthor{\bsnm{Zynger},~\bfnm{Debra~L}\binits{D.~L.}},
  \bauthor{\bsnm{Machiraju},~\bfnm{Raghu}\binits{R.}},
  \bauthor{\bsnm{Shapiro},~\bfnm{Charles~L}\binits{C.~L.}} \AND
  \bauthor{\bsnm{Huang},~\bfnm{Kun}\binits{K.}}
(\byear{2013}).
\btitle{Identifying survival associated morphological features of triple
  negative breast cancer using multiple datasets}.
\bjournal{Journal of the American Medical Informatics Association}
\bvolume{20}
\bpages{680--687}.
\end{barticle}
\endbibitem

\bibitem[\protect\citeauthoryear{Wang et~al.}{2016}]{wang2016deep}
\begin{barticle}[author]
\bauthor{\bsnm{Wang},~\bfnm{Dayong}\binits{D.}},
  \bauthor{\bsnm{Khosla},~\bfnm{Aditya}\binits{A.}},
  \bauthor{\bsnm{Gargeya},~\bfnm{Rishab}\binits{R.}},
  \bauthor{\bsnm{Irshad},~\bfnm{Humayun}\binits{H.}} \AND
  \bauthor{\bsnm{Beck},~\bfnm{Andrew~H}\binits{A.~H.}}
(\byear{2016}).
\btitle{Deep learning for identifying metastatic breast cancer}.
\bjournal{arXiv preprint arXiv:1606.05718}.
\end{barticle}
\endbibitem

\bibitem[\protect\citeauthoryear{Waskom et~al.}{2018}]{waskom2018seaborn}
\begin{bmisc}[author]
\bauthor{\bsnm{Waskom},~\bfnm{Michael}\binits{M.}},
  \bauthor{\bsnm{Botvinnik},~\bfnm{Olga}\binits{O.}},
  \bauthor{\bsnm{O'Kane},~\bfnm{Drew}\binits{D.}},
  \bauthor{\bsnm{Hobson},~\bfnm{Paul}\binits{P.}},
  \bauthor{\bsnm{Ostblom},~\bfnm{Joel}\binits{J.}},
  \bauthor{\bsnm{Lukauskas},~\bfnm{Saulius}\binits{S.}},
  \bauthor{\bsnm{Gemperline},~\bfnm{David~C}\binits{D.~C.}},
  \bauthor{\bsnm{Augspurger},~\bfnm{Tom}\binits{T.}},
  \bauthor{\bsnm{Halchenko},~\bfnm{Yaroslav}\binits{Y.}},
  \bauthor{\bsnm{Cole},~\bfnm{John~B.}\binits{J.~B.}},
  \bauthor{\bsnm{Warmenhoven},~\bfnm{Jordi}\binits{J.}},
  \bauthor{\bparticle{de} \bsnm{Ruiter},~\bfnm{Julian}\binits{J.}},
  \bauthor{\bsnm{Pye},~\bfnm{Cameron}\binits{C.}},
  \bauthor{\bsnm{Hoyer},~\bfnm{Stephan}\binits{S.}},
  \bauthor{\bsnm{Vanderplas},~\bfnm{Jake}\binits{J.}},
  \bauthor{\bsnm{Villalba},~\bfnm{Santi}\binits{S.}},
  \bauthor{\bsnm{Kunter},~\bfnm{Gero}\binits{G.}},
  \bauthor{\bsnm{Quintero},~\bfnm{Eric}\binits{E.}},
  \bauthor{\bsnm{Bachant},~\bfnm{Pete}\binits{P.}},
  \bauthor{\bsnm{Martin},~\bfnm{Marcel}\binits{M.}},
  \bauthor{\bsnm{Meyer},~\bfnm{Kyle}\binits{K.}},
  \bauthor{\bsnm{Miles},~\bfnm{Alistair}\binits{A.}},
  \bauthor{\bsnm{Ram},~\bfnm{Yoav}\binits{Y.}},
  \bauthor{\bsnm{Brunner},~\bfnm{Thomas}\binits{T.}},
  \bauthor{\bsnm{Yarkoni},~\bfnm{Tal}\binits{T.}},
  \bauthor{\bsnm{Williams},~\bfnm{Mike~Lee}\binits{M.~L.}},
  \bauthor{\bsnm{Evans},~\bfnm{Constantine}\binits{C.}},
  \bauthor{\bsnm{Fitzgerald},~\bfnm{Clark}\binits{C.}}, \bauthor{\bsnm{Brian}}
  \AND \bauthor{\bsnm{Qalieh},~\bfnm{Adel}\binits{A.}}
(\byear{2018}).
\btitle{Seaborn (v0.9.0)}.
\bdoi{10.5281/zenodo.1313201}
\end{bmisc}
\endbibitem

\bibitem[\protect\citeauthoryear{Weigelt et~al.}{2009}]{weigelt2009mucinous}
\begin{barticle}[author]
\bauthor{\bsnm{Weigelt},~\bfnm{Britta}\binits{B.}},
  \bauthor{\bsnm{Geyer},~\bfnm{Felipe~C}\binits{F.~C.}},
  \bauthor{\bsnm{Horlings},~\bfnm{Hugo~M}\binits{H.~M.}},
  \bauthor{\bsnm{Kreike},~\bfnm{Bas}\binits{B.}},
  \bauthor{\bsnm{Halfwerk},~\bfnm{Hans}\binits{H.}} \AND
  \bauthor{\bsnm{Reis-Filho},~\bfnm{Jorge~S}\binits{J.~S.}}
(\byear{2009}).
\btitle{Mucinous and neuroendocrine breast carcinomas are transcriptionally
  distinct from invasive ductal carcinomas of no special type}.
\bjournal{Modern Pathology}
\bvolume{22}
\bpages{1401}.
\end{barticle}
\endbibitem

\bibitem[\protect\citeauthoryear{Wein et~al.}{2017}]{wein2017clinical}
\begin{barticle}[author]
\bauthor{\bsnm{Wein},~\bfnm{Lironne}\binits{L.}},
  \bauthor{\bsnm{Savas},~\bfnm{Peter}\binits{P.}},
  \bauthor{\bsnm{Luen},~\bfnm{Stephen~J}\binits{S.~J.}},
  \bauthor{\bsnm{Virassamy},~\bfnm{Balaji}\binits{B.}},
  \bauthor{\bsnm{Salgado},~\bfnm{Roberto}\binits{R.}} \AND
  \bauthor{\bsnm{Loi},~\bfnm{Sherene}\binits{S.}}
(\byear{2017}).
\btitle{Clinical validity and utility of tumor-infiltrating lymphocytes in
  routine clinical practice for breast cancer patients: current and future
  directions}.
\bjournal{Frontiers in oncology}
\bvolume{7}
\bpages{156}.
\end{barticle}
\endbibitem

\bibitem[\protect\citeauthoryear{Whitfield
  et~al.}{2002}]{whitfield2002identification}
\begin{barticle}[author]
\bauthor{\bsnm{Whitfield},~\bfnm{Michael~L}\binits{M.~L.}},
  \bauthor{\bsnm{Sherlock},~\bfnm{Gavin}\binits{G.}},
  \bauthor{\bsnm{Saldanha},~\bfnm{Alok~J}\binits{A.~J.}},
  \bauthor{\bsnm{Murray},~\bfnm{John~I}\binits{J.~I.}},
  \bauthor{\bsnm{Ball},~\bfnm{Catherine~A}\binits{C.~A.}},
  \bauthor{\bsnm{Alexander},~\bfnm{Karen~E}\binits{K.~E.}},
  \bauthor{\bsnm{Matese},~\bfnm{John~C}\binits{J.~C.}},
  \bauthor{\bsnm{Perou},~\bfnm{Charles~M}\binits{C.~M.}},
  \bauthor{\bsnm{Hurt},~\bfnm{Myra~M}\binits{M.~M.}},
  \bauthor{\bsnm{Brown},~\bfnm{Patrick~O}\binits{P.~O.}} \betal{et~al.}
(\byear{2002}).
\btitle{Identification of genes periodically expressed in the human cell cycle
  and their expression in tumors}.
\bjournal{Molecular biology of the cell}
\bvolume{13}
\bpages{1977--2000}.
\end{barticle}
\endbibitem

\bibitem[\protect\citeauthoryear{Williams
  et~al.}{2019}]{williams2019differences}
\begin{barticle}[author]
\bauthor{\bsnm{Williams},~\bfnm{Lindsay~A}\binits{L.~A.}},
  \bauthor{\bsnm{Hoadley},~\bfnm{Katherine~A}\binits{K.~A.}},
  \bauthor{\bsnm{Nichols},~\bfnm{Hazel~B}\binits{H.~B.}},
  \bauthor{\bsnm{Geradts},~\bfnm{Joseph}\binits{J.}},
  \bauthor{\bsnm{Perou},~\bfnm{Charles~M}\binits{C.~M.}},
  \bauthor{\bsnm{Love},~\bfnm{Michael~I}\binits{M.~I.}},
  \bauthor{\bsnm{Olshan},~\bfnm{Andrew~F}\binits{A.~F.}} \AND
  \bauthor{\bsnm{Troester},~\bfnm{Melissa~A}\binits{M.~A.}}
(\byear{2019}).
\btitle{Differences in race, molecular and tumor characteristics among women
  diagnosed with invasive ductal and lobular breast carcinomas}.
\bjournal{Cancer Causes \& Control}
\bvolume{30}
\bpages{31--39}.
\end{barticle}
\endbibitem

\bibitem[\protect\citeauthoryear{Wold}{1985}]{wold1985partial}
\begin{bmisc}[author]
\bauthor{\bsnm{Wold},~\bfnm{H}\binits{H.}}
(\byear{1985}).
\btitle{Partial least squares. S. Kotz and NL Johnson (Eds.), Encyclopedia of
  statistical sciences (vol. 6)}.
\end{bmisc}
\endbibitem

\bibitem[\protect\citeauthoryear{Yang and Michailidis}{2015}]{yang2015non}
\begin{barticle}[author]
\bauthor{\bsnm{Yang},~\bfnm{Zi}\binits{Z.}} \AND
  \bauthor{\bsnm{Michailidis},~\bfnm{George}\binits{G.}}
(\byear{2015}).
\btitle{A non-negative matrix factorization method for detecting modules in
  heterogeneous omics multi-modal data}.
\bjournal{Bioinformatics}
\bvolume{32}
\bpages{1--8}.
\end{barticle}
\endbibitem

\bibitem[\protect\citeauthoryear{Yosinski
  et~al.}{2014}]{yosinski2014transferable}
\begin{binproceedings}[author]
\bauthor{\bsnm{Yosinski},~\bfnm{Jason}\binits{J.}},
  \bauthor{\bsnm{Clune},~\bfnm{Jeff}\binits{J.}},
  \bauthor{\bsnm{Bengio},~\bfnm{Yoshua}\binits{Y.}} \AND
  \bauthor{\bsnm{Lipson},~\bfnm{Hod}\binits{H.}}
(\byear{2014}).
\btitle{How transferable are features in deep neural networks?}
In \bbooktitle{Advances in neural information processing systems}
\bpages{3320--3328}.
\end{binproceedings}
\endbibitem

\bibitem[\protect\citeauthoryear{Zack, Rogers and
  Latt}{1977}]{zack1977automatic}
\begin{barticle}[author]
\bauthor{\bsnm{Zack},~\bfnm{GW}\binits{G.}},
  \bauthor{\bsnm{Rogers},~\bfnm{WE}\binits{W.}} \AND
  \bauthor{\bsnm{Latt},~\bfnm{SA}\binits{S.}}
(\byear{1977}).
\btitle{Automatic measurement of sister chromatid exchange frequency.}
\bjournal{Journal of Histochemistry \& Cytochemistry}
\bvolume{25}
\bpages{741--753}.
\end{barticle}
\endbibitem

\bibitem[\protect\citeauthoryear{Zeiler and
  Fergus}{2014}]{zeiler2014visualizing}
\begin{binproceedings}[author]
\bauthor{\bsnm{Zeiler},~\bfnm{Matthew~D}\binits{M.~D.}} \AND
  \bauthor{\bsnm{Fergus},~\bfnm{Rob}\binits{R.}}
(\byear{2014}).
\btitle{Visualizing and understanding convolutional networks}.
In \bbooktitle{European conference on computer vision}
\bpages{818--833}.
\bpublisher{Springer}.
\end{binproceedings}
\endbibitem

\end{thebibliography}


\begin{thebibliography}{6}

\bibitem[\protect\citeauthoryear{Benjamini and
  Hochberg}{1995}]{benjamini1995controlling}
\begin{barticle}[author]
\bauthor{\bsnm{Benjamini},~\bfnm{Yoav}\binits{Y.}} \AND
  \bauthor{\bsnm{Hochberg},~\bfnm{Yosef}\binits{Y.}}
(\byear{1995}).
\btitle{Controlling the false discovery rate: a practical and powerful approach
  to multiple testing}.
\bjournal{Journal of the Royal statistical society: series B (Methodological)}
\bvolume{57}
\bpages{289--300}.
\end{barticle}
\endbibitem

\bibitem[\protect\citeauthoryear{Elmore et~al.}{2015}]{elmore2015diagnostic}
\begin{barticle}[author]
\bauthor{\bsnm{Elmore},~\bfnm{Joann~G}\binits{J.~G.}},
  \bauthor{\bsnm{Longton},~\bfnm{Gary~M}\binits{G.~M.}},
  \bauthor{\bsnm{Carney},~\bfnm{Patricia~A}\binits{P.~A.}},
  \bauthor{\bsnm{Geller},~\bfnm{Berta~M}\binits{B.~M.}},
  \bauthor{\bsnm{Onega},~\bfnm{Tracy}\binits{T.}},
  \bauthor{\bsnm{Tosteson},~\bfnm{Anna~NA}\binits{A.~N.}},
  \bauthor{\bsnm{Nelson},~\bfnm{Heidi~D}\binits{H.~D.}},
  \bauthor{\bsnm{Pepe},~\bfnm{Margaret~S}\binits{M.~S.}},
  \bauthor{\bsnm{Allison},~\bfnm{Kimberly~H}\binits{K.~H.}},
  \bauthor{\bsnm{Schnitt},~\bfnm{Stuart~J}\binits{S.~J.}} \betal{et~al.}
(\byear{2015}).
\btitle{Diagnostic concordance among pathologists interpreting breast biopsy
  specimens}.
\bjournal{Jama}
\bvolume{313}
\bpages{1122--1132}.
\end{barticle}
\endbibitem

\bibitem[\protect\citeauthoryear{Feng et~al.}{2018}]{feng2018angle}
\begin{barticle}[author]
\bauthor{\bsnm{Feng},~\bfnm{Qing}\binits{Q.}},
  \bauthor{\bsnm{Jiang},~\bfnm{Meilei}\binits{M.}},
  \bauthor{\bsnm{Hannig},~\bfnm{Jan}\binits{J.}} \AND
  \bauthor{\bsnm{Marron},~\bfnm{JS}\binits{J.}}
(\byear{2018}).
\btitle{Angle-based joint and individual variation explained}.
\bjournal{Journal of multivariate analysis}
\bvolume{166}
\bpages{241--265}.
\end{barticle}
\endbibitem

\bibitem[\protect\citeauthoryear{Rosen}{2001}]{rosen2001rosen}
\begin{bbook}[author]
\bauthor{\bsnm{Rosen},~\bfnm{Paul~Peter}\binits{P.~P.}}
(\byear{2001}).
\btitle{Rosen's breast pathology}.
\bpublisher{Lippincott Williams \& Wilkins}.
\end{bbook}
\endbibitem

\bibitem[\protect\citeauthoryear{Schnitt and Collins}{2009}]{schnitt2009biopsy}
\begin{bbook}[author]
\bauthor{\bsnm{Schnitt},~\bfnm{Stuart~J}\binits{S.~J.}} \AND
  \bauthor{\bsnm{Collins},~\bfnm{Laura~C}\binits{L.~C.}}
(\byear{2009}).
\btitle{Biopsy interpretation of the breast}.
\bpublisher{Lippincott Williams \& Wilkins}.
\end{bbook}
\endbibitem

\bibitem[\protect\citeauthoryear{Seabold and
  Perktold}{2010}]{seabold2010statsmodels}
\begin{binproceedings}[author]
\bauthor{\bsnm{Seabold},~\bfnm{Skipper}\binits{S.}} \AND
  \bauthor{\bsnm{Perktold},~\bfnm{Josef}\binits{J.}}
(\byear{2010}).
\btitle{Statsmodels: Econometric and statistical modeling with python}.
In \bbooktitle{Proceedings of the 9th Python in Science Conference}
\bvolume{57}
\bpages{61}.
\bpublisher{Scipy}.
\end{binproceedings}
\endbibitem

\end{thebibliography}

\end{document}


\begin{frontmatter}
\title{Supplementary material for: joint and individual analysis of breast cancer histologic images and genomic covariates} 
\runtitle{Joint image and genetic analysis}

\begin{aug}
\author{\fnms{Iain} \snm{Carmichael}\thanksref{m1}\ead[label=e1]{idc9@uw.edu}},
\author{\fnms{Benjamin C.} \snm{Calhoun}\thanksref{m2}\ead[label=e2]{ben.calhoun@unchealth.unc.edu}},
\author{\fnms{Katherine A.} \snm{Hoadley}\thanksref{m2}\ead[label=e3]{hoadley@med.unc.edu}},
\author{\fnms{Melissa A.} \snm{Troester}\thanksref{m2}\ead[label=e4]{troester@unc.edu}},
\author{\fnms{Joseph} \snm{Geradts}\thanksref{m3}\ead[label=e5]{jgeradts@coh.org}},
\author{\fnms{Heather D.} \snm{Couture}\thanksref{m4}\ead[label=e6]{heather@pixelscientia.com}},
\author{\fnms{Linnea} \snm{Olsson}\thanksref{m2}\ead[label=e7]{lolsson@live.unc.edu}},
\author{\fnms{Charles M.} \snm{Perou}\thanksref{m2}\ead[label=e8]{cperou@med.unc.edu}},
\author{\fnms{Marc} \snm{Niethammer}\thanksref{m2}\ead[label=e9]{mn@cs.unc.edu}},
\author{\fnms{Jan} \snm{Hannig}\thanksref{m2}\ead[label=e10]{jan.hannig@unc.edu}},
\and
\author{\fnms{J.S.} \snm{Marron}\thanksref{m2}\ead[label=e11]{marron@unc.edu}}

\runauthor{I. Carmichael et al.}

\affiliation{University of Washington\thanksmark{m1}, University of North Carolina at Chapel Hill\thanksmark{m2}, City of Hope National Medical Center \thanksmark{m3}, and Pixel Scientia Labs\thanksmark{m4}}

\address{I. Carmichael\\
Department of Statistics\\
University of Washington\\
Seattle, WA, 98195 \\
\phantom{E-mail:\ }\printead*{e1}}

\address{B.C. Calhoun\\
Department of Pathology and Laboratory Medicine\\
University of North Carolina at Chapel Hill\\
Chapel Hill, NC, 27599 \\
\phantom{E-mail:\ }\printead*{e2}}

\address{K.A. Hoadley\\
Department of Genetics\\
Lineberger Comprehensive Cancer Center\\
Computational Medicine Program\\
University of North Carolina at Chapel Hill\\
Chapel Hill, NC, 27599 \\
\phantom{E-mail:\ }\printead*{e3}}

\address{M.A. Troester\\
Department of Epidemiology\\
Department of Pathology and Laboratory Medicine\\
University of North Carolina at Chapel Hill\\
Chapel Hill, NC, 27599 \\
\phantom{E-mail:\ }\printead*{e4}}

\address{J. Geradts\\
Department of Population Sciences \\
City of Hope National Medical Center\\
Duarte, CA 91010\\
\phantom{E-mail:\ }\printead*{e5}}

\address{H.D. Couture\\
Pixel Scientia Labs\\
Raleigh, NC, 27615\\
\phantom{E-mail:\ }\printead*{e6}}

\address{L. Olsson\\
Department of Epidemiology\\
University of North Carolina at Chapel Hill\\
Chapel Hill, NC, 27599 \\
\phantom{E-mail:\ }\printead*{e7}}

\address{C.M. Perou\\
Department of Genetics\\
Department of Pathology and Laboratory Medicine\\
University of North Carolina at Chapel Hill\\
Chapel Hill, NC, 27599 \\
\phantom{E-mail:\ }\printead*{e8}}

\address{M. Niethammer\\
Department of Computer Science\\
University of North Carolina at Chapel Hill\\
Chapel Hill, NC, 27599 \\
\phantom{E-mail:\ }\printead*{e9}}

\address{J. Hannig\\
Department of Statistics\\
University of North Carolina at Chapel Hill\\
Chapel Hill, NC, 27599 \\
\phantom{E-mail:\ }\printead*{e10}}

\address{J.S. Marron\\
Department of Statistics\\
University of North Carolina at Chapel Hill\\
Chapel Hill, NC, 27599\\
\phantom{E-mail:\ }\printead*{e11}}

\end{aug}


%
\end{frontmatter}

Section \ref{app:tumor_structure_examples} discusses some of the common tissue structures that play important roles in the results section of the paper.
Section \ref{app:ajive_diagnostics} provides the AJIVE diagnostic plot for the CBCS data.
Section \ref{ss:clinical_interpretation_methods} discusses the statistical methods used to compare the AJIVE scores with the clinical variables (including multiple testing control).
Additional visualizations can be found at \url{https://marronwebfiles.sites.oasis.unc.edu/AJIVE-Hist-Gene/}

\appendix

\section{Common tissue structures} \label{app:tumor_structure_examples}

Below we give examples of some of the tissue structures which are relevant to this paper.
Histopathology images are quite complex and pathologists are trained for years to interpret them\footnote{And pathologists don't always agree with each other about their interpretations \cite{elmore2015diagnostic}.}.
For a more in-depth discussion of breast cancer pathology see \cite{rosen2001rosen, schnitt2009biopsy}.

\begin{figure}[H]
\begin{subfigure}[t]{.25\textwidth}
\centering
\includegraphics[width=1\linewidth]{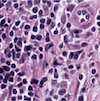}
\caption{
Lymphocytes
}
\label{fig:example_lymphocytes}
\end{subfigure}%
\begin{subfigure}[t]{.25\textwidth}
\centering
\includegraphics[width=1\linewidth]{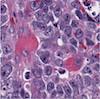}
\caption{
Tumor cells
}
\label{fig:example_high_nuclear_grade}
\end{subfigure}%
\begin{subfigure}[t]{.25\textwidth}
\centering
\includegraphics[width=1\linewidth]{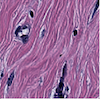}
\caption{
Collagenous stroma (1)
}
\label{fig:example_stroma_1}
\end{subfigure}%
\begin{subfigure}[t]{.25\textwidth}
\centering
\includegraphics[width=1\linewidth]{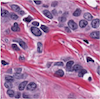}
\caption{
Collagenous stroma (2)
}
\label{fig:example_stroma_2}
\end{subfigure}
\begin{subfigure}[t]{.25\textwidth}
\centering
\includegraphics[width=1\linewidth]{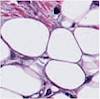}
\caption{
Fat cells (adipocytic stroma)
}
\label{fig:example_fat}
\end{subfigure}%
\begin{subfigure}[t]{.25\textwidth}
\centering
\includegraphics[width=1\linewidth]{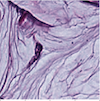}
\caption{
Mucin
}
\label{fig:example_mucin}
\end{subfigure}%
\begin{subfigure}[t]{.25\textwidth}
\centering
\includegraphics[width=1\linewidth]{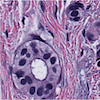}
\caption{
Duct
}
\label{fig:example_duct}
\end{subfigure}
\caption{
Examples of some of the common histology structures discussed in this paper.
}
\label{fig:LABEL}
\end{figure}

\begin{itemize}


\item Figure \ref{fig:example_lymphocytes} shows tumor infiltrating lymphocytes (TILs), the nuclei of which are basophilic (dark blue or purple) in conventional hematoxylin and eosin-stained (H\&E) tissue sections.
Hematoxylin quantitatively stains nucleic acids (DNA and RNA).
Stromal TILs are more common in certain subtypes of breast cancer and may be associated with prognosis and response to treatment. 

\item Figure \ref{fig:example_high_nuclear_grade} shows mostly high nuclear grade tumor cells (basophilic, dark blue or purple nuclei on H\&E) as well as some stroma (eosinophilic or pink on H\&E).
Nuclear grade describes how abnormal the tumor cells look: ``low grade" means the nuclei resemble those of normal cells and ``high grade" means the nuclei are enlarged, hyperchromatic (more basophilic, darker blue/purple staining than normal nuclei), irregularly shaped and may contain multiple nucleoli.
Nucleoli are small intranuclear organelles that contain DNA, RNA and protein and are responsible for the synthesis of ribosomes (ribosomes are the organelles that synthesize proteins).


\item Figure \ref{fig:example_stroma_1} shows collagenous stroma which is synthesized by fibroblasts and myofibroblasts and appears as eosinophilic (pink) fibrillar extracellular material on H\&E.
Collagenous stroma is the connective tissue that provides the scaffolding and support for epithelial structures.
The nuclei (basophilic, dark blue or purple on H\&E) of a few fibroblasts are visible in a collagenous stroma. 

\item Figure \ref{fig:example_stroma_2} shows clusters of tumor cells separated by areas or eosinophilic (pink) collagenous stroma.
On H\&E-stained tissue sections, the tumor cell nuclei and their contents are basophilic (blue/purple) and the tumor cell cytoplasm varies from pale to eosinophilic (pink).
This is a common histologic appearance of breast cancer as it invades into the stroma as aggregates or sheets of tumor cells.

\item The large white spaces in Figure  \ref{fig:example_fat}  are the cytoplasm of adipocytes, cells that synthesize lipids (i.e, fat). 
The lipid-filled cytoplasm of adipocytes appears as these optically clear areas because the solvents used in the routine preparation of H\&E tissue sections dissolve the lipids, leaving blank spaces where the lipids in the cytoplasm were.
The ratio of fatty stoma to collagenous or fibrous stroma varies with age.
Older patients will have more fatty stroma than younger patients.
This age-related decrease in breast stromal density accounts for the increased accuracy of mammography in older patients.


\item
Figure \ref{fig:example_mucin}  shows extracellular mucin, a glycoprotein produced by epithelial cells which can be present in both normal and tumor tissue.
Mucin appears almost clear or pale pink or blue in H\&E-stained tissue sections. 
Invasive breast cancers with pure mucinous histology are often low-grade and are thought to have a better prognosis than invasive ductal carcinoma of no special type. 

\item  Figure \ref{fig:example_duct}  shows a normal duct in the lower left and low cellularity invasive carcinoma in the upper right part of the image.
The benign cells contain nuclei that lack the enlargement, irregular shape and multiple nucleoli often seen in tumor cell nuclei.
The benign cells have ample pale eosinophilic (pink) cytoplasm.
The cells rest on a thin basement membrane which appears as a circumferential eosinophilic (pink) band around the periphery of the duct.
The optically clear space in the middle of the duct is the lumen. 


\end{itemize}

\begin{remark} \label{rem:term_overload}
There are a couple of terms important to this paper which are similar but have different meanings.
Clinical HER2 and molecular HER2 are two separate classifications used in breast cancer; the former is a immunohistochemical classification used in the clinic to determine clinical decision making while the latter is a genetic subtype.
High nuclear grade refers to individual cancer cells; high tumor grade is based on a composite index including nuclear grade, tubule formation and mitotic activity.
Collagenous stroma refers to (the pink) connective tissue; adipocytic stroma refers to (the white) fat cells.
\end{remark}

\section{AJIVE diagnostics} \label{app:ajive_diagnostics}

This section presents the AJIVE diagnostic plot based on the singular values of the concatenated basis matrix.
This plot is described in Section 2 of \citep{feng2018angle}.

The initial signal ranks are 81 (image features) and 30 (genes).
There were chosen by inspection of the the difference of the log-singular values and airing on the side of picking too high a rank.

Note the Wedin bound does not provide any value for these data.
This is likely due to known conservativeness of the Wedin bound for non-square matrices.
The random direction bound -- which can be seen to be equivalent to the classical Roy's latent root test for CCA rank selection -- estimates the joint rank to be 7.
This estimated joint rank was fairly robust to moderate changes in the initial signal ranks.
The image individual rank is estimated to be 76 and genetic individual rank is estimated to be 25.
These are likely overestimates, however, we focus only on the first few individual components.

\begin{figure}[h]
\centering
\includegraphics[scale=.4]{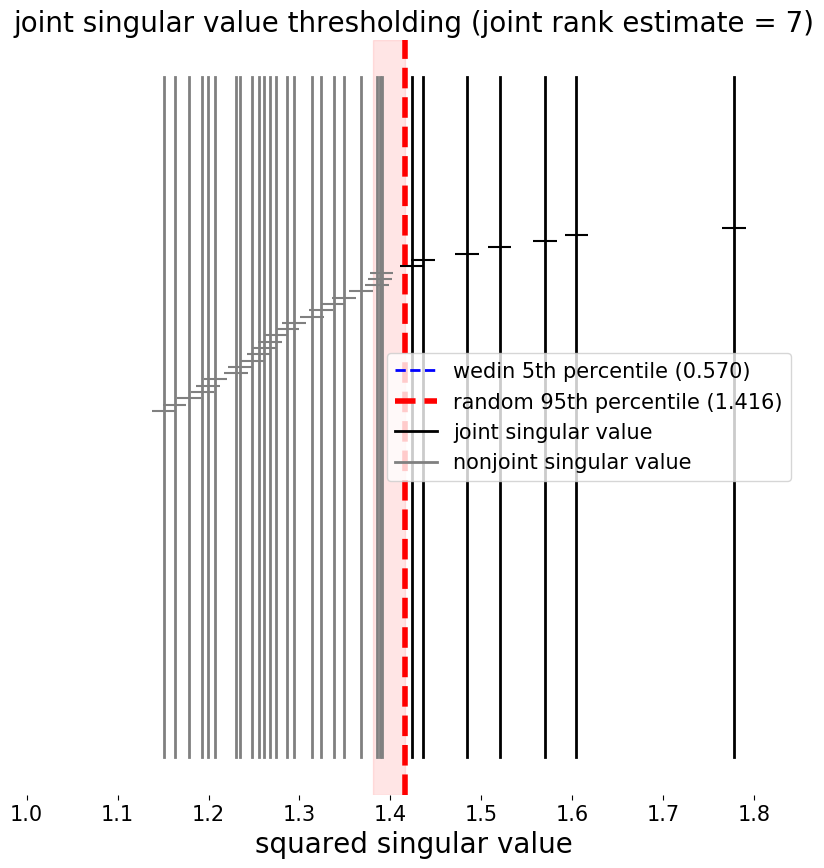}
\caption{
The AJIVE diagnostic plot shows that 7 possible joint directions are closer together than random.
The vertical black/gray line segments show the principal angles between the $X$ and $Y$ signal subspaces (on the squared singular value scale).
The vertical, red, shaded bar shows the 5-95th percentile of the distribution of leading principal angles between two random subspaces with the same dimensions as $X$ and $Y$. The dashed, vertical, red line shows the random direction cutoff.
The Wedin bound is less than one (off the figure) and is vacuous for this dataset.
The gray line segments correspond to angles larger than the cutoff while the black line segments correspond to the angles smaller than the cutoff.
}
\label{fig:patch_rep}
\end{figure}

\section{Clinical data interpretation methods}\label{ss:clinical_interpretation_methods}

In addition to the H\&E images and gene expression data, we have a variety of clinical variables which can be used to interpret the different AJIVE components (e.g. PAM50 subtype).
We compare each clinical variable of interest with the AJIVE scores for each component (i.e. the common normalized scores, image individual scores and genetic individual scores).

For continuous variables (e.g. proliferation score) we create a scatter plot, report the Pearson correlation and use the standard t-test test to determine if the association is statistically significant.
For example, Figure \ref{fig:common_1_scores_v_proliferation} shows the first joint component (x-axis) compared to the proliferation scores (y-axis).
The text in the top left reports the Pearson correlation and is bolded if the correlation is statistically significant (after correction for multiple testing).

For categorical variables we show a conditional histogram of the scores and report difference in distribution tests for each possible class comparison using the Mann Whitney U test.
This test test is used because it looks for location differences and because its test statistic is equivalent to the AUC statistic which gives an interpretable measure of how well separated two classes are.
For categorical variables with more than 2 classes (e.g. there are five PAM50 subtypes) we do all one-vs-one comparisons. 

For example, Figure \ref{fig:common_1_scores_v_pam50} shows the first joint component (x-axis) conditioned on PAM50 subtype.
The legend lists the classes, number of subjects in each class, and the the type of test.
For a given class, the legend lists the other classes which were statistically significantly separated (after multiple testing adjustment) and reports the test statistics (AUC score) in parentheses. 
The class name is bold if at least one other class is statistically significantly separated.
For example, in Figure \ref{fig:common_1_scores_v_pam50} molecular Her2 is statistically significantly separated from Basal (AUC = 0.876 ), Luminal A (AUC = 0.897) and Normal (AUC = 0.827), but not Luminal B.

We compare each of the joint and individual AJIVE components to 33 variables.
Additionally, for each multi-class categorical variable we do all of the one vs. one tests (e.g. for 5 classes we do ${5 \choose 2}$ tests).
Therefore adjustment for multiple testing is necessary to avoid spurious results.
For each of the AJIVE joint, image individual and genetic individual components we use the Benjamini-Hochberg procedure \cite{benjamini1995controlling} which is implemented in Statsmodels \cite{seabold2010statsmodels}.

Some of the clinical variables (e.g. proliferation scores, PAM50 molecular subtype) compared to the AJIVE scores are derived directly from the PAM50 gene expression which were used in the AJIVE analysis.
This raises issues related to \textit{post selection inference} when we compute p-values for these comparisons.
Because the focus of this paper is on exploratory analysis we leave these issues for follow up work. 
 
All joint, image individual and genetic individual clinical data comparisons are shown provided in \ref{supp_viz}.

\bibliographystyle{imsart-nameyear}
\bibliography{refs}